\documentclass[aps,floats]{revtex4-2}
\usepackage{amsmath,amssymb}
\usepackage{graphicx,epsfig}
\usepackage[greek,english]{babel}
\usepackage{bbold}

\begin{document}
\bibliographystyle {plain}

\pdfoutput=1
\def\oppropto{\mathop{\propto}} 
\def\opsimeq{\mathop{\simeq}}
\def\opoverderline{\mathop{\overline}}
\def\operarrow{\mathop{\longrightarrow}}
\def\opsim{\mathop{\sim}}

\def\opmin{\mathop{\min}} 
\def\opmax{\mathop{\max}} 
\def\oplim{\mathop{\lim}}

%%%%%%%%%%%%%%%%%%%%%%%%%%%%%%%%%%%%%%%%%%%%%%%%%%%%%%%%%%%%%%%%%%%%%%%%%%%%
\title{ Statistical properties of non-linear observables of fractional Gaussian fields  \\
with a focus on spatial-averaging observables and on composite operators } 

%%%%%%%%%%%%%%%%%%%%%%%%%%%%%%%%%%%%%%%%%%%%%%%%%%%%%%%%%%%%%%%%%%%%%%%%%%%%

\author{C\'ecile Monthus}
\affiliation{Universit\'e Paris-Saclay, CNRS, CEA, Institut de Physique Th\'eorique, 91191 Gif-sur-Yvette, France}

%%%%%%%%%%%%%%%%%%%%%%%%%%%%%%%%%%%%%%%%%%%%%%%%%%%%%%%%%%%%%%%%%%%%%%%%%%%%

\begin{abstract}

The statistical properties of non-linear observables of the fractional Gaussian field $\phi(\vec x)$ of negative Hurst exponent $H<0$ in dimension $d$  are revisited with a focus on spatial-averaging observables and on the properties of the finite parts $\phi_n(\vec x)$ of the ill-defined composite operators $\phi^n(\vec x) $. For the special case $n=2$ of quadratic observables, explicit results include the cumulants of arbitrary order, the L\'evy-Khintchine formula for the characteristic function and the anomalous large deviations properties. The case of observables of arbitrary order $n>2$ is analyzed via the Wiener-Ito chaos-expansion for functionals of the white noise: the multiple stochastic Ito integrals are useful to identify the finite parts $\phi_n(\vec x)$ of the ill-defined composite operators $\phi^n(\vec x) $ and to compute their correlations involving the Hurst exponents $H_n=nH$.

\end{abstract}

\maketitle

\section{ Introduction }

Scale-invariant random fields appear in many areas of statistical physics and stochastic processes
\cite{Mandelbrot_Books,Book_FF,Lamperti_processes,Lamperti_fields,reviewFGF}.
The goal of the present paper is to revisit some of their statistical properties
 via a pedagogical perspective for physicists, as will be explained in more details in subsection I.C
after recalling some basic facts in the two following subsections I.A and I.B.

\subsection{ Scale-invariant random fields of Hurst exponent $H$ either positive $H>0$ or negative $H<0$}

The Hurst exponent $H$ that characterizes the fractal scaling properties can be either positive or negative
with very different properties with respect to continuity and stationarity as we now recall.

\subsubsection{ Fractal random fields with Hurst exponent $H>0$ : 
continuous processes with stationary increments}

The positive Hurst exponent $H>0$ directly governs the H\"older regularity on short distances
so that the field is continuous, but the field cannot be statistically invariant via translation, 
and only its increments can be stationary. The simplest example in dimension $d$
is the fractional Brownian field $B_H(\vec x)$ of Hurst exponent $0<H<1$,
that can be defined as the Gaussian process with vanishing average 
${\mathbb E} \left( B_H(\vec x) \right) =0$ and with the correlation 
\begin{eqnarray}
 {\mathbb E} \left( B_H(\vec x) B_H(\vec y) \right)  = \frac{1}{2} \left( \vert \vec x \vert^{2H}
 +\vert \vec y \vert^{2H} - \vert \vec x -\vec y\vert^{2H}
 \right)
  \label{FracBrowniandd}
\end{eqnarray}
Then the variance of the increment $[B_H(  \vec x) - B_H( \vec y) ] $ 
grows as the power-law of the distance $\vert   \vec x - \vec y \vert $ 
with the positive exponent $(2H)>0$
\begin{eqnarray}
 {\mathbb E} \left( [B_H(  \vec x) - B_H(  \vec y) ]^2 \right)   
 = \gamma^2 \vert  \vec x - \vec y\vert^{2H}
  \label{FracBrowniandincrementsd}
\end{eqnarray}
As recalled in Appendix \ref{app_positiveH}, the most well-known examples 
are in dimension $d=1$
with the Brownian motion $B(x)=B_{H=\frac{1}{2}}(x)$ of Hurst exponent $H=\frac{1}{2}$
and the fractional Brownian motion $B_H(x)$ of Hurst exponent $0<H<1$
\cite{Mandelbrot1968,Mandelbrot75,fBMHistory} that has remained a very active area over the years
(see the recent works \cite{fBM_GeoOptics,fBM_PathIntegral,fBM_Benichou,fBM_Hartman,fOUconfining_Meerson,fOU_Meerson} and references therein). 

%%%%%%%%%%%%%%%%%%%%%%%%%%%%%%%%%%%%

\subsubsection{ Fractal random fields with Hurst exponent $H<0$ : 
stationary fields defined as distributions (and not pointwise)}

 When the Hurst exponent is negative $H<0$, the field can be statistically invariant via translation
 but becomes very singular on short distances and cannot be defined pointwise. 
 As recalled in Appendix \ref{app_positiveH}, the simplest example is the 
 one-dimensional fractional Gaussian noise $\frac{dB_H(x)}{dx}$ of Hurst exponent
 $H'=(H-1) \in ]-1,0[ $ obtained from the derivative the fractional Brownian motion $B_H(x)$
 of Hurst exponent $0<H<1$ discussed above. In particular for $H=\frac{1}{2}$, the derivative of the 
 Brownian motion $B(x)=B_{H=\frac{1}{2}}(x)$ of Hurst exponent $H=\frac{1}{2}$
 is the white noise $W(x)= \frac{dB(x)}{dx}$ of Hurst exponent $H'=H-1=-\frac{1}{2}$,
 that cannot be defined pointwise, since its correlation 
 \begin{eqnarray}
{\mathbb E} \left( W(x) W(y) \right) = \delta(x-y)
\label{correwhitenoised1}
\end{eqnarray}
  is the delta function
 with its well-known properties by physicists, while the appropriate rigorous mathematical framework
   is the Schwartz theory of tempered distributions.

 Another important example in statistical physics
 is the critical point of ferromagnetic models in dimension $d$, 
 where the field $\phi(\vec x) $ representing the local continuous-spin
 has a vanishing averaged value  
\begin{eqnarray}
{\mathbb E} \left( \phi(\vec x) \right) =0
\label{AvFieldZero}
\end{eqnarray}
and displays power-law decaying correlations 
 with respect to the distance $\vert \vec x-\vec y \vert $
\begin{eqnarray}
C(\vec x,\vec y)\equiv {\mathbb E} \left( \phi(\vec x)\phi(\vec y) \right)
=
\frac{ \kappa}{ \vert \vec x-\vec y \vert^{d-2+\eta}} \equiv \frac{ \kappa}{ \vert \vec x-\vec y \vert^{(-2 H) }}
\label{power-lawCorre}
\end{eqnarray}
The exponent $\eta$ is a standard notation in the area of critical phenomena to compare with the Mean-Field value $\eta^{MF}=0$,
while the corresponding negative Hurst exponent 
$ H= - \frac{d-2+\eta}{2}  <0 $ is more convenient to characterize 
directly the scaling properties of the field.
The correlation $C(\vec x,\vec y)$ of Eq. \ref{power-lawCorre}
is a function that diverges at coinciding points $ \vec x \to  \vec y $  
\begin{eqnarray}
 C(\vec x,\vec x) \equiv  
{\mathbb E} \left( \phi^2(\vec x) \right)
= + \infty
\label{power-lawCorreHDV}
\end{eqnarray}
so that the field $\phi(\vec x)$ cannot be defined pointwise and one should be careful 
when discussing the properties of the composite operators $\phi^n(\vec x) $ with $n=2,3,4...$

In summary, the fractal random fields with negative Hurst exponent $H<0$
cannot be defined as pointwise functions and 
should be interpreted mathematically as Schwartz tempered distributions.
From a more physical point of view, this means that it is important
 to focus on observables, especially those corresponding to spatial-averaging over some region
 as discussed in the next subsection.

%%%%%%%%%%%%%%%%%%%%%%%%%%%%%%%%%%%%%%%

\subsection{ Spatial-averaged observables involving scale-invariant fields of Hurst exponent $H<0$}

%%%%%%%%%%%%%%%%%%%%%%%%%%%%%%%%%%%%

\subsubsection{ The important example of the empirical magnetization $m_e$ associated to the volume $L^d$
in spin models}
 
In the context of spin models, the simplest spatial-averaged observable is the empirical magnetization associated to the volume $L^d$
 \begin{eqnarray}
m_e \equiv  \frac{1}{L^d} \int_{L^d}  d^d \vec x \ \phi(\vec x) 
\ \ \ \ \text{continuous conterpart of } \ \ \ \  m_e^{Lattice} \equiv  \frac{1}{L^d} \sum_{i=1}^{L^d} S_i 
\label{empime}
\end{eqnarray}
that belongs to the important area of sums of correlated variables, where the goal is to study generalizations of the Central-Limit-Theorem valid
for sums of independent variables (see the reviews \cite{JonaReview,JPBreview,BertinReview,SurveyQuadraticFunctionals} and references therein).
The probability distribution $p_L(m_e)$ of the empirical magnetization $m_e$ 
plays an essential role in numerical studies of phase transitions to locate the critical point
 via the famous Binder cumulant method \cite{Binder1,Binder2,BinderReview1985}.
 At the level of large deviations 
 (see the reviews \cite{oono,ellis,review_touchette} and references therein),
 the usual behavior of $p_L(m_e)$ involving the volume $L^d$ and some rate function $I(m_e)$
 characterizing how rare it is to see a given value $m_e$ different from the typical value $m_e^{typ}$
 satisfying $I(m_e^{typ} )=0 $
  \begin{eqnarray} 
 {\mathbb P}_L \left( m_e  \right) 
  \oppropto_{L \to + \infty}  e^{\displaystyle  - L^d I(m_e)  } \ \ \ \text{ for short-ranged-correlations} 
  \label{UsualLDVolume}
\end{eqnarray}
 is not valid anymore at criticality where the correlations become long-ranged with the power-laws of Eq. \ref{power-lawCorre}. These anomalous large deviations properties at criticality have been discussed in detail in a series of recent papers 
 \cite{Delamotte2022,Delamotte2024per,Delamotte2024uni,Delamotte2025,Sahu}
 based on the exact functional RG of field theory, in particular to identify the universal and non-universal properties in the critical region \cite{Delamotte2024uni,stella}.
 
 However, since the Ising critical point is exactly solvable only in dimension $d=2$,
 it is useful in arbitrary dimension $d$ to consider the simpler Gaussian models displaying scale invariance,
 in contrast to most critical points where scale-invariance appears in combination with non-Gaussian statistics.
 So in order not to mislead the readers, let us emphasize that the example of critical phenomena was mentioned
 above just to illustrate the meaning of negative Hurst exponent $H<0$ in Eq. \ref{power-lawCorre}
 and the physical relevance of observables defined via spatial-averaging in Eq. \ref{empime},
 and that from now on, we will only focus on scale-invariant Gaussian fields
 in order to analyze the properties of various observables as discussed in the next subsection.
 
 %%%%%%%%%%%%%%%%%%%%%%%%%%%%%%%%%%%%%%%%%
 
 \subsubsection{Non-linear observables involving scale-invariant Gaussian fields }
 
For the fractional Gaussian Field of Hurst exponent $H<0 $ in dimension $d$
  (see the review \cite{reviewFGF} and references therein), 
  the correlation $C(\vec x,\vec y) $ of Eq. \ref{power-lawCorre}  
  is sufficient to define the full statistics.
  Then the empirical magnetization $m_e$ of Eq. \ref{empime}
  and more generally all observables that are linear in the field $\phi$
  inherits the Gaussian character.
   It is also interesting to study non-linear observables in the field $\phi$ \cite{Dobrushin1971,Dobrushin1979,DobrushinMajor,HermiteProcesses},
   in particular quadratic observables  
  where many explicit results have been written in relation with the one-dimensional Rosenblatt process
  \cite{Taqq1975,Fox,Rosenblatt_history,Taqq2013,Rosenblatt_fields}.
  These studies use the mathematical theory of multiple stochastic integrals 
  and of the Wiener-Ito chaos expansion for functionals of the white noise
  \cite{MeyerYan,HuMeyer1988,ItoStrato} that leads to the Hida-product of distributions 
  with better properties than the Wick-product introduced in the physical field-theory literature
  (see the reviews \cite{ReviewWick1993,ReviewWick2009}).

%%%%%%%%%%%%%%%%%%%%%%%%%%%%%%%%%%%%%%%

\subsection{ Goals and organization of the paper}

Since the above results concerning non-linear observables of fractional Gaussian fields
have been obtained in the mathematical literature with the corresponding mathematical vocabulary and methods, it seems useful to revisit them via self-contained pedestrian calculations
   for statistical physicists familiar with stochastic processes.
  The goal of the present paper is thus to give an elementary unified perspective in dimension $d$, 
  and to focus on the observables corresponding to spatial-averaging
and on the properties of the finite parts $\phi_n(\vec x)$ of the ill-defined
composite operators $\phi^n(\vec x) $.

The paper is organized as follows:

$\bullet$ In section \ref{sec_corre}, we introduce the bra-ket notations familiar from quantum mechanics 
in order to analyze the scale-invariance of fields of negative Hurst exponent $H<0$ in dimension $d$
both in real-space and in Fourier-space, while the correlation matrix $\bold C$
with the power-law matrix-elements of Eq. \ref{power-lawCorre}
can be interpreted as a fractional Laplacian.
We describe the properties of linear and quadratic observables
 that depend only on the correlation matrix $\bold C$, in particular those corresponding to spatial-averaging with the kernel introduced Eq. \ref{ARkernel}.

$\bullet$ In section \ref{sec_gauss}, we turn to the case of the Fractional-Gaussian-Field of Hurst exponent $H<0$ in dimension $d$, 
where the Gaussian measure involves the inverse ${\bold C}^{-1}$ of the correlation matrix ${\bold C}$. The linear observables are then also Gaussian,
in particular the empirical magnetization associated to the volume $R^d$
that displays the anomalous large deviation behavior of Eq. \ref{GaussLinearObsmR}
with respect to the standard behavior recalled in Eq. \ref{UsualLDVolume}.

$\bullet$ In section  \ref{sec_quadratic},
we analyze the statistical properties of quadratic observables
via their generating function of Eq. \ref{Generating2fWTraces},
via the series of their cumulants (Eqs \ref{CumulantnTracesBC} \ref{CumulantnTracesBCRealSpace}
\ref{CumulantnTracesBCFourierSpace})
and via the L\'evy-Khintchine formula for their characteristic functions (Eq. \ref{Generating2fWTracesLevyKhint}).
We study the consequences for the special case of the spatial-averaging
of the finite part $\phi_2(x) \equiv \phi^2(\vec x) - {\mathbb E} \left( \phi^2(\vec x) \right) $ 
of the ill-defined composite operator $\phi^2(\vec x) $, with its 
cumulants of Eq. \ref{CumulantnTracesBCRealSpaceAR}
and its anomalous large deviation behavior of Eq. \ref{AB2expDecay}.
The conclusion is that $\phi_2(.) $ is a non-Gaussian scale-invariant field with the Hurst exponent $H_2 = 2 H$ of
 Eq. \ref{RescalFieldRealSpaceH2} and with the power-law correlation of Eq. \ref{Correphi2}.

$\bullet$ In section \ref{sec_higher}, we focus on observables of higher order $n>2$
that are rewritten as observables of order $n$ of the white noise
in order to use the theory of multiple Ito stochastic integrals summarized in Appendix \ref{app_WienerIto}.
The application to the spatial-averging of the ill-defined composite operators $\phi^n(\vec x) $
leads to the identification of their finite parts $\phi_n(\vec x)$ in Eq. \ref{phinComposite}
(with the special cases of Eq. \ref{phinComposite3} for $n=3$
and \ref{phinComposite4} for $n=4$) and to their scale invariances with the Hurst exponents $H_n=nH$
of Eq. \ref{RescalFieldRealSpaceHn} with their correlations of Eq. \ref{correphincomposite}.

$\bullet$ Our conclusions are summarized in section \ref{sec_conclusion},
while two appendices contain useful reminders :

(a) Appendix \ref{app_positiveH} contains a reminder on fractional Gaussian fields with positive Hurst exponents $H>0$ and negative Hurst exponents $H<0$ in dimension $d=1$
 in order to make the link with
the case of arbitrary dimension $d$ discussed in the main text;

(b) Appendix \ref{app_WienerIto} contains a reminder on the Wiener-Ito chaos 
expansion for functionals of the white noise
that is used in section \ref{sec_higher} of the main text.
 
%%%%%%%%%%%%%%%%%%%%%%%%%%%%%%%%%%

\section{ Correlation matrix ${\bold C}$ for scale-invariant fields with Hurst exponents $H<0$}

\label{sec_corre}

To discuss the scale-invariance properties of fields, it is useful
 to write what happens both in real-space and in Fourier-space,
so that it is convenient to use the bra-ket notations familiar from quantum mechanics 
as described in the next section.

\subsection{ Bra-ket notations to decompose the field $\vert \phi \rangle $ 
either in real-space or in Fourier-space}

As in quantum mechanics, it is convenient to use the bra-ket notations  
to denote
the real-space basis $\vert \vec x \rangle $ 
and the Fourier-basis $\vert \vec q \rangle $
satisfying the orthonormalizations
\begin{eqnarray}
 \langle \vec x \vert \vec y \rangle && = \delta^{(d)} (\vec x-\vec y)
\nonumber \\
 \langle \vec k \vert \vec q \rangle && = \delta^{(d)} (\vec k-\vec q)
\label{orthonorma}
\end{eqnarray}
and the closure relations
\begin{eqnarray}
 \int d^d \vec x \vert \vec x \rangle \langle \vec x \vert = {\mathbb 1} 
= \int d^d \vec q \vert \vec q \rangle \langle \vec q \vert
\label{closure}
\end{eqnarray}
while the unitary transformation between the two basis involves the scalar products
\begin{eqnarray}
  \langle \vec x \vert \vec q \rangle && = e^{i 2 \pi \vec q . \vec x}
\nonumber \\
 \langle \vec q \vert \vec x \rangle && =  \langle \vec x \vert \vec q \rangle^* = e^{-i 2 \pi \vec q . \vec x}
\label{scalarproducts}
\end{eqnarray}

The field $\vert \phi \rangle $ can be then expanded either in the real-space basis
\begin{eqnarray}
\vert \phi \rangle = \int d^d \vec x \vert \vec x \rangle \langle \vec x \vert \phi \rangle 
\equiv \int d^d \vec x \vert \vec x \rangle  \phi (\vec x)
\label{ketreal}
\end{eqnarray}
or in the Fourier-basis
\begin{eqnarray}
\vert \phi \rangle = \int d^d \vec q \vert \vec q \rangle \langle \vec q \vert \phi \rangle 
\equiv \int d^d \vec q \vert \vec q \rangle  {\hat \phi} (\vec q)
\label{ketfourier}
\end{eqnarray}
where the real components $ \phi (\vec x) \equiv \langle \vec x \vert \phi \rangle$ in real-space
and the complex components ${\hat \phi} (\vec q) \equiv \langle \vec q \vert \phi \rangle $ in Fourier-space
are related via the Fourier transformations based on the scalar products of Eq. \ref{scalarproducts}
\begin{eqnarray}
\phi (\vec x) && = \langle \vec x \vert \phi \rangle 
=  \int d^d \vec q \langle \vec x\vert \vec q \rangle \langle \vec q \vert \phi \rangle 
\equiv  \int d^d \vec q e^{i 2 \pi \vec q . \vec x} {\hat \phi} (\vec q) 
\nonumber \\
{\hat \phi} (\vec q) && = \langle \vec q \vert \phi \rangle
= \int d^d \vec x \langle \vec q \vert \vec x \rangle \langle \vec x \vert \phi \rangle
= \int d^d \vec x e^{- i 2 \pi \vec q . \vec x} \phi (\vec x) = {\hat \phi}^* (-\vec q)
\label{Fourierphit}
\end{eqnarray}

The correlation then corresponds to the operator
\begin{eqnarray} 
{\bold C} \equiv  {\mathbb E} \left( \vert \phi \rangle  \langle \phi \vert  \right)   
   \label{CorreOperator}
\end{eqnarray}
that can be projected either in real-space with the matrix-elements
\begin{eqnarray} 
\langle \vec x \vert {\bold C} \vert \vec y \rangle 
=  {\mathbb E} \left( \langle \vec x \vert \phi \rangle  \langle \phi \vert \vec y \rangle \right)   
=  {\mathbb E} \left( \phi(\vec x) \phi(\vec y)  \right)   \equiv C(\vec x, \vec y)
   \label{CorreOperatorRealSpace}
\end{eqnarray}
or in Fourier-space with the matrix-elements
\begin{eqnarray} 
\langle \vec q \vert {\bold C} \vert \vec k \rangle  = {\mathbb E} \left( \langle \vec q \vert \phi \rangle  \langle \phi \vert \vec k \rangle \right)   
= {\mathbb E} \left({\hat  \phi} ( \vec q ) {\hat  \phi}^* ( \vec k ) \right) \equiv {\hat C}( \vec q, \vec k)
   \label{CorreOperatorFourierSpace}
\end{eqnarray}

%%%%%%%%%%%%%%%%%%%%%%%%%%%%%%%%%

\subsection{Definition of a scale-invariant field $\phi$ with negative Hurst exponent $H \in ]-\frac{d}{2}  ,0[$ in dimension d}

 One would like the random field $\phi$ to be statistically scale-invariant
when one rescales by a factor $b$
in real-space
\begin{eqnarray} 
 \phi ( \vec x ) \opsim_{law} b^H \phi\left(\vec X = \frac{ \vec x}{ b}  \right)    
 \ \ \ \text{ with the negative Hurst exponent } \ \ H  <0
  \label{RescalFieldRealSpace}
\end{eqnarray}
or equivalently by a factor $b^{-1}$ in Fourier space 
\begin{eqnarray} 
 {\hat  \phi} ( \vec q ) \opsim_{law} b^{- \hat{H}}  {\hat  \phi} \left(\vec Q =b \vec q  \right)
 \ \ \ \text{ with the negative Hurst exponent } \ \ \hat{H} \equiv -d-H   <0
  \label{RescalFieldFourierSpace}
\end{eqnarray}

For the correlation matrix ${\bold C}$, 
this means that the real-space matrix elements $ C(\vec x,\vec y)$
should display the power-law decay of Eq. \ref{power-lawCorre},
with the corresponding behavior for
the Fourier matrix elements $ {\hat C}(\vec q,\vec k) $
\begin{eqnarray}
 {\hat C}(\vec q,\vec k) && \equiv 
 {\mathbb E} \left( {\hat  \phi} ( \vec q ) {\hat  \phi}^* ( \vec k ) \right) 
= \int d^d \vec x e^{- i 2 \pi \vec q . \vec x }  \int d^d \vec y e^{i 2 \pi \vec k . \vec y }  
   \frac{ \kappa}{ \vert \vec x-\vec y \vert^{- 2 H}}
\nonumber \\
&&    = { \hat \kappa} \frac{   \delta^{(d)} (\vec k-\vec q )}{ \vert \vec q \vert^{d+2H}}
    = { \hat \kappa}    \delta^{(d)} (\vec k-\vec q ) \vert \vec q \vert^{d+2\hat{H}}
\label{power-lawCorreFourierq1q2H}
\end{eqnarray}

In the present perspective where the real-space correlation $C(\vec x,\vec y) $ should decay as the power-law of Eq. \ref{power-lawCorre}, the natural interval for the Hurst exponent $H$ is 
\begin{eqnarray}
 - \frac{d}{2}<H<0
\label{HFourierDomain}
\end{eqnarray}
where the two boundaries are easy to understand : 

(i) The case $H= - \frac{d}{2}= \hat{H}$ corresponds to the case where the Fourier-correlation of Eq. \ref{power-lawCorreFourierq1q2H} reduces to the delta function $\delta^{(d)} (\vec k-\vec q ) $ coinciding 
with the well-known correlations of the White-Noise $W(\vec x)$ in dimension $d$
\begin{eqnarray}
 C_W (\vec x, \vec y) && = {\mathbb E} \left( W(\vec x) W(\vec y) \right)    = \delta^{(d)}(\vec x-\vec y) \ \ \ \text{ with } \ H=-\frac{d}{2}
\nonumber \\
 {\hat C}_W (\vec q, \vec k) && = {\mathbb E} \left( {\hat W}(\vec q)   {\hat W}^*(\vec k)  \right)  = \delta^{(d)}(\vec q-\vec k)\ \ \ \text{ with } \ {\hat H}=-\frac{d}{2}
  \label{CorreWhiteNoise}
\end{eqnarray}
that satisfies the rescaling properties of Eq. \ref{RescalFieldRealSpace}
and Eq. \ref{RescalFieldFourierSpace}
but that does not correspond to correlations decaying on large distances.

(ii) The strict case $H = 0$ means that the real-space-correlation $C(\vec x,\vec y) $ of Eq. \ref{power-lawCorre} does not decay anymore with the distance. 
Note that the vanishing Hurst exponent $H = 0$ can be also interpreted as
 the world of logarithmic correlations with its own very specific interesting properties 
(see the reviews \cite{reviewGaussMultiplicative,reviewlogCorre} and references therein)
that will not be discussed here,
while the case of positive Hurst exponent $H>0$ produces completely different properties 
as already mentioned in the Introduction around Eq. \ref{FracBrowniandincrementsd}
and as discussed in more details in Appendix \ref{app_positiveH} on the example of the dimension $d=1$.

%%%%%%%%%%%%%%%%%%%%%%%%%%%%%%%%%

\subsection{ Interpretation of the correlation Matrix ${\bold C}  
 \equiv {\mathbb E} \left( \vert \phi \rangle  \langle \phi \vert  \right) 
 =  (-{\bold \Delta})^{- \frac{d}{2}-H}$ as a fractional-Laplacian}
 
\label{sec_correMatrix}
 
Since the correlation matrix ${\bold C}   \equiv {\mathbb E} \left( \vert \phi \rangle  \langle \phi \vert  \right) $ 
is diagonal in Fourier-space in Eq. \ref{power-lawCorreFourierq1q2H}, with eigenvalues given by 
${ \hat \kappa}   \vert \vec q \vert^{d+2\hat{H} }$, it can be interpreted as a fractional Laplacian
as we now recall.

\subsubsection{ Reminder on the Laplacian operator ${\bold \Delta}$ and on its fractional-powers $(-{\bold \Delta})^{- \frac{\alpha}{2}} $}

The Laplacian operator ${\bold \Delta}$ is a local differential operator in real-space
\begin{eqnarray} 
\langle \vec x  \vert {\bold \Delta} \vert \vec y \rangle = \left( \sum_{\mu=1}^{d} \frac{\partial^2}{\partial x_{\mu}^2 } \right) \delta^{(d)}(\vec x-\vec y)
  \label{LaplacianReal}
\end{eqnarray}
that becomes diagonal in Fourier-space
\begin{eqnarray} 
\langle \vec k  \vert {\bold \Delta} \vert \vec q \rangle && = 
 \int d^d \vec x  \int d^d \vec y 
 \langle \vec k \vert \vec x \rangle \langle \vec x \vert {\bold \Delta} \vert \vec y \rangle \langle \vec y \vert \vec q \rangle
 =  \int d^d \vec x   e^{- i 2 \pi \vec k . \vec x}  \left( \sum_{\mu=1}^{d} \frac{\partial^2}{\partial x_{\mu}^2 } \right) 
  e^{ i 2 \pi \vec q . \vec x}
 = - 4 \pi^2 \vec q^{\ 2}  \delta^{(d)}(\vec k-\vec q)
  \label{LaplacianFourier}
\end{eqnarray}
with the negative eigenvalues $(- 4 \pi^2 \vec q^{\ 2}) \leq 0$.
As a consequence, the fractional power $ (-{\bold \Delta})^{- \frac{\alpha}{2}} $ of the opposite Laplacian $(-{\bold \Delta})$
can be defined via its diagonal matrix elements in the Fourier basis
\begin{eqnarray} 
\langle \vec k  \vert (-{\bold \Delta})^{- \frac{\alpha}{2}} \vert \vec q \rangle 
=  ( 4 \pi^2 \vec q^{\ 2} )^{- \frac{\alpha}{2}} \delta^{(d)}(\vec k-\vec q)
= \frac{ \delta^{(d)}(\vec k-\vec q) }{\vert 2 \pi \vec q \vert^{\alpha} }
  \label{LaplacianFouriers}
\end{eqnarray}

Another useful perspective for any $\alpha >0$ is the integral representation based 
on the definition of the Gamma function
\begin{eqnarray} 
(-{\bold \Delta})^{- \frac{\alpha}{2}}  
= \frac{1}{ \Gamma \left(\frac{\alpha}{2}\right) }
\int_0^{+\infty} dt t^{ \frac{\alpha}{2} -1} e^{ t {\bold \Delta} }
   \label{PowerLaplacianHeatKernel}
\end{eqnarray}
where the Heat-kernel $e^{ t {\bold \Delta} } $ 
is characterized by its diagonal matrix elements in Fourier-space
\begin{eqnarray} 
\langle \vec k  \vert e^{ t {\bold \Delta} } \vert \vec q \rangle && 
 = e^{ - t  4 \pi^2 \vec q^{\ 2} }  \delta^{(d)}(\vec k-\vec q)
  \label{HeatKernelFourier}
\end{eqnarray}
and by its well-known matrix elements in real-space
\begin{eqnarray}
\langle \vec y  \vert e^{ t {\bold \Delta} } \vert \vec x \rangle 
&& = \int d^d \vec k  \int d^d \vec q  \langle \vec y  \vert  \vec k \rangle
\langle \vec k  \vert e^{ t {\bold \Delta} } \vert \vec q \rangle 
\langle \vec q  \vert  \vec x \rangle
= \int d^d \vec k  \int d^d \vec q  e^{ i 2 \pi \vec k .  \vec y }
e^{ - t  4 \pi^2 \vec q^{\ 2} }  \delta^{(d)}(\vec k-\vec q) 
 e^{- i 2 \pi \vec q .  \vec x }
 \nonumber \\
 &&  =
  \int d^d \vec q e^{ - t  4 \pi^2 \vec q^{\ 2} }  e^{- i 2 \pi \vec q . (\vec x- \vec y) } 
  = \frac{1}{ ( 4 \pi t)^{\frac{d}{2}} } e^{ - \frac{(\vec y- \vec x)^2}{4 t } }
\label{HeatKernelRealSpace}
\end{eqnarray}

The real-space matrix-elements of the fractional Laplacian $ (-{\bold \Delta})^{- \frac{\alpha}{2}} $  
can be obtained via the Fourier transformation of the Fourier matrix-elements of Eq. \ref{LaplacianFouriers}
\begin{eqnarray}
\langle \vec y  \vert (-{\bold \Delta})^{- \frac{\alpha}{2}} \vert \vec x \rangle 
&& = \int d^d \vec k  \int d^d \vec q  \langle \vec y  \vert  \vec k \rangle
\langle \vec k  \vert (-{\bold \Delta})^{- \frac{\alpha}{2}} \vert \vec q \rangle 
\langle \vec q  \vert  \vec x \rangle
= \int d^d \vec k  \int d^d \vec q  e^{ i 2 \pi \vec k .  \vec y }
\frac{ \delta^{(d)}(\vec k-\vec q) }{\vert 2 \pi \vec q \vert^{\alpha} }
 e^{- i 2 \pi \vec q .  \vec x }
 \nonumber \\
 &&  =
  \int d^d \vec q \frac{ e^{- i 2 \pi \vec q . (\vec x- \vec y) } }{ \vert 2 \pi \vec q \vert^{\alpha} }
\label{FourierRadialPowerLaplacian}
\end{eqnarray}
In the region $0<\alpha<d $, one can instead
use the integral representation of Eq. \ref{PowerLaplacianHeatKernel}
in terms of the heat kernel with its real-space matrix elements of Eq. \ref{HeatKernelRealSpace}
and the change of variable $ u= \frac{(\vec x- \vec y)^2}{4 t }$ to obtain an explicit power-law
\begin{eqnarray}
\langle \vec y  \vert (-{\bold \Delta})^{- \frac{\alpha}{2}} \vert \vec x \rangle 
&& =  \frac{1}{ \Gamma \left(\frac{\alpha}{2}\right) }
\int_0^{+\infty} dt t^{ \frac{\alpha}{2} -1} \langle \vec y  \vert e^{ t {\bold \Delta} } \vert \vec x \rangle 
=  \frac{1}{( 4 \pi )^{\frac{d}{2}} \Gamma \left(\frac{\alpha}{2}\right) }
\int_0^{+\infty} dt t^{ \frac{\alpha-d}{2} -1}   e^{ - \frac{(\vec x- \vec y)^2}{4 t } }
\nonumber \\
&&=  \frac{1}{ 2^d \pi^{\frac{d}{2}} \Gamma \left(\frac{\alpha}{2}\right) }
\int_0^{+\infty} \frac{du}{u}  \left( \frac{(\vec x- \vec y)^2}{4 u }\right)^{ \frac{\alpha-d}{2}}   e^{ - u }
=  \frac{ 1 }{\vert \vec x -\vec y\vert^{d- \alpha} 
2^{\alpha} \pi^{\frac{d}{2}} \Gamma \left(\frac{\alpha}{2}\right) }
\int_0^{+\infty} du  u^{ \frac{d-\alpha}{2}-1}   e^{ - u }
\nonumber \\
&& 
 =
   \frac{  \gamma(d,\alpha) }{ \vert \vec x - \vec y \vert^{d- \alpha} }
\ \ \ \text{ for } \ \ 0<\alpha<d
\ \ \ \text{ with } \ \ \gamma(d,\alpha) =  \frac{  \Gamma \left( \frac{d- \alpha}{2} \right) }
{  2^{\alpha} \pi^{\frac{d}{2}} \Gamma \left( \frac{ \alpha}{2} \right)}
\label{FourierRadialPowerLaplacianalpha}
\end{eqnarray}

%%%%%%%%%%%%%%%%%%%%%%%%%%%

\subsubsection{  Correspondance with the correlation matrix ${\bold C}$ of the scale-invariant field via $\alpha=d+2H $ }

 The Fourier-space correlation ${\hat C}(\vec q,\vec k)  $ of Eq. \ref{power-lawCorreFourierq1q2H}
 is thus directly related to the Fourier matrix-elements of Eq. \ref{LaplacianFouriers} for the fractional Laplacian $(-{\bold \Delta})^{- \frac{\alpha}{2}} $ with $\alpha=d+2H $
 \begin{eqnarray} 
\langle \vec q \vert {\bold C} \vert \vec k \rangle
=  {\hat C}(\vec q,\vec k) 
 = {\hat \kappa} (2 \pi)^{d+2H} \frac{\delta^{(d)}(\vec q-\vec k)}{\vert 2 \pi \vec q \vert^{d +2 H} }
 = {\hat \kappa} (2 \pi)^{d+2H} \langle \vec q  \vert (-{\bold \Delta})^{- \frac{d}{2}-H} \vert \vec k \rangle 
  \label{CorreFracFourierHurstkappa}
\end{eqnarray}
It is then convenient to choose the prefactor
\begin{eqnarray} 
{\hat \kappa} = \frac{1}{(2 \pi)^{d+2H} }
  \label{kappaCorreFracFourierHurst}
\end{eqnarray}
in order to have the direct correspondance 
\begin{eqnarray} 
{\bold C} \equiv 
 {\mathbb E} \left( \vert \phi \rangle  \langle \phi \vert  \right) 
 =  (-{\bold \Delta})^{- \frac{d}{2}-H}
   \label{CorreFracFourier}
\end{eqnarray}
 with the Fourier matrix-elements
 \begin{eqnarray} 
\langle \vec q \vert {\bold C} \vert \vec k \rangle
=  {\hat C}(\vec q,\vec k) = \frac{\delta^{(d)}(\vec q-\vec k)}{\vert 2 \pi \vec q \vert^{d +2 H} }
  \label{CorreFracFourierHurst}
\end{eqnarray}
while the real-space correlations 
are then given by Eq. \ref{FourierRadialPowerLaplacianalpha}
 with $\alpha=d+2H$
 \begin{eqnarray}
C(\vec y, \vec x) && = \langle \vec y  \vert (-{\bold \Delta})^{- \frac{d}{2}-H} \vert \vec x \rangle 
 =
  \int d^d \vec q \frac{ e^{- i 2 \pi \vec q . (\vec x- \vec y) } }{ \vert 2 \pi \vec q \vert^{d+2H} } 
  \nonumber \\
  && = \frac{\kappa }{\vert \vec x -\vec y \vert^{-2H} }
\ \ \ \text{ for } \ \ \ \ -\frac{d}{2} < H < 0
\ \ \ \text{ with } \ \ \kappa \equiv \gamma(d,\alpha=d+2H) = \frac{  \Gamma \left( -H \right) }
{  2^{d +2 H} \pi^{\frac{d}{2}} \Gamma \left( \frac{ d }{2} +H\right)}
\label{CorreFracRealSpace}
\end{eqnarray}

%%%%%%%%%%%%%%%%%%%%%%%%%%%%%%%%%%%%%%%%%%

\subsection{ Karhunen–Loeve theorem based on the spectral decomposition of the correlation matrix ${\bold C}$ }

The Karhunen–Loève theorem based on the spectral decomposition of the correlation matrix ${\bold C}$
can be rephrased as a change of fields based on the square-root of the correlation-matrix $ {\bold C}$,
that is given by the fractional Laplacian of Eq. \ref{CorreFracFourier} in our present case
\begin{eqnarray} 
\vert \phi \rangle \equiv  
  ( {\bold C}  )^{\frac{1}{2} } \vert \varphi \rangle
=   (-{\bold \Delta})^{-\frac{d}{4}-\frac{H}{2}} \vert \varphi  \rangle 
   \label{Loeve}
\end{eqnarray}
so that the new field $\vert \varphi \rangle $ 
\begin{eqnarray} 
\vert \varphi \rangle
=  ( {\bold C}  )^{-\frac{1}{2} }\vert \phi \rangle
 =   (-{\bold \Delta})^{\frac{d}{4}+\frac{H}{2}} \vert  \phi \rangle
   \label{Loeveinv}
\end{eqnarray}
is characterized by the correlation-matrix
  \begin{eqnarray} 
{\mathbb E} \left(  \vert \varphi \rangle \langle  \varphi \vert  \right)
 =  ( {\bold C}  )^{-\frac{1}{2} } {\mathbb E} \left(  \vert \phi \rangle \langle  \phi \vert  \right) ( {\bold C}  )^{-\frac{1}{2} }
=  ( {\bold C}  )^{-\frac{1}{2} } {\bold C} ( {\bold C}  )^{-\frac{1}{2} }
=  {\mathbb 1} = {\bold C}_W
   \label{LoeveCoefCorre}
\end{eqnarray}
that coincides with the white-noise correlation matrix ${\bold C}_W $ 
whose matrix-elements in real-space and in Fourier-space have already been discussed in Eq. \ref{CorreWhiteNoise}.
 
Let us stress that one needs to distinguish two cases :

(i)  if the statistics of the scale-invariant field $\vert \phi \rangle $ is not Gaussian, 
then the field $\vert \varphi \rangle $ defined via the Karhunen–Loève expansion of Eq. \ref{Loeve} 
is not Gaussian either, so that it is different from the Gaussian White-Noise $\vert W \rangle$
even if they have the same correlation-matrix of Eq. \ref{LoeveCoefCorre}
i.e. they differ via higher-order-correlations.

(ii) if the statistics of the scale-invariant field $\vert \phi \rangle $ is Gaussian, 
then the field $\vert \varphi \rangle $ defined via the Karhunen–Loève expansion of Eq. \ref{Loeve} coincides with the White-Noise $\vert W \rangle$,
so that all the statistical properties of the Gaussian scale-invariant field $\vert \phi \rangle $
can be reformulated in terms of the statistical properties of the White-Noise $\vert W \rangle$,
as will be discussed in more details in section \ref{sec_gauss}.

%%%%%%%%%%%%%%%%%%%%%%%%%%%%%%%%%

\subsection{ Consequences for linear and quadratic observables related to spatial-averaging over a volume $R^d$ }

\subsubsection{ General properties of linear and quadratic observables   }

An observable ${\cal L}$ that is linear with respect to the field $\phi$
can be parametrized by a real function 
$L(\vec x)=\langle \vec x \vert L \rangle =\langle L \vert \vec x \rangle$ in real-space
and can be rewritten as a scalar product $\langle L \vert \phi \rangle $
\begin{eqnarray}
{\cal L}  \equiv  \langle L \vert \phi \rangle
=\int d^d \vec x L(\vec x) \phi(\vec x)  
= \int   d^d \vec q  {\hat L}^*(\vec q) {\hat \phi}(\vec q)
\label{LinearObs}
\end{eqnarray}
Its averaged value vanishes as a consequence of Eq. \ref{AvFieldZero}
\begin{eqnarray}
 {\mathbb E} \left( {\cal L} \right) && =0
 \label{O1Avzero}
\end{eqnarray}
while its variances 
can be computed in terms of the correlation matrix 
${\bold C}= {\mathbb E} \left( \vert \phi \rangle \langle \phi \vert \right)$ discussed previously
and can be evaluated either in real-space or in Fourier-space
\begin{eqnarray}
{\mathbb E} \left( {\cal L}^2 \right) && 
= {\mathbb E} \left(\langle L \vert \phi \rangle \langle \phi \vert L \rangle  \right)
 = \langle L \vert {\bold C} \vert L \rangle
 \nonumber \\
&& =  \int d^d \vec x \int d^d \vec y L(\vec x) C(\vec x, \vec y)  L(\vec y)
=  \kappa  \int d^d \vec x \int d^d \vec y  \frac{ L(\vec x) L(\vec y)}{\vert \vec x -\vec y \vert^{-2H}} 
\nonumber \\
&& = \int d^d \vec q \int d^d \vec k   {\hat L}^*(\vec q)   {\hat C}(\vec q,\vec k)  {\hat L} (\vec k) 
=   \int d^d \vec q  \frac{ {\hat L}^*(\vec q) {\hat L} (\vec q) }{\vert 2 \pi \vec q \vert^{d +2 H} }
\label{O1Var}
\end{eqnarray}

An observable ${\cal B}$ that is quadratic with respect to the field $\phi$
can be parametrized by a matrix ${\bold B} $
with its real-space matrix elements $B(\vec x, \vec y)\equiv \langle \vec x \vert {\bold B}  \vert \vec y \rangle $
or its Fourier matrix elements ${\hat B}(\vec k, \vec q) \equiv \langle \vec k \vert {\bold B}  \vert \vec q \rangle$
\begin{eqnarray} 
{\cal B} \equiv \langle \phi \vert  {\bold B}  \vert  \phi \rangle
&& =  \int d^d \vec x \int d^d \vec y \langle \phi \vert \vec x \rangle 
\langle \vec x \vert {\bold B}  \vert \vec y \rangle \langle \vec y \vert \phi \rangle
= \int d^d \vec x \int d^d \vec y \phi(\vec x) B(\vec x, \vec y) \phi(\vec y)
\nonumber \\
&& = \int d^d \vec k \int d^d \vec q \langle \phi \vert \vec k \rangle \langle \vec k \vert {\bold B}  \vert \vec q \rangle \langle \vec q \vert \phi \rangle
= \int d^d \vec k \int d^d \vec q {\hat \phi}^*(\vec k) {\hat B}(\vec k, \vec q) {\hat \phi}(\vec q)
  \label{B2quadratic}
\end{eqnarray}
The averaged value involves the correlation matrix ${\bold C}
= {\mathbb E} \left( \vert \phi \rangle \langle \phi \vert \right)$ discussed previously
and can be evaluated either in real-space or in Fourier-space
\begin{eqnarray}
 {\mathbb E} \left( {\cal B} \right) && 
 = {\mathbb E} \left( \langle \phi \vert  {\bold B}  \vert  \phi \rangle \right) 
  ={\mathbb E} \bigg( {\rm Trace} ( {\bold B}  \vert \phi \rangle \langle \phi \vert ) \bigg) 
 = {\rm Trace} ( {\bold B} {\bold C}) 
\nonumber \\
&& =  \int d^d \vec x \int d^d \vec y B(\vec x, \vec y) C(\vec y, \vec x)
=  \kappa  \int d^d \vec x \int d^d \vec y  \frac{B(\vec x, \vec y)}{\vert \vec x -\vec y \vert^{-2H}} 
\nonumber \\
&& = \int d^d \vec q \int d^d \vec k   {\hat B} (\vec k, \vec q)   {\hat C}(\vec q,\vec k) 
=   \int d^d \vec q  \frac{ {\hat B} (\vec q, \vec q)}{\vert 2 \pi \vec q \vert^{d +2 H} }
 \label{B2quadraticav}
\end{eqnarray}

Note that the difference between the quadratic observable ${\cal B}  $ 
and its averaged value ${\mathbb E}( {\cal B}  ) $
can be rewritten in terms of the difference between the operator 
$\vert  \phi \rangle\langle \phi \vert $ and its averaged value
corresponding to the correlation matrix ${\bold C} = {\mathbb E} \left( \vert \phi \rangle \langle \phi \vert \right)$
\begin{eqnarray} 
{\cal B} - {\mathbb E}( {\cal B}  )
=   {\rm Trace} \bigg( {\bold B} \big( \vert  \phi \rangle\langle \phi \vert -  {\bold C} \big) \bigg)
&& =  \int d^d \vec x \int d^d \vec y
  B(\vec y, \vec x) \bigg( \phi(\vec x) \phi(\vec y) -  C(\vec x,\vec y) \bigg)  
\nonumber \\
&& =    \int d^d \vec q \int d^d \vec k
   {\hat B} (\vec k, \vec q) \bigg( {\hat \phi}(\vec q) {\hat \phi}^*(\vec k) -  {\hat C}(\vec q,\vec k) \bigg)
 \label{ObserB2minusAv}
\end{eqnarray}

Among these linear and quadratic observables, one is particularly interested into observables 
corresponding to spatial-averaging over a volume scaling as $R^d$ as discussed in the following 
subsections.

%%%%%%%%%%%%%%%%%%%%%%%%%%%%%%%%%%%

\subsubsection{ Spatial-averaging-kernel $A_R(\vec x) =  \frac{1}{R^d} A \left( \frac{ \vec x}{R} \right) $ associated to a volume scaling as $R^d$ and to the shape $A(\vec X) $}

 Let us introduce the spatial-averaging-kernel $A_R(\vec x) $ over a volume scaling as $R^d$
 based on the shape $A(\vec x)$ centered around the origin $\vec 0$
\begin{eqnarray}
 A_R(\vec x) \equiv  \frac{1}{R^d} A \left( \frac{ \vec x}{R} \right) 
\label{ARkernel}
\end{eqnarray}
with the normalization
\begin{eqnarray}
1= \int d^d \vec x A_R(\vec x) =\int   \frac{d^d \vec x}{R^d} A \left( \frac{ \vec x}{R} \right) 
= \int   d^d \vec X A \left( \vec X \right)
\label{Avnorma}
\end{eqnarray}
and its Fourier transform
\begin{eqnarray} 
{\hat A}_R(\vec q  )  =  \int d^d \vec x    e^{- i 2 \pi \vec q  .  \vec x }A_R(\vec x) 
=   \int d^d \vec x    e^{- i 2 \pi \vec q  .  \vec x } \frac{1}{R^d} A \left( \frac{ \vec x}{R} \right) 
=  \int d^d \vec X    e^{- i 2 \pi (R \vec q ) .  \vec X } A \left( \vec X \right) = {\hat A}( R \vec q  )
 \label{ARkernelFourier}
\end{eqnarray}

Even if we will keep an arbitrary shape $A(\vec X) $ in the discussions of the present paper,
let us mention two simple examples for the shape $A(\vec X) $ centered around the origin $\vec 0$  :

(i) the shape associated to the unit box where $X_{\mu} \in ]-\frac{1}{2},+ \frac{1}{2}[$ for $\mu=1,..,d$
\begin{eqnarray}
A^{Box}(\vec X) = \prod_{\mu=1}^d  \theta \left( -\frac{1}{2} \leq X_{\mu} \leq \frac{1}{2}\right)
\label{VHard}
\end{eqnarray}
with its Fourier transform
\begin{eqnarray}
{\hat A}^{Box}(\vec Q) =\int d^d \vec X e^{- i 2 \pi \vec Q . \vec X} A^{Box}(\vec X)  
=  \prod_{\mu=1}^d  \left[ \int_{-\frac{1}{2}}^{+\frac{1}{2} } dX_{\mu} e^{- i 2 \pi Q_{\mu}  X_{\mu} } \right] 
=  \prod_{\mu=1}^d  \left[ \frac{ \sin( \pi Q_{\mu}  ) }{ \pi Q_{\mu}}  \right] 
\label{VHardFourier}
\end{eqnarray}

(ii) the Gaussian shape 
\begin{eqnarray}
A^{Gauss}(\vec X) = e^{- \pi \vec X^2} 
\label{VSmooth}
\end{eqnarray}
that has many technical advantages over the box-shape of Eq. \ref{VHard} : 
it is smooth, rotation-invariant and its Fourier transform has the same Gaussian shape
\begin{eqnarray}
{\hat A}^{Gauss}(\vec Q) =\int d^d \vec X e^{- i 2 \pi \vec Q . \vec X}  e^{- \pi \vec X^2} =  e^{- \pi \vec Q^2}
\label{VSmoothFourier}
\end{eqnarray}
The comparison with the heat-kernel $\langle \vec y  \vert e^{ t {\bold \Delta} } \vert \vec x \rangle $
of Eq. \ref{HeatKernelRealSpace} 
shows that the spatial-averaging kernel $A_R^{Gauss}(\vec x-\vec y) $
on a volume $R^d$ around the point $\vec y$
can be interpreted as the heat-kernel with the correspondance $4 \pi t = R^2 $
\begin{eqnarray}
A_R^{Gauss}(\vec x-\vec y) = \frac{1}{R^d} e^{- \pi \frac{(\vec x-\vec y)^2}{R^2}}  
  =   \langle \vec x  \vert e^{ \frac{R^2}{4 \pi}  {\bold \Delta} } \vert \vec y \rangle
\label{AHeatKernelRealSpace}
\end{eqnarray}

In the next subsections, we discuss how
the spatial-averaging-kernel $A_R(\vec x)  $ 
can be then used to construct 
spatial-averaged observables.

%%%%%%%%%%%%%%%%%%%%%%%%%%%%%%%%%%%%%%%

\subsubsection{ Empirical magnetization ${\cal M}_R$ associated to the volume $R^d$ around the origin }

The empirical magnetization ${\cal M}_R$ corresponds to the linear observable of Eq. \ref{LinearObs}
based on the spatial-averaging-kernel $L(\vec x) \to A_R(\vec x)$ 
of Eq. \ref{ARkernel}
\begin{eqnarray}
{\cal M}_R  =  \langle A_R \vert \phi \rangle = \int d^d \vec x A_R(\vec x) \phi(\vec x)  
= \int   \frac{d^d \vec x}{R^d} A \left( \frac{ \vec x}{R} \right) \phi(\vec x)
= \int   d^d \vec X A \left( \vec X \right) \phi( R \vec X)
\label{EmpiricalMagnetization}
\end{eqnarray}

The rescaling property of Eq. \ref{RescalFieldRealSpace} for the field $\phi(.)$  
yields that the empirical magnetization
\begin{eqnarray}
{\cal M}_R  && 
= \int   d^d \vec X A \left( \vec X \right) \phi( R \vec X)
 \opsim_{law} R^H \int   d^d \vec X A \left( \vec X \right) \phi(  \vec X) \equiv R^H {\cal M}_{1}
\label{EmpiricalMagnetizationScaling}
\end{eqnarray}
is statistically scale-invariant
with the same Hurst exponent $H$ as the field $\phi(.)$.
In particular, the variance of Eq. \ref{O1Var}
reads in terms of the shape $A(.)$ and of the correlation $C(.,.)$ of Eq. \ref{CorreFracRealSpace}
\begin{eqnarray}
{\mathbb E} \left( {\cal M}_R^2 \right) && 
 = \int   d^d \vec X_1 A ( \vec X_1 ) \int   d^d \vec X_2 A ( \vec X_2 )
  {\mathbb E} \left( \phi( R \vec X_1)\phi(R \vec X_2) \right)
  =  \int   d^d \vec X_1 A ( \vec X_1 ) \int   d^d \vec X_2 A ( \vec X_2 )
 C( R \vec X_1, R \vec X_2)
  \nonumber \\
 && 
  = R^{2H} \ \kappa \int   d^d \vec X_1  \int   d^d \vec X_2 
 \frac{ A ( \vec X_1 ) A ( \vec X_2 ) }{\vert \vec X_1 -\vec X_2  \vert^{- 2 H}}
  \nonumber \\
&& 
=  R^{2H} \int d^d \vec Q  \frac{ {\hat A}^*(\vec Q) {\hat A} (\vec Q) }{\vert 2 \pi \vec Q \vert^{d +2 H} }
\equiv R^{2H} {\mathbb E} \left( {\cal M}_1^2 \right) 
\label{EmpiricalMagnetizationVarC}
\end{eqnarray}

This means that the Hurst exponent $H$ is stable via spatial-averaging, 
and that the empirical magnetization ${\cal M}_{\epsilon}  $ associated to the short distance $R =\epsilon$
\begin{eqnarray}
{\cal M}_{\epsilon} &&  =  \langle A_{\epsilon} \vert \phi \rangle = \int d^d \vec x A_{\epsilon}(\vec x) \phi(\vec x)  
= \int   \frac{d^d \vec x}{\epsilon^d} A \left( \frac{ \vec x}{\epsilon} \right) \phi(\vec x)
= \int   d^d \vec X A \left( \vec X \right) \phi( \epsilon \vec X)
\nonumber \\
&& \opsim_{law} \epsilon^H \int   d^d \vec X A \left( \vec X \right) \phi(  \vec X) \equiv \epsilon^H {\cal M}_{1}
\label{EmpiricalMagnetizationScalingEpsilon}
\end{eqnarray}
can be considered as an appropriate regularization of the field $\phi$.
So whenever one encounters difficulties, one can always consider the regularization of Eq. \ref{EmpiricalMagnetizationScalingEpsilon} to understand what is going on.
However the goal is more to learn how to make computations
involving the scale-invariant $\phi$ without regularization, 
as in the area of Brownian motion where one knows how to make calculations without
returning to regularizations.

%%%%%%%%%%%%%%%%%%%%%%%%%%%%%%%%%%%%%%%%%%%

\subsubsection{ Empirical spatial-average of the fluctuating part $\phi_2(\vec x)=\phi^2(\vec x)  - {\mathbb E}( \phi^2(\vec x)  ) $ of the ill-defined composite operator $\phi^2(\vec x)$  }

The divergence of Eq. \ref{power-lawCorreHDV} means that the averaged value 
${\mathbb E} \left( \phi^2(\vec x) \right) =+\infty$
  of the composite operator $\phi^2(\vec x) $ is infinite.
It is thus convenient to introduce
 the fluctuating part of the ill-defined composite operator $\phi^2(\vec x) $ as the new-field
\begin{eqnarray}
\phi_2(x) \equiv \phi^2(\vec x) - {\mathbb E} \left( \phi^2(\vec x) \right)
\equiv \lim_{ \vec y \to \vec x} \bigg( \phi(\vec x)\phi(\vec y) - {\mathbb E} \left( \phi(\vec x)\phi(\vec y) \right) \bigg)
\label{phi2defcomposite}
\end{eqnarray}
where one recognizes the fluctuating part of the product $\phi(\vec x)\phi(\vec y) $
around its averaged value ${\mathbb E} \left( \phi(\vec x)\phi(\vec y) \right) = C(\vec x, \vec y)$
that appears in quadratic observables as discussed in Eq. \ref{ObserB2minusAv}.
So if one chooses the special case where the matrix ${\bold B} $ is diagonal in real-space
and involves the spatial-averaging kernel $A_R(\vec x)$ introduced in Eq. \ref{ARkernel}
\begin{eqnarray} 
 B_R (\vec x, \vec y) = A_R(\vec x) \delta^{(d)}( \vec x- \vec y) 
 =  \frac{1}{R^{2d} } A \left( \frac{ \vec x}{R} \right) \delta^{(d)}\left( \frac{ \vec x -\vec y}{R} \right)
 \label{BdiagralSpace}
\end{eqnarray}
then the observable of Eq. \ref{ObserB2minusAv}
\begin{eqnarray} 
{\cal B}_R - {\mathbb E}( {\cal B}_R  )
&& =  \int d^d \vec x 
  A_R(\vec x) \bigg( \phi^2(\vec x)  -  C(\vec x,\vec x) \bigg)  
  \equiv  \int d^d \vec x 
  A_R(\vec x) \phi_2(\vec x) = \langle A_R \vert \phi_2 \rangle
 \label{Generating2fWminusavBdiag}
\end{eqnarray}
represents the spatial-average of the fluctuating part $\phi_2(\vec x)$ 
of the composite operator $\phi^2(\vec x)$ over a volume scaling as $R^d$.

\subsection{ Discussion }

In this section, we have discussed some properties that depend only on the matrix correlation 
${\bold C}$. However, many other interesting issues involve higher correlations,
so that it is useful in the following sections
to focus on the case of Fractional-Gaussian-Fields in order to obtain explicit results.

%%%%%%%%%%%%%%%%%%%%%%%%%%%%%%%%%%%%%%%%%%%

\section{ Fractional-Gaussian-Field of Hurst exponent $H$ in dimension $d$ }

\label{sec_gauss}

The case where the statistics of the scale-invariant field $ \phi $ is Gaussian
is a huge simplification: 
the correlation matrix ${\bold C} = (-{\bold \Delta})^{ -\frac{d}{2}-H} $ described in the previous section
then determines the full statistics as recalled in the present section together with some important consequences.

%%%%%%%%%%%%%%%%

\subsection{ Gaussian probability distribution based on the inverse ${\bold C}^{-1} = (-{\bold \Delta})^{ \frac{d}{2}+H}  $ 
of the correlation matrix ${\bold C} $}

Since the correlation matrix corresponds to the fractional Laplacian ${\bold C} = (-{\bold \Delta})^{- \frac{d}{2}-H} $, the inverse 
${\bold C}^{-1} =  (-{\bold \Delta})^{ \frac{d}{2}+H} $ that governs
the Gaussian probability $G^{[d]}_H \left( \phi \right) $ of the field $\phi$ 
\begin{eqnarray} 
G^{[d]}_H \left( \phi \right) && \propto  \frac{1}{\sqrt{ \det [{\bold C}] } }e^{\displaystyle  - \frac{1}{2}   \langle \phi \vert {\bold C}^{-1} \vert  \phi \rangle }
= \frac{1}{\sqrt{ \det [ (-{\bold \Delta})^{ -\frac{d}{2}-H} ] } }
e^{\displaystyle  - \frac{1}{2}   \langle \phi \vert (-{\bold \Delta})^{\frac{d}{2}+H} \vert  \phi \rangle }
  \label{FracGauss}
\end{eqnarray}
 is also a fractional Laplacian that is diagonal in Fourier-space
\begin{eqnarray} 
G^{[d]}_H \left( \phi \right) &&  \propto e^{\displaystyle  - \frac{1}{2}  \int d^d \vec k
\int d^d \vec q     \langle \phi \vert \vec k \rangle \langle \vec k  \vert (-{\bold \Delta})^{\frac{d}{2}+H} \vert \vec q \rangle 
\langle  \phi \vert   }
= e^{\displaystyle  - \frac{1}{2}  \int d^d \vec q    \vert 2 \pi \vec q \vert^{d +2 H}  {\hat \phi}^*(\vec q) {\hat \phi}(\vec q)   } 
  \label{FracGaussFourier}
\end{eqnarray}
The Gaussian probability in real-space
\begin{eqnarray} 
G^{[d]}_H \left( \phi \right)   \propto  e^{\displaystyle  - \frac{1}{2}   \int d^d \vec x \int d^d \vec y  \phi ( \vec x )
 \langle \vec x  \vert (-{\bold \Delta})^{\frac{d}{2}+H} \vert \vec y \rangle  \phi ( \vec y )}
  \label{FracGaussRealSpace}
\end{eqnarray}
involves the real-space matrix elements $\langle \vec x  \vert (-{\bold \Delta})^{\frac{d}{2}+H} \vert \vec y \rangle $ obtained via the Fourier transformation
\begin{eqnarray} 
\langle \vec x  \vert (-{\bold \Delta})^{\frac{d}{2}+H} \vert \vec y \rangle
=  \int d^d \vec q     \langle \vec x \vert \vec q \rangle \vert 2 \pi \vec q \vert^{d +2 H} \langle \vec q \vert \vec y \rangle
=   \int d^d \vec q  \vert 2 \pi \vec q \vert^{d +2 H} e^{i 2 \pi  \vec q . (\vec x -\vec y) }
  \label{FracGaussKernelRealSpace}
\end{eqnarray}

In the region $-d<H< - \frac{d}{2} $, one can use the formulas of Eqs \ref{FourierRadialPowerLaplacian}
and \ref{FourierRadialPowerLaplacianalpha}
 with $\alpha=-d-2H$ and $d-\alpha=2d+2H $ to obtain the power-law-kernel
\begin{eqnarray}
\langle \vec x  \vert (-{\bold \Delta})^{\frac{d}{2}+H} \vert \vec y \rangle
&& = 
  \int d^d \vec q \frac{ e^{i 2 \pi \vec q . (\vec x -\vec y)} }{ \vert 2 \pi \vec q \vert^{-d-2H} } 
= \frac{ 2^{d +2 H} \Gamma \left( d+H \right) }
{ \vert \vec x - \vec y\vert^{2d+2H}  \pi^{\frac{d}{2}} \Gamma \left( - \frac{ d}{2} -H\right)}
\ \ \ \text{ for }  \ \ -d<H< - \frac{d}{2}
\label{FourierRadialPowerFinalappli2}
\end{eqnarray}
Note that the region $-d<H< - \frac{d}{2} $ of this simple power-law 
is complementary to the validity region $-\frac{d}{2} < H < 0 $ of 
the simple power-law for the real-space correlation function of Eq. \ref{CorreFracRealSpace}
as expected since the correlation matrix ${\bold C}$ and the Gaussian kernel ${\bold C}^{-1}$
are inverse to each-other.

%%%%%%%%%%%%%%%%%%%%%%%%%%%%%%%%%%%%%%%%%%%

\subsection{ Link with the White-Noise $W(.)$ in dimension $d$ with the negative Hurst exponents
$H=-\frac{d}{2} = {\hat H}$ }

 The special case $H=-\frac{d}{2}$ in Eq. \ref{FracGauss} 
corresponds to the Gaussian White Noise $W( \vec x)$ in dimension $d$ 
\begin{eqnarray} 
P^{[d]}_{WhiteNoise}(W) \equiv G^{[d]}_{H=-\frac{d}{2}} \left( W \right) && \propto 
e^{\displaystyle  - \frac{1}{2}   \langle W \vert  W \rangle }
 =  e^{\displaystyle  - \frac{1}{2}   \int d^d \vec x  W^2 ( \vec x ) }
= e^{\displaystyle  - \frac{1}{2}  \int d^d \vec q   {\hat W}^*(\vec q)  {\hat W}(\vec q)   }
  \label{FracGausszerowhitenoise}
\end{eqnarray}
The delta-correlations in real-space and in Fourier-space of Eq. \ref{CorreWhiteNoise}
correspond to the identity for the correlation matrix 
\begin{eqnarray}
 {\bold C}_W  = {\mathbb E} \left( \vert W \rangle \langle W \vert  \right)   && = {\mathbb 1}
   \label{CorreWhiteNoiseOp}
\end{eqnarray}

The comparison between the Gaussian distributions of
Eqs \ref{FracGauss} and \ref{FracGausszerowhitenoise} shows that the Fractional-Gaussian-field
$\vert  \phi \rangle $ can be rewritten in terms of the White-Noise $\vert  W \rangle $ as
\begin{eqnarray} 
\vert \phi \rangle =  \sqrt{ {\bold C} } \vert W \rangle
=   (-{\bold \Delta})^{-\frac{d}{4}-\frac{H}{2}} \vert  W \rangle
   \label{LoeveW}
\end{eqnarray}
in agreement with the Karhunen–Loeve theorem of Eq. \ref{Loeve} 
when $\vert  \varphi \rangle =\vert  W \rangle $.
So the White-Noise $W(.)$
can be considered as the basic building block from which all the other
Fractional-Gaussian-fields can be constructed via the application of the appropriate fractional Laplacian 
$(-{\bold \Delta})^{-\frac{d}{4}-\frac{H}{2}} $.

In the Fourier-space where the fractional Laplacian is diagonal, Eq. \ref{LoeveW}
reduces to the rescaling of the Fourier components
 \begin{eqnarray} 
{\hat \phi}(\vec q) = \langle \vec q \vert  \phi \rangle=  (4 \pi^2 \vec q^{\ 2})^{-\frac{d}{4}-\frac{H}{2}} \langle \vec q \vert  w \rangle
= \vert 2 \pi q \vert^{-\frac{d}{2}-H} {\hat W}(\vec q)
  \label{FracGaussFromWhiteNoiseFourier}
\end{eqnarray}
The real-space field $\phi(\vec x)$ can be then reconstructed either 
from the White-Noise Fourier-components ${\hat W}(\vec q) $ via the Fourier transformation
\begin{eqnarray}
\phi (\vec x)  =  \int d^d \vec q e^{i 2 \pi \vec q . \vec x} {\hat \phi} (\vec q) 
= \int d^d \vec q e^{i 2 \pi \vec q . \vec x} \vert 2 \pi q \vert^{-\frac{d}{2}-H} {\hat W}(\vec q)
\label{phixfromFourierWhiteNoise}
\end{eqnarray}
or from the White-Noise Real-space-components $ W( \vec y)$
via the convolution
 \begin{eqnarray} 
  \phi (\vec x)= \int d^d \vec y \langle \vec x \vert (-{\bold \Delta})^{-\frac{d}{4}-\frac{H}{2}} \vert \vec y \rangle W( \vec y)
  \label{FracGaussFromWhiteNoise}
\end{eqnarray}
where the matrix element of the fractional power of the opposite Laplacian
can be computed from Eq. \ref{FourierRadialPowerLaplacianalpha} with $\alpha=\frac{d}{2}+H$
and $d-\alpha=\frac{d}{2}-H$
\begin{eqnarray}
\langle \vec x \vert (-{\bold \Delta})^{-\frac{d}{4}-\frac{H}{2}} \vert \vec y \rangle
&& 
=  \int d^d \vec q \frac{ e^{i 2 \pi \vec q . (\vec x- \vec y)} }{ \vert 2 \pi \vec q \vert^{\frac{d}{2}+H} } 
   = \frac{1}{\vert \vec x \vert^{\frac{d}{2}-H}} \left(  \frac{  \Gamma \left( \frac{d}{4} -\frac{ H}{2}\right) }
{  2^{\frac{d}{2}+H} \pi^{\frac{d}{2}} \Gamma \left( \frac{d}{4} +\frac{ H}{2} \right)}
\right)
\ \ \ \text{ for } \ \ - \frac{d}{2} <H<0
\label{FourierRadialPowerFinalappli3}
\end{eqnarray}

%%%%%%%%%%%%%%%%%%%%%%%%%%%

%%%%%%%%%%%%%%%%%%%%%%%%%%%%%%%%%%%%%%%

\subsection{ Gaussian statistics of linear observables with the example of the empirical magnetization ${\cal M}_R$}

When the field $\phi$ is Gaussian,
the linear observable ${\cal L} \equiv \langle L \vert \phi \rangle $ of Eq. \ref{LinearObs}
inherits its Gaussian statistics : its probability distribution
\begin{eqnarray} 
{\mathbb P} \left({\cal L}   \right) 
  = \frac{1}{ \sqrt{ 2 \pi {\mathbb E} \left({\cal L}^2 \right)} } e^{\displaystyle  - \frac{{\cal L}^2}{2 {\mathbb E} \left({\cal L}^2 \right)  }  }
  \label{GaussLinearObs}
\end{eqnarray}
and its generating function  
\begin{eqnarray} 
{\mathbb E} \left(e^{ \displaystyle \lambda {\cal L}  } \right) 
  = e^{\displaystyle   \frac{\lambda^2}{2} {\mathbb E} \left({\cal L}^2 \right)    }
  \label{FracGaussGeneAbstract}
\end{eqnarray}
only involve the variance ${\mathbb E}\left({\cal L}^2 \right) $
already computed in Eq. \ref{O1Var}.

For the special case of the empirical magnetization ${\cal M}_R $ of Eq. \ref{EmpiricalMagnetization}
characterized by the variance ${\mathbb E} \left( {\cal M}_R^2 \right)
= R^{2H} \sigma_1^2 $ with $\sigma_1^2 \equiv  {\mathbb E} \left( {\cal M}_1^2 \right)$ of Eq. \ref{EmpiricalMagnetizationVarC},
the Gaussian probability of Eq. \ref{GaussLinearObs} 
means that the probability  ${\mathbb P}_R \left({\cal M}_R = m  \right)$ to see the empirical magnetization $ {\cal M}_R = m$ when averaging over a volume scaling
as $R^d$ reads
\begin{eqnarray} 
{\mathbb P}_R \left({\cal M}_R = m  \right) 
  = \frac{R^{-H}}{ \sqrt{ 2 \pi  \sigma_1^2} } 
  e^{\displaystyle  - R^{- 2 H} \frac{m^2}{2 \sigma_1^2  }  }
  \label{GaussLinearObsmR}
\end{eqnarray}
This scaling property means that the large deviations properties for large $R$
are governed by the unusual exponent $R^{- 2 H} $ in the exponential,
instead of the volume $R^d$ that appear in the usual large deviations of Eq. \ref{UsualLDVolume}
that is recovered here only for the Hurst exponent $H=-\frac{d}{2}$
of the White-Noise of Eq. \ref{FracGausszerowhitenoise}.

%%%%%%%%%%%%%%%%%%%%%%%%%%%%%%%%%%%%%%%%

\subsection{ Example of the free Gaussian field $\phi_f$ of Hurst exponent $H=1 - \frac{d}{2} <0$ in dimension $d>2$  }

 The case $H=1 - \frac{d}{2} <0$ involves the Laplacian ${\bold \Delta}$ with exponent unity
 in the Gaussian measure of Eq. \ref{FracGauss} 
 and thus corresponds to the well-known Gaussian-Free-Field $\phi_f$ in dimension $d>2$
\begin{eqnarray} 
P^{[d]}_{free}(\phi_f) \equiv G^{[d]}_{H=1-\frac{d}{2}} \left( \phi_f \right) 
 && \propto 
e^{\displaystyle  - \frac{1}{2}   \langle\phi_f \vert (-{\bold \Delta}) \vert \phi_f \rangle }
 =  e^{\displaystyle  - \frac{1}{2}   \int d^d \vec x  \phi_f ( \vec x )  (- {\bold \Delta}) \phi_f ( \vec x )}
 = e^{\displaystyle  - \frac{1}{2}   \int d^d \vec x  \left( \vec \nabla\phi_f ( \vec x ) \right)^2}
\nonumber \\
&& =  e^{\displaystyle  - \frac{1}{2}  \int d^d \vec q \ (4 \pi^2 \vec q^{\ 2} ){\hat\phi_f}^*(\vec q) {\hat\phi_f}(\vec q)   }
  \label{FracGaussFrees1}
\end{eqnarray}
while in $d=1$, the free Gaussian field corresponds to the 
 Brownian motion $B(x)$ of positive Hurst exponent $H=\frac{1}{2}$ as recalled around Eqs
  \ref{PBrown} and \ref{PBrownFourier}.

The correlation matrix of the free Gaussian field $\phi_f$ corresponds to the inverse $(-{\bold \Delta})^{-1} $ of the opposite Laplacian $(-{\bold \Delta}) $
i.e. the real-space matrix elements can be obtained from the special case $\alpha=2$  in Eq. \ref{FourierRadialPowerLaplacian}
\begin{eqnarray} 
 {\mathbb E} \left(\phi_f(\vec x)\phi_f(\vec y) \right) 
 =  \langle \vec x \vert (-{\bold \Delta})^{-1} \vert \vec y \rangle
  = \int d^d \vec q   \    \frac{e^{i 2 \pi \vec q . (\vec y - \vec x) }  }{ 4 \pi^2 \vec q^{\ 2} } 
  \label{FracGaussEqcorreRealSpaceFree}
\end{eqnarray}
and are given by the power-law of Eq. \ref{CorreFracRealSpace}
\begin{eqnarray}
{\mathbb E} \left(\phi_f(\vec x)\phi_f(\vec y) \right) 
&&  =    \frac{  \Gamma \left( \frac{d}{2}-1 \right) }
{ \vert \vec  x -\vec y \vert^{d-2 } 4   \pi^{1+\frac{d}{2}} }
\ \ \ \text{ for } \ \  d>2
\label{FourierRadialPowerFinalappli1Free}
\end{eqnarray}
 corresponding to the case $\eta=0$ in Eq. \ref{power-lawCorre}.

Then the probability distribution
of the empirical magnetization ${\cal M}_R $ of Eq. \ref{GaussLinearObsmR} reads
\begin{eqnarray} 
\text{ Special case } H= 1 - \frac{d}{2} \ \  : {\mathbb P}_R \left({\cal M}_R = m  \right) 
  = \frac{R^{ \frac{d-2}{2}}}{ \sqrt{ 2 \pi  \sigma_1^2} } 
  e^{\displaystyle  - R^{d-2} \frac{m^2}{2 \sigma_1^2  }  }
  \label{GaussLinearObsmRfree}
\end{eqnarray}
that involves the unusual exponent $R^{d-2} $ in the exponential,
instead of the volume $R^d$ that appear in the usual large deviations of Eq. \ref{UsualLDVolume}.

%%%%%%%%%%%%%%%%%%%%%%%%%%%

\section{ Statistics of quadratic observables ${\cal B} = \langle \phi \vert  {\bold B}  \vert  \phi \rangle $ 
of the Fractional-Gaussian-Field  }

\label{sec_quadratic}

For the Fractional-Gaussian-Field $\phi$ of Hurst exponent $H$ described in the previous section,
it is interesting to analyze the statistical properties
of quadratic observables of Eq. \ref{B2quadratic}
that can be parametrized by a symmetric operator ${\bold B}$.

%%%%%%%%%%%%%%%%%%%%%%%%%%

\subsection{ Computation of the generating function ${\mathbb E}\left( e^{ \displaystyle \lambda {\cal B} }  \right) $  }

Via the change of variables of Eq. \ref{LoeveW} towards the white-noise $W$,
 the quadratic 
observable ${\cal B} = \langle \phi \vert  {\bold B}  \vert  \phi \rangle$ of Eq. \ref{B2quadratic} 
\begin{eqnarray} 
  {\cal B}  = \langle \phi \vert  {\bold B}  \vert  \phi \rangle  
  = \langle W \vert \sqrt{ {\bold C} } {\bold B} \sqrt{ {\bold C} } \vert  W \rangle 
 \equiv
   \langle W \vert {\bold F}  \vert  W \rangle 
  \label{B2phiW}
\end{eqnarray}
becomes the quadratic observable $\langle W \vert {\bold F}  \vert  W \rangle $
of the White Noise $W(.)$ that involves the operator
\begin{eqnarray} 
{\bold F} \equiv \sqrt{ {\bold C} } {\bold B} \sqrt{ {\bold C} }
  \label{operatorF}
\end{eqnarray}
So the generating function of the observable $  {\cal B}$
 can be evaluated in terms of the ratio of two Gaussian integrals concerning the white-noise $W$
 to obtain 
\begin{eqnarray} 
  {\mathbb E}\left( e^{ \displaystyle \lambda {\cal B} }  \right) =  {\mathbb E}( e^{ \displaystyle \lambda \langle W \vert {\bold F} \vert  W \rangle }  )
 &&  = \frac{ \int {\cal D} W e^{ - \frac{1}{2}  \langle W \vert \bigg( {\mathbb 1} - \lambda 2 {\bold F} \bigg)\vert  W \rangle } }
  { \int {\cal D} W e^{ - \frac{1}{2} \langle W \vert W \rangle  } }
  = \sqrt \frac{ \det ({\mathbb 1}) }{  { \det ({\mathbb 1} - 2 \lambda {\bold F}) }}
  = e^{ \displaystyle  - \frac{1}{2} {\rm Trace}  \ln ({\mathbb 1} - 2 \lambda {\bold F})}  
  \label{Generating2fWTraces}
\end{eqnarray}
This formula involving the trace of the logarithm of the operator $({\mathbb 1} - 2 \lambda {\bold F}) $
will be used in the two next subsections,
 to compute the cumulants of arbitrary order and to analyse the infinite-divisibility properties.

%%%%%%%%%%%%%%%%%%%%%%%%%%%%%%%%%%%%%%%%%%

\subsection{ Cumulants ${\cal C}_n({\cal B} ) $ of the variable ${\cal B} $ in terms of the correlation matrix ${\bold C}$ and of the observable matrix ${\bold B}$ }

Plugging the series expansion of the logarithmic function
\begin{eqnarray} 
 - \ln (1-z) =\sum_{n=1}^{+\infty} \frac{z^n}{n}
  \label{logseries}
\end{eqnarray}
into the generating function
of Eq. \ref{Generating2fWTraces} leads to the series expansion in $\lambda$ in the exponential
\begin{eqnarray} 
  {\mathbb E}\left( e^{ \displaystyle \lambda {\cal B} }  \right)   = e^{ \displaystyle  - \frac{1}{2} {\rm Trace}  \ln ({\mathbb 1} - 2 \lambda {\bold F})}  
   && = e^{  \displaystyle \frac{1}{2}  \sum_{n=1}^{+\infty} \frac{ (2\lambda)^n}{n} {\rm Trace} ({\bold F}^n) }   
  \nonumber \\
  &&   
  = e^{  \displaystyle  \lambda {\rm Trace} ({\bold F})
  +  \lambda^2 {\rm Trace} ({\bold F}^2)
  + \frac{1}{2}  \sum_{n=3}^{+\infty} \frac{ (2\lambda)^n}{n} {\rm Trace} ({\bold F}^n) } 
  \label{Generating2fWTracesSeries}
\end{eqnarray}
 that corresponds to the expansion
 in terms of the cumulants $c_n({\cal B} ) $ of the variable ${\cal B} $
\begin{eqnarray} 
  {\mathbb E}\left( e^{ \displaystyle \lambda {\cal B} }  \right)  
     = e^{  \displaystyle    \sum_{n=1}^{+\infty} \frac{ \lambda^n}{n !} c_n({\cal B} ) } 
  \label{defcumulants}
\end{eqnarray}
The identification between the two formulas
yields that the cumulant $c_n({\cal B} ) $ of order $n$ 
involves the trace of the power $n$ of the operator $F=\sqrt{ {\bold C} } {\bold B} \sqrt{ {\bold C} }$ of Eq. \ref{operatorF} 
that can be rewritten, using the cyclic property of the trace 
\begin{eqnarray} 
c_n({\cal B} ) && =  (n-1)!   2^{n-1} {\rm Trace} ({\bold F}^n)  
= (n-1)!   2^{n-1} {\rm Trace} ( [ \sqrt{ {\bold C} } {\bold B} \sqrt{ {\bold C} } ]^n)
\nonumber \\
&& = (n-1)!   2^{n-1} {\rm Trace} ( [ {\bold B} {\bold C}]^n) 
  \label{CumulantnTracesBC}
\end{eqnarray}
in terms of the trace of the power $[ {\bold B} {\bold C}]^n $.

The evaluation based on the real-space correlation of Eq. \ref{CorreFracRealSpace}
\begin{eqnarray} 
c_n({\cal B} ) && = (n-1)!   2^{n-1} 
\int d^d \vec x_1 \int d^d \vec x_2 ...  \int d^d \vec x_{2n}
  B(\vec x_1,\vec x_2) C(\vec x_2,\vec x_3)  B(\vec x_3,\vec x_4) ...  B(\vec x_{2n-1},\vec x_{2n})C(\vec x_{2n},\vec x_1)
  \nonumber \\
  && = (n-1)!   (2 )^{n-1} \kappa^n
\int d^d \vec x_1 \int d^d \vec x_2 ...  \int d^d \vec x_{2n}
  \frac{ B(\vec x_1,\vec x_2)   B(\vec x_3,\vec x_4) ...  B(\vec x_{2n-1},\vec x_{2n}) }
  { \vert \vec x_2 -\vec x_3 \vert^{-2H} \vert \vec x_4 -\vec x_5 \vert^{-2H}...\vert \vec x_{2n} -\vec x_1 \vert^{-2H} }
  \label{CumulantnTracesBCRealSpace}
\end{eqnarray}
involves an integral over $(2n)$ variables,
while the evaluation based on the Fourier-space diagonal correlation of Eq. \ref{CorreFracFourierHurst}
\begin{eqnarray} 
c_n({\cal B} ) && = (n-1)!   2^{n-1} \int d^d \vec q_1 \int d^d \vec q_2 ...  \int d^d \vec q_{2n}
  {\hat B}(\vec q_1,\vec q_2)  {\hat C}(\vec q_2,\vec q_3)   {\hat B}(\vec q_3,\vec q_4) ...   {\hat B}(\vec q_{2n-1},\vec q_{2n}) {\hat C}(\vec q_{2n},\vec q_1)
  \nonumber \\
  && =  (n-1)!   2^{n-1} \int d^d \vec q_2 \int d^d \vec q_4 ...  \int d^d \vec q_{2n}
 \frac{ {\hat B}(\vec q_{2n},\vec q_2)    {\hat B}(\vec q_2,\vec q_4) ...   {\hat B}(\vec q_{2n-2},\vec q_{2n}) }
 { \vert 2 \pi \vec q_2 \vert^{d +2 H} \vert 2 \pi \vec q_4 \vert^{d +2 H} ... \vert 2 \pi \vec q_{2n} \vert^{d +2 H}} 
  \label{CumulantnTracesBCFourierSpace}
\end{eqnarray}
involves an integral over $n$ variables.

In particular, the first cumulant $c_{n=1}({\cal B} ) $ 
corresponding to the averaged value ${\mathbb E}( {\cal B}  ) $ 
was already mentioned in Eq. \ref{B2quadraticav},
while the second cumulant $c_{n=2}({\cal B} ) $ 
corresponding to the variance of ${\cal B} $ 
\begin{eqnarray} 
&& c_{n=2}({\cal B} ) =  {\mathbb E}(  {\cal B}^2   ) - [ {\mathbb E}( {\cal B} )]^2  
  = 2 {\rm Trace} ( {\bold B} {\bold C} {\bold B} {\bold C})
  \label{Generating2fWvarTrace}
\end{eqnarray}
can be evaluated either in real-space via an integral over four variables
\begin{eqnarray} 
 c_{n=2}({\cal B} )  
% && = 2 \int d^d \vec x_1 \int d^d \vec x_2 \int d^d \vec x_3 \int d^d \vec x_4
%  B(\vec x_1,\vec x_2) C(\vec x_2,\vec x_3)  B(\vec x_3,\vec x_4)  C(\vec x_4,\vec x_1) 
% \nonumber \\  &&  
  =  2 \kappa^2
\int d^d \vec x_1 \int d^d \vec x_2 \int d^d \vec x_3 \int d^d \vec x_4
  \frac{ B(\vec x_1,\vec x_2)   B(\vec x_3,\vec x_4)  }
  { \vert \vec x_2 -\vec x_3 \vert^{-2H} \vert \vec x_4 -\vec x_1 \vert^{-2H} }
  \label{Generating2fWvarTraceRealSpace}
\end{eqnarray}
or in Fourier-space via an integral over two variables
\begin{eqnarray} 
 c_{n=2}({\cal B} ) 
%  && = 2\int d^d \vec q_1 \int d^d \vec q_2 \int d^d \vec q_3 \int d^d \vec q_4
%  {\hat B}(\vec q_1,\vec q_2) {\hat C}(\vec q_2,\vec q_3)  {\hat B}(\vec q_3,\vec q_4)  {\hat C}(\vec q_4,\vec q_1) 
%    \nonumber \\
  && = 2\int d^d \vec q_2 \int d^d \vec q_4 
    \frac{ {\hat B}(\vec q_4,\vec q_2) {\hat B}(\vec q_2,\vec q_4)}
    {\vert 2 \pi \vec q_2 \vert^{d +2 H} \vert 2 \pi \vec q_4 \vert^{d +2 H} }
  \label{Generating2fWvarTraceFourierSpace}
\end{eqnarray}

%%%%%%%%%%%%%%%%%%%%%%%%%%%%%%%%%%%%%%%%%%%%%%%%%

%%%%%%%%%%%%%%%%%%%%%%%%%%%%%%%%%%%%%%%%%%

\subsection{ Infinite-divisibility and the L\'evy-Khintchine formula for the characteristic function 
${\mathbb E}\left( e^{ \displaystyle i \theta {\cal B} }  \right) $ }

Instead of the series expansion of Eq. \ref{logseries}, one can use the integral representation of 
the logarithmic function
\begin{eqnarray} 
 - \ln (1- i z) = \int_0^{+\infty} dv \ e^{-v} \left( \frac{ e^{i z v} -1}{v} \right)
  \label{logasintegral}
\end{eqnarray}
in the generating function of Eq. \ref{Generating2fWTraces} with $\lambda= i \theta$
to obtain that the characteristic function ${\mathbb E}\left( e^{ \displaystyle i \theta {\cal B} }  \right) $ of the quadratic observable ${\cal B} $ reads
\begin{eqnarray} 
  {\mathbb E}\left( e^{ \displaystyle i \theta {\cal B} }  \right) 
  = e^{ \displaystyle  - \frac{1}{2} {\rm Trace}  \ln ({\mathbb 1} - i 2 \theta {\bold F})}  
  =  e^{ \displaystyle   \frac{1}{2}  \int_0^{+\infty} dv \ e^{-v} \ 
  {\rm Trace} \left( \frac{ e^{i 2 v \theta {\bold F}} -1}{v} \right)}  
  \label{Generating2fWTracesLevyK}
\end{eqnarray}

Let us assume that the spectral decomposition of the symmetric operator ${\bold F} $ of Eq. \ref{operatorF}
involves discrete positive eigenvalues $f_1>f_2>... >0$ 
with the corresponding orthonormalized basis of eigenvectors $ \vert f_{\alpha} \rangle$
\begin{eqnarray} 
  {\bold F} = \sum_{\alpha=1 }^{+\infty} f_{\alpha} \vert f_{\alpha} \rangle \langle   f_{\alpha} \vert
  \label{Fspectral}
\end{eqnarray}
Then the characteristic function of Eq. \ref{Generating2fWTracesLevyK} 
can be rewritten in the L\'evy-Khintchine form
\begin{eqnarray} 
  {\mathbb E}\left( e^{ \displaystyle i \theta {\cal B} }  \right) 
&&  =  e^{ \displaystyle    \frac{1}{2}  \int_0^{+\infty} dv e^{-v} 
\sum_{\alpha=1 }^{+\infty}  \left( \frac{ e^{i 2 v \theta f_{\alpha} } -1}{v} \right)}  
\nonumber \\
&& = e^{ \displaystyle    \frac{1}{2}  \sum_{\alpha=1 }^{+\infty}  \int_0^{+\infty} du e^{- \frac{ u}{2 f_{\alpha} } }
 \left( \frac{ e^{i  \theta u } -1}{u} \right)}  
 \equiv e^{ \displaystyle     \int_0^{+\infty} du \nu_{\bold F}(u) \left(  e^{i  \theta u } -1 \right)}  
  \label{Generating2fWTracesLevyKhint}
\end{eqnarray}
where the density
\begin{eqnarray} 
 \nu_{\bold F}(u) \equiv \frac{1}{2 u }  \sum_{\alpha=1 }^{+\infty} e^{- \frac{ u}{2 f_{\alpha} } } 
  \label{DensityLevyK}
\end{eqnarray}
involves the eigenvalues $  f_{\alpha}$ of the symmetric operator ${\bold F} $ of Eq. \ref{operatorF}. 

Here the cumulants $c_n({\cal B} ) $ of Eq. \ref{CumulantnTracesBC}
are rewritten in terms of the eigenvalues $f_{\alpha} $ of $ {\bold F} = \sqrt{ {\bold C} } {\bold B} \sqrt{ {\bold C} }$ 
\begin{eqnarray} 
c_n({\cal B} ) = (n-1)!   2^{n-1} {\rm Trace} ( [ {\bold F} ]^n) 
= (n-1)!   2^{n-1} \sum_{\alpha=1 }^{+\infty} f_{\alpha}^n 
  \label{CumulantnTraceF}
\end{eqnarray}
and correspond to the moments of the density $ \nu_{\bold F}(u)  $
\begin{eqnarray} 
 \int_0^{+\infty} du \ u^n \ \nu_{\bold F}(u) 
&& = \frac{1}{2  } \sum_{\alpha=1 }^{+\infty} \int_0^{+\infty} du \ u^{n-1}  \ e^{- \frac{ u}{2 f_{\alpha} } } 
 = 2^{n-1} \sum_{\alpha=1 }^{+\infty} f_{\alpha}^n 
   \int_0^{+\infty} dv \ v^{n-1}  \ e^{- v } 
   \nonumber \\
   && = (n-1)!   2^{n-1} \sum_{\alpha=1 }^{+\infty} f_{\alpha}^n  =c_n({\cal B} )
  \label{DensityLevyKMoments}
\end{eqnarray}

The largest eigenvalue $f_{\alpha=1}$ governs the growth of the cumulants $c_n({\cal B} ) $ for large $n$
\begin{eqnarray} 
c_n({\cal B} ) \opsimeq_{n \to + \infty}  (n-1)!   2^{n-1}  f_1^n 
  \label{CumulantnTraceFlargen}
\end{eqnarray}
and the asymptotic of the density $\nu_{\bold F}(u) $ of Eq. \ref{DensityLevyK}
for large $u$
\begin{eqnarray} 
 \nu_{\bold F}(u) \opsimeq_{u \to + \infty} \frac{1}{2 u }  e^{- \frac{ u}{2 f_1 } } 
  \label{DensityLevyKlargeu}
\end{eqnarray}
so that it also governs the exponential decay of the probability distribution ${\mathbb P}\left(  {\cal B}   \right) $
for the observable ${\cal B} $ 
\begin{eqnarray} 
  {\mathbb P}\left(  {\cal B}   \right) \oppropto_{ {\cal B} \to + \infty } e^{-   \frac{ {\cal B}}{2 f_1 }}
    \label{B2expDecay}
\end{eqnarray}

Since one is only interested into the eigenvalues $f_{\alpha}$ and not in the eigenfunctions $f_{\alpha}(\vec x)$ of $ {\bold F} = \sqrt{ {\bold C} } {\bold B} \sqrt{ {\bold C} }$, one can rewrite the eigenvalue equation
for the ket $\vert f_{\alpha} \rangle $
\begin{eqnarray} 
   f_{\alpha}  \vert f_{\alpha} \rangle = {\bold F} \vert  f_{\alpha} \rangle
   = \sqrt{ {\bold C} } {\bold B} \sqrt{ {\bold C} } \vert  f_{\alpha} \rangle
  \label{Feigen}
\end{eqnarray}
as the eigenvalue equation for the new ket $\vert h_{\alpha} \rangle \equiv \sqrt{ {\bold C} } \vert  f_{\alpha} \rangle$
\begin{eqnarray} 
   f_{\alpha}  \vert h_{\alpha} \rangle 
   =  {\bold C}  {\bold B}  \vert  h_{\alpha} \rangle
  \label{Feigentilde}
\end{eqnarray}
that involves the product ${\bold C}  {\bold B} $.
The projection in real-space then involves the real-space correlation $C(\vec x, \vec y) $
of Eq. \ref{CorreFracRealSpace}
\begin{eqnarray} 
   f_{\alpha}  h_{\alpha} (\vec x) 
   = \int d^d \vec y \int d^d \vec z C(\vec x, \vec y) B(\vec y, \vec z)h_{\alpha} (\vec z)
   =  \kappa \int d^d \vec y \int d^d \vec z \frac{B(\vec y, \vec z) }{\vert \vec x -\vec y \vert^{-2H} } h_{\alpha} (\vec z)
  \label{Feigenrealspace}
\end{eqnarray}
while the projection in Fourier-space involves the diagonal Fourier-space correlation $ {\hat C}(\vec q_1,\vec q_2) $
of Eq. \ref{CorreFracFourierHurst}
\begin{eqnarray} 
   f_{\alpha}  {\hat h}_{\alpha} (\vec q) 
   = \frac{1}{\vert 2 \pi \vec q \vert^{d +2 H} }  \int d^d \vec k {\hat B}(\vec q, \vec k)   {\hat h}_{\alpha} (\vec k)   
  \label{FeigenFourier}
\end{eqnarray}

%%%%%%%%%%%%%%%%%%%%%%%%%%%%%%%%%%%%%%%%%%%%%%%

\subsection{ Rephrasing for the fluctuations of the operator $\vert  \phi \rangle\langle \phi \vert  $ 
around its averaged-value ${\mathbb E}\left(\vert  \phi \rangle\langle \phi \vert \right)={\bold C} $  }

As explained in Eq. \ref{ObserB2minusAv}, 
the difference between the observable ${\cal B}  $ and its averaged value ${\mathbb E}( {\cal B}  ) $
can be rewritten in terms of the difference between $\vert  \phi \rangle\langle \phi \vert $ and its averaged value ${\mathbb E}\left(\vert  \phi \rangle\langle \phi \vert \right)={\bold C} $
corresponding to the correlation matrix.
So the generating function of this difference $\big( \vert  \phi \rangle\langle \phi \vert -  {\bold C} \big)  $
involves the expansion of Eqs \ref{defcumulants} and \ref{CumulantnTracesBC}
in terms of the cumulants of order $n \geq 2$
\begin{eqnarray} 
  {\mathbb E}\left( e^{ \displaystyle \lambda  {\rm Trace} \bigg( {\bold B} \big( \vert  \phi \rangle\langle \phi \vert -  {\bold C} \big) \bigg) }  \right)
   = e^{  \displaystyle    \lambda^2 {\rm Trace} ([ {\bold B} {\bold C}]^2)
  + \frac{1}{2}  \sum_{n=3}^{+\infty} \frac{ (2\lambda)^n}{n} {\rm Trace} ( [ {\bold B} {\bold C}]^n) } 
  \label{Generating2fWminusav}
\end{eqnarray}
while the L\'evy-Khintchine formula
of Eq. \ref{Generating2fWTracesLevyKhint} becomes 
\begin{eqnarray} 
  {\mathbb E}\left( e^{ \displaystyle i \theta {\rm Trace} \bigg( {\bold B} \big( \vert  \phi \rangle\langle \phi \vert -  {\bold C} \big) \bigg)}  \right) 
&& = e^{ \displaystyle    \frac{1}{2}  \sum_{\alpha=1 }^{+\infty}  \int_0^{+\infty} du e^{- \frac{ u}{2 f_{\alpha} } }
 \left( \frac{ e^{i  \theta u } -1 - i \theta u}{u} \right)}  
\nonumber \\
&& \equiv e^{ \displaystyle     \int_0^{+\infty} du \nu_{\bold F}(u) \left(  e^{i  \theta u } -1 - i \theta u\right)}  
  \label{Generating2fWTracesLevyKhintav}
\end{eqnarray}
with the density $\nu_{\bold F}(u) $ of Eq. \ref{DensityLevyK}.

%%%%%%%%%%%%%%%%%%%%%%%%%%%

\subsection{ Application to the spatial-average of the fluctuating part $\phi_2(\vec x)$ of the composite operator $\phi^2(\vec x)$  }

In order to obtain the statistical properties of the observable
 $\langle A_R \vert \phi_2 \rangle $ of Eq. \ref{Generating2fWminusavBdiag}
 associated to the spatial-averaging 
 of the fluctuating part $\phi_2(\vec x)$ of the composite operator $\phi^2(\vec x)$,
 we apply the previous results to the special case of the matrix ${\bold B}_R  $ 
 with the real-space matrix elements of Eq. \ref{BdiagralSpace}
\begin{eqnarray} 
 B_R (\vec x, \vec y) = A_R(\vec x) \delta^{(d)}( \vec x- \vec y) 
 =  \frac{1}{R^{2d} } A \left( \frac{ \vec x}{R} \right) \delta^{(d)}\left( \frac{ \vec x -\vec y}{R} \right)
 \label{BdiagralSpacebis}
\end{eqnarray}
while the Fourier matrix elements read using Eq. \ref{ARkernelFourier}
\begin{eqnarray} 
{\hat B}_R (\vec q, \vec k)  
&& =  \int d^d \vec x  \int d^d \vec y  e^{- i 2 \pi \vec q .  \vec x }B_R (\vec x, \vec y)
 e^{ i 2 \pi \vec k .  \vec y }
 =  \int d^d \vec x    e^{- i 2 \pi (\vec q -\vec k) .  \vec x }A_R(\vec x) 
\nonumber \\
&& = {\hat A}_R(\vec q -\vec k ) = {\hat A}\big( R( \vec q -\vec k) \big)
 \label{BdiaginFourier}
\end{eqnarray}

%%%%%%%%%%%%%%%%%%%%%%%%%%
 
 \subsubsection{ Scaling properties of the cumulants }

For $n \geq 2$, the cumulant of order $n$ of Eq. \ref{CumulantnTracesBCRealSpace} 
\begin{eqnarray} 
c_n(\langle A_R \vert \phi_2 \rangle ) && = (n-1)!   (2 )^{n-1} \kappa^n
\int d^d \vec x_1 \int d^d \vec x_2 ...  \int d^d \vec x_{2n}
  \frac{ B_R(\vec x_1,\vec x_2)   B_R(\vec x_3,\vec x_4) ...  B_R(\vec x_{2n-1},\vec x_{2n}) }
  { \vert \vec x_2 -\vec x_3 \vert^{-2H} \vert \vec x_4 -\vec x_5 \vert^{-2H}...\vert \vec x_{2n} -\vec x_1 \vert^{-2H} }
  \nonumber \\
  && =  (n-1)!   (2 )^{n-1} \kappa^n
 \int d^d \vec x_2  \int d^d \vec x_4 ...  \int d^d \vec x_{2n}
  \frac{ A_R(\vec x_2)   A_R(\vec x_4) ...  A_R(\vec x_{2n}) }
  { \vert \vec x_2 -\vec x_4 \vert^{-2H} \vert \vec x_4 -\vec x_6 \vert^{-2H}...\vert \vec x_{2n} -\vec x_2 \vert^{-2H} }
 \nonumber \\
  && = \frac{1}{ R^{- n 2H} }  (n-1)!   (2 )^{n-1} \kappa^n
 \int d^d \vec X_1  \int d^d \vec X_2 ...  \int d^d \vec X_n
  \frac{ A(\vec X_1)   A(\vec X_2) ...  A(\vec X_n) }
  { \vert \vec X_1 -\vec X_2 \vert^{-2H} \vert \vec X_2 -\vec X_3 \vert^{-2H}...\vert \vec X_n -\vec X_1 \vert^{-2H} }  \ \ \ \ 
  \label{CumulantnTracesBCRealSpaceAR}
\end{eqnarray}
scales as $R^{n 2H}$ with respect to the size $R$,
while the prefactor involves an integral over $n$ variables $(\vec X_1,..,\vec X_n )$.

In particular for $n=2$, the second cumulant reduces to
\begin{eqnarray} 
c_{n=2}(\langle A_R \vert \phi_2 \rangle ) =  {\mathbb E}(  \langle A_R \vert \phi_2 \rangle 
\langle \phi_2 \vert A_R \rangle   )
&& = 2  \kappa^2
 \int d^d \vec x_1  \int d^d \vec x_2 
  \frac{ A_R(\vec x_1)   A_R(\vec x_2)  }
  { \vert \vec x_1 -\vec x_2 \vert^{-4H}  }
\nonumber \\
  &&   = \frac{2  \kappa^2}{ R^{- 4 H} }  \int d^d \vec X_1  \int d^d \vec X_2   \frac{ A(\vec X_1)   A(\vec X_2)  }
  { \vert \vec X_1 -\vec X_2 \vert^{-4H}  }   \ \ 
  \label{Cumulant2TracesBCRealSpaceAR}
\end{eqnarray}
This means that the real-space correlation of the field $\phi_2(.)$ 
\begin{eqnarray} 
   {\mathbb E}( \phi_2(\vec x) \phi_2(\vec y)   )  
= \langle \vec x \vert {\mathbb E}( \vert \phi_2 \rangle \langle  \phi_2 \vert   )  \vert \vec y \rangle
  = \frac {2 \kappa^2  }{ \vert \vec x-\vec y \vert^{- 4 H}} = 2 \left[ C(\vec x, \vec y)\right]^2
    \label{Correphi2}
\end{eqnarray}
is simply the square of the real-space correlation $C(\vec x, \vec y)= \frac{\kappa }{\vert \vec x -\vec y \vert^{-2H} }$ of Eq. \ref{CorreFracRealSpace}
concerning the initial field $\phi(.)$.

In conclusion, the scaling properties with $R$ of the cumulants $c_n(\langle A_R \vert \phi_2 \rangle ) $ of arbitrary order $n \geq 2$
of Eq. \ref{CumulantnTracesBCRealSpaceAR} show
that $\phi_2(.) $ is a non-Gaussian scale-invariant field 
\begin{eqnarray} 
 \phi_2 ( \vec x ) \opsim_{law} b^{H_2} \phi_2\left( \frac{ \vec x}{ b}  \right)    
 \ \ \ \text{ with the Hurst exponent } \ \ \ H_2 = 2 H
  \label{RescalFieldRealSpaceH2}
\end{eqnarray}
The  convergence region of Eq. \ref{CorreFracRealSpace}
\begin{eqnarray} 
-\frac{d}{2} <H_2=2H <0 \ \ \ \ \text{ corresponds to } \ \ -\frac{d}{4} <H <0
  \label{regioncomposite2}
\end{eqnarray}
i.e. to the upper-half of the convergence region $-\frac{d}{2} <H<0 $ associated to the correlation $C(\vec x,\vec y) $ of Eq. \ref{CorreFracRealSpace},
while in the lower-half $-\frac{d}{2} <H< -\frac{d}{4} $, 
the integral in Eq. \ref{Cumulant2TracesBCRealSpaceAR} diverges.

%%%%%%%%%%%%%%%%%%%%%%%%%%%%%

 \subsubsection{ Properties of the Levy-Khintchine formula for the characteristic function of 
 $\langle A_R \vert \phi_2 \rangle$ }

 Plugging $B_R (\vec x, \vec y) $ of Eq. \ref{BdiagralSpacebis}
 into the real-space eigenvalue Equation \ref{Feigenrealspace} yields
\begin{eqnarray} 
   f_{\alpha}^{[R]}  h^{[R]}_{\alpha} (\vec x) 
 && =  \kappa \int d^d \vec y \int d^d \vec z \frac{B_R(\vec y, \vec z) }{\vert \vec x -\vec y \vert^{-2H} }
   h^{[R]}_{\alpha} (\vec z)
  =  \kappa \int d^d \vec y \int d^d \vec z \frac{A_R(\vec y) \delta^{(d)}( \vec y- \vec z) }{\vert \vec x -\vec y \vert^{-2H} }   h^{[R]}_{\alpha} (\vec z)
  \nonumber \\
  && = \kappa \int d^d \vec y  \frac{A_R(\vec y)  }{\vert \vec x -\vec y \vert^{-2H} }   h^{[R]}_{\alpha} (\vec y)
  =  \kappa \int d^d \vec y  \frac{ \frac{1}{R^{d} } A \left( \frac{ \vec y}{R} \right)  }{\vert \vec x -\vec y \vert^{-2H} }   h^{[R]}_{\alpha} (\vec y)
  =  \kappa \int d^d \vec Y  \frac{ A (\vec Y)  }{\vert \vec x - R \vec Y \vert^{-2H} }   h^{[R]}_{\alpha} (R \vec Y)
  \label{FeigenrealspaceBR}
\end{eqnarray}
As a consequence, the dependance with respect to the scale $R$ can be taken into account via
the rescaling of the eigenvalues $ f^{[R]}_{\alpha}$ and of the eigenfunctions $h_{\alpha}^{[R]} (\vec x) $
\begin{eqnarray} 
   f^{[R]}_{\alpha} && =  R^{2H} F_{\alpha}
   \nonumber \\
   h^{[R]}_{\alpha} (\vec x) && =  H_{\alpha}  \left( \frac{ \vec x}{R} \right) 
  \label{Feigenrescalx}
\end{eqnarray}
 and one obtains the $R$-independent eigenvalue equations for the rescaled eigenvalues $F_{\alpha} $ and the rescaled eigenfunctions $H_{\alpha}  ( \vec X ) $
\begin{eqnarray} 
 F_{\alpha} H_{\alpha}  ( \vec X )  
  = \kappa \int d^d \vec Y  \frac{ A (\vec Y)  }{\vert  \vec X -  \vec Y \vert^{-2H} }   H_{\alpha}  ( \vec Y) 
  \label{FeigenrealspaceBRrescal}
\end{eqnarray}

Equivalently, one can plug ${\hat B}_R (\vec q, \vec k) = {\hat A}\big( R( \vec q -\vec k) \big)$ of Eq. \ref{BdiaginFourier}
into the Fourier-space eigenvalue Equation \ref{FeigenFourier}
\begin{eqnarray} 
   f_{\alpha}^{[R]}  {\hat h}^{[R]}_{\alpha} (\vec q) 
   =\frac{1}{\vert 2 \pi \vec q \vert^{d +2 H} }
     \int d^d \vec k {\hat B}_R(\vec q, \vec k)   {\hat h}^{[R]}_{\alpha} (\vec k)  
   = \frac{1}{\vert 2 \pi \vec q \vert^{d +2 H} }  \int d^d \vec k  {\hat A}\big( R( \vec q -\vec k) \big)  {\hat h}^{[R]}_{\alpha} (\vec k) 
  \label{FeigenFourierBR}
\end{eqnarray}
to obtain, via the rescaling of Eq. \ref{Feigenrescalx} that translates into
\begin{eqnarray} 
{\hat h}^{[R]}_{\alpha} (\vec q) = \int d^d \vec x e^{- i 2 \pi \vec q .  \vec x }
   h^{[R]}_{\alpha} (\vec x)  = \int d^d \vec x e^{- i 2 \pi \vec q .  \vec x } H_{\alpha}  \left( \frac{ \vec x}{R} \right) 
   = R^d \int d^d \vec X e^{- i 2 \pi (R \vec q) .  \vec X } H_{\alpha}  \left( \vec X \right) 
   = R^d {\hat H}_{\alpha} (R \vec q )
  \label{Feigenrescalq}
\end{eqnarray}
that ${\hat H}_{\alpha} ( \vec Q ) $ satisfies the $R$-independent eigenvalue equation
\begin{eqnarray} 
 F_{\alpha} {\hat H}_{\alpha}  ( \vec Q )  
  = \frac{ 1}{\vert 2 \pi \vec Q \vert^{d +2 H} }   \int d^d \vec K  {\hat A}\big(  \vec Q -\vec K \big)   {\hat H}_{\alpha} (\vec K) 
  \label{FeigenFourierBRrescal}
\end{eqnarray}
associated to the rescaled eigenvalue $F_{\alpha} $.

In conclusion, the eigenvalues $ f^{[R]}_{\alpha}  =  R^{2H} F_{\alpha}$ can be plugged into the cumulants
of Eq. \ref{CumulantnTraceF}
 \begin{eqnarray} 
c_n(\langle A_R \vert \phi_2 \rangle ) 
= (n-1)!   2^{n-1} \sum_{\alpha=1 }^{+\infty} \left[ f_{\alpha}^{[R]} \right]^n 
=R^{n 2H}  (n-1)!   2^{n-1} \sum_{\alpha=1 }^{+\infty} F_{\alpha}^n 
  \label{CumulantnTraceFR}
\end{eqnarray}
to recover the scaling as $R^{n 2H} $ discussed in Eq. \ref{CumulantnTracesBCRealSpaceAR},
while the prefactors now involve the rescaled eigenvalues $F_{\alpha} $.

The eigenvalues $ f^{[R]}_{\alpha}  =  R^{2H} F_{\alpha}$ can also be plugged into the characteristic function of Eq. \ref{Generating2fWTracesLevyKhintav}
to obtain via the change of variables $ u=R^{2H} v$
\begin{eqnarray} 
  {\mathbb E}\left( e^{ \displaystyle i \theta \langle A_R \vert \phi_2 \rangle}  \right) 
&& = e^{ \displaystyle    \frac{1}{2}  \sum_{\alpha=1 }^{+\infty}  \int_0^{+\infty} du e^{- \frac{ u}{2 f_{\alpha}^{[R]} } }
 \left( \frac{ e^{i  \theta u } -1 - i \theta u}{u} \right)}  
 = e^{ \displaystyle    \frac{1}{2}  \sum_{\alpha=1 }^{+\infty}  \int_0^{+\infty} du 
e^{- \frac{ u}{2 R^{2H} F_{\alpha} } }
 \left( \frac{ e^{i  \theta u } -1 - i \theta u}{u} \right)}
 \nonumber \\
&& =   e^{ \displaystyle    \frac{1}{2}  \sum_{\alpha=1 }^{+\infty}  \int_0^{+\infty} dv 
e^{- \frac{ v}{2  F_{\alpha} } }
 \left( \frac{ e^{i ( \theta R^{2H}) v } -1 - i (\theta R^{2H}) v}{v} \right)} 
 =  {\mathbb E}\left( e^{ \displaystyle i ( \theta R^{2H} ) \langle A \vert \phi_2 \rangle}  \right) 
  \label{Generating2fWTracesLevyKhintavBR}
\end{eqnarray}
that the characteristic function ${\mathbb E}\left( e^{ \displaystyle i \theta \langle A_R \vert \phi_2 \rangle}  \right) $
depends only on the rescaled variable $\Theta= \theta R^{2H} $,
in agreement with the scaling properties of Eq. \ref{RescalFieldRealSpaceH2}
the field $  \phi_2 ( \vec x ) $.

Equivalently, the probability ${\mathbb P}_R\left(  \langle A_R \vert \phi_2 \rangle =\varphi \right) $
 to see the value $\langle A_R \vert \phi_2 \rangle=\varphi$ follows the scaling form
 \begin{eqnarray} 
  {\mathbb P}_R\left(  \langle A_R \vert \phi_2 \rangle =\varphi \right) =  \frac{1}{R^{2H} } {\mathbb P}_1\left(  \frac{\varphi}{R^{2H}} \right)
    \label{ScalingAphi2}
\end{eqnarray}
with the characteristic function of Eq. \ref{Generating2fWTracesLevyKhintavBR}
\begin{eqnarray} 
  {\mathbb E}\left( e^{ \displaystyle i \theta \langle A_R \vert \phi_2 \rangle}  \right) 
&& = \int d\varphi   {\mathbb P}_R\left( \varphi \right) e^{i \theta \varphi}
= \int  \frac{d\varphi}{R^{2H} } {\mathbb P}_1\left(  \frac{\varphi}{R^{2H}} \right)e^{i \theta \varphi}
=  \int  d \Phi {\mathbb P}_1\left(  \Phi \right)e^{i (\theta R^{2H}) \Phi}
  \label{ScalingAphi2Fourier}
\end{eqnarray}
leading to the characteristic function of the function ${\mathbb P}_1 $
\begin{eqnarray} 
\int  d \Phi {\mathbb P}_1\left(  \Phi \right)e^{i \Theta  \Phi} 
 =   e^{ \displaystyle    \frac{1}{2}  \sum_{\alpha=1 }^{+\infty}  \int_0^{+\infty} dv 
e^{- \frac{ v}{2  F_{\alpha} } }
 \left( \frac{ e^{i  \Theta v } -1 - i \Theta  v}{v} \right)} 
  \label{{ScalingAphi2}LevyKhint}
\end{eqnarray}

The exponential decay of Eq. \ref{B2expDecay} that involves the eigenvalue
$ f^{[R]}_1  =  R^{2H} F_1$ yields
\begin{eqnarray} 
  {\mathbb P}_R\left(  \langle A_R \vert \phi_2 \rangle =\varphi \right) \oppropto_{ \varphi \to + \infty } 
  e^{ \displaystyle -   \frac{ \varphi}{2 f_1^{[R]} }} = e^{\displaystyle - R^{-2H}  \frac{ \varphi}{2 F_1 }}
    \label{AB2expDecay}
\end{eqnarray}
This scaling property means that the large deviations properties for large $R$
are governed by the same unusual exponent $R^{- 2 H} $ in the exponential as in Eq. \ref{GaussLinearObsmR} concerning the empirical magnetization,
while the usual volume-scaling $R^d$ is recovered here only for the Hurst exponent $H=-\frac{d}{2}$
corresponding to the White-Noise of Eq. \ref{FracGausszerowhitenoise}.

%%%%%%%%%%%%%%%%%%%%%%%%%%%%%%%%%%%%%%%%%%%

%%%%%%%%%%%%%%%%%%%%%%%%%%%%%%%%%%%%%%%%

\subsection{ Discussion }

In summary, $\phi_2(\vec x)$ is a scale-invariant non-Gaussian field with Hurst exponent $H_2=2 H$,
characterized by its cumulants of arbitrary order $n$ or by the Levy-Khintchine formula for its characteristic function.
In the next section, it is thus interesting
 to discuss what can be said about observables of higher order $n>2$.

%%%%%%%%%%%%%%%%%%%%%%%%%%%%%%%%%%%

\section{ Statistics of observables of order $n>2$
of the Fractional-Gaussian-Field }

\label{sec_higher}

After the previous section concerning the special case of quadratic observables $n=2$,
this section is devoted to the observables of higher order $n>2$.

\subsection{ Observable ${\cal B}^{(n)} =\langle B^{(n)}  \vert \phi^{\otimes n} \rangle $ of order $n$ parametrized by the symmetric tensor $B^{(n)}(\vec x_1,..,\vec x_n)  $ } 

For arbitrary $n$, it is convenient to write an observable ${\cal B}^{(n)}$ of order $n$
as the scalar product between a symmetric n-tensor $B^{(n)} $
and the tensor-product of $n$ fields $\vert \phi^{\otimes n} \rangle$
\begin{eqnarray} 
{\cal B}^{(n)} \equiv \langle B^{(n)}  \vert \phi^{\otimes n} \rangle  
= \int d^d \vec x_1 ...  \int d^d \vec x_n B^{(n)}( \vec x_1,..,\vec x_n) \phi(\vec x_1) ... \phi(\vec x_n)
  \label{Bnf}
\end{eqnarray}
Note that in the previous section concerning the special case $n=2$ for quadratic observables,
we have replaced the 2-tensor $\langle B^{(n=2)} \vert $ by the matrix ${\bold B}$,
i.e. the 2-tensor-elements $B^{(2)}( \vec x_1,\vec x_2)= \langle B^{(2)}  \vert \vec x_1, \vec x_2\rangle$ 
 by the matrix-elements $ \langle \vec x_1 \vert {\bold B} \vert \vec x_2 \rangle$ to obtain instead
\begin{eqnarray} 
{\cal B}_2 \equiv \langle B^{(2)}  \vert \phi^{\otimes 2} \rangle 
  = \int d^d \vec x_1   \int d^d \vec x_2 B^{(2)}( \vec x_1,\vec x_2) \phi(\vec x_1)  \phi(\vec x_2)
  = \langle \phi \vert {\bold B} \vert \phi \rangle
  \label{B2f}
\end{eqnarray}
since it was technically more convenient to use the properties of matrices.
Similarly in the previous sections, it was technically more convenient to represent the 2-field correlation
by the matrix ${\bold C} = {\mathbb E} \left( \vert \phi \rangle  \langle \phi \vert  \right)  $,
while for arbitrary $n$, the correlation of arbitrary order $n$ can be considered as the n-tensor
\begin{eqnarray} 
 \vert C^{(n)} \rangle  \equiv {\mathbb E} \bigg(\vert \phi^{\otimes n} \rangle \bigg)
   \label{Zntensor}
\end{eqnarray}
with its real-space elements
\begin{eqnarray} 
C^{(n)}( \vec x_1, \vec x_2,..,\vec x_n) =  \bigg( \langle \vec x_1 \vert \otimes \langle \vec x_2 \vert ... \otimes \langle \vec x_n \vert \bigg)\vert C^{(n)} \rangle  
= {\mathbb E} \left( \phi (\vec x_1)\phi (\vec x_2) ... \phi (\vec x_n) \right) 
   \label{ZntensorReal}
\end{eqnarray}
so that the averaged value of the observable ${\cal B}^{(n)} $ of Eq. \ref{Bnf} corresponds to the scalar product
\begin{eqnarray} 
{\mathbb E} \left( {\cal B}^{(n)} \right) \equiv \langle B^{(n)}  \vert C^{(n)} \rangle  
= \int d^d \vec x_1 ...  \int d^d \vec x_n B^{(n)}( \vec x_1,..,\vec x_n) C^{(n)}(\vec x_1, ..., \vec x_n)
  \label{Bnfav}
\end{eqnarray}

%%%%%%%%%%%%%%%%%%%%%%%%%%%%%%%%%%%%%%%%

\subsection{ Rewriting of ${\cal B}^{(n)}  =\langle B^{(n)}  \vert \phi^{\otimes n} \rangle
=\langle F^{(n)}  \vert W^{\otimes n} \rangle $ 
as an observable of order $n$ for the White Noise $W(.)$ } 

As in Eq. \ref{B2phiW} concerning quadratic observables,
it is convenient to make the change of variables $\vert \phi \rangle =  \sqrt{ {\bold C} } \vert W \rangle $ of Eq. \ref{LoeveW} towards the white-noise $W$
in order to rewrite the observable ${\cal B}^{(n)} $ of Eq. \ref{Bnf} 
\begin{eqnarray} 
{\cal B}^{(n)}   
&& = \int d^d \vec x_1 ...  \int d^d \vec x_n B^{(n)}( \vec x_1,..,\vec x_n) 
\prod_{i=1}^n \langle \vec x_i \vert \phi \rangle
=  \int d^d \vec x_1 ...  \int d^d \vec x_n B^{(n)}( \vec x_1,..,\vec x_n) 
\prod_{i=1}^n \langle \vec x_i \vert \sqrt{ {\bold C} } \vert W \rangle
\nonumber \\
&& \equiv \int d^d \vec y_1 ...  \int d^d \vec y_n F^{(n)}( \vec y_1,..,\vec y_n) 
\prod_{i=1}^n W( \vec y_i )
= \langle F^{(n)}  \vert W^{\otimes n} \rangle
  \label{BnfW}
\end{eqnarray}
as an observable of order $n$
for the White Noise $W(.)$ that involves the tensor $ \langle F^{(n)}  \vert $ with the real-space elements
\begin{eqnarray} 
F^{(n)}( \vec y_1,..,\vec y_n) 
=  \int d^d \vec x_1 ...  \int d^d \vec x_n B^{(n)}( \vec x_1,..,\vec x_n) 
\prod_{i=1}^n \langle \vec x_i \vert \sqrt{ {\bold C} } \vert \vec y_i \rangle
  \label{TensorFn}
\end{eqnarray}
that can be considered as the n-tensor generalization of Eq. \ref{operatorF}.

%%%%%%%%%%%%%%%%%%%%%%%%%%%%

\subsection{ Rewriting of ${\cal B}^{(n)} =\langle B^{(n)}  \vert \phi^{\otimes n} \rangle $ 
in terms of Ito integrals of order $(n-2l)$ with $1 \leq l \leq \frac{n}{2} $ } 

The properties of multiple stochastic integrals involving the white noise $W(.)$
 like ${\cal B}^{(n)}  = \langle F^{(n)}  \vert W^{\otimes n} \rangle $  of Eq. \ref{BnfW}
are discussed in detail in Appendix \ref{app_WienerIto} 
with the notation ${\cal B}^{(n)}=S_n(F^{(n)}) $ of Eq. \ref{Snbraket},
while the corresponding Ito integrals $I_n(.) $ of Eq. \ref{Itonbraket}
correspond to the modified observables
\begin{eqnarray} 
I_n (F^{(n)})
 \equiv \int d^d \vec y_1 ...  \int d^d \vec y_n F^{(n)}( \vec y_1,..,\vec y_n) \left( \prod_{1 \leq i<j \leq n}  \theta( \vec y_i \ne \vec y_j) \right)
\prod_{i=1}^n W( \vec y_i )
  \label{BnfWIto}
\end{eqnarray}

In particular, the Wiener-Ito chaos-expansion of Eq. \ref{Stratofinite} reads
\begin{eqnarray} 
{\cal B}^{(n)}   = S_n(F^{(n)})
=I_{n}(F^{(n)}) +   \sum_{1 \leq l \leq \frac{n}{2} }   \frac{n!}{ l! (n-2l)! 2^l  }  I_{n-2l}( f^{(n-2l)})
  \label{BnStratofinite}
\end{eqnarray}
where the functions $f^{(n-2l)} ( \vec y_1, ... ,\vec y_{n-2l})$ of $(n-2l)$ variables
are obtained from the tensor $F^{(n)}$ of Eq. \ref{TensorFn}
via the following integrations over $l$ variables
(as explained in more details around Eq. \ref{traceofOrderk})
  \begin{eqnarray} 
&&f^{(n-2l)}( \vec y_1, ... ,\vec y_{n-2l})  =
\int d^d \vec z_1 ... \int d^d\vec z_l F^{(n)}( \vec y_1, \vec y_2,.. \vec y_{n-2l},  
\vec z_1,\vec z_1, \vec z_2,\vec z_2,...,\vec z_l, \vec z_l) 
\nonumber \\
&& = 
\int d^d \vec x_1 ...  \int d^d \vec x_n B^{(n)}( \vec x_1,..,\vec x_n) 
\left[ \prod_{i=1}^{n-2l} \langle \vec x_i \vert \sqrt{ {\bold C} } \vert \vec y_i \rangle \right]
\int d^d \vec z_1 ... \int d^d\vec z_l 
\left[ \prod_{j=1}^l \langle \vec x_{n-2l+2j-1} \vert \sqrt{ {\bold C} } \vert \vec z_j \rangle
\langle \vec x_{n-2l+2j} \vert \sqrt{ {\bold C} } \vert \vec z_j \rangle \right]
\nonumber \\
&& = 
\int d^d \vec x_1 ...  \int d^d \vec x_n B^{(n)}( \vec x_1,..,\vec x_n) 
\left[ \prod_{i=1}^{n-2l} \langle \vec x_i \vert \sqrt{ {\bold C} } \vert \vec y_i \rangle \right]
\prod_{j=1}^l \left[ \int d^d \vec z_j \langle \vec x_{n-2l+2j-1} \vert \sqrt{ {\bold C} } \vert \vec z_j \rangle
\langle \vec z_j \vert \sqrt{ {\bold C} } \vert \vec x_{n-2l+2j} \rangle \right]
\nonumber \\
&& = 
\int d^d \vec x_1 ...  \int d^d \vec x_n B^{(n)}( \vec x_1,..,\vec x_n) 
\left[ \prod_{i=1}^{n-2l} \langle \vec x_i \vert \sqrt{ {\bold C} } \vert \vec y_i \rangle \right]
\prod_{j=1}^l \left[  \langle \vec x_{n-2l+2j-1} \vert  {\bold C}  \vert \vec x_{n-2l+2j} \rangle \right]
\nonumber \\
&& = 
\int d^d \vec x_1 ...  \int d^d \vec x_n B^{(n)}( \vec x_1,..,\vec x_n) 
\left[ \prod_{i=1}^{n-2l} \langle \vec x_i \vert \sqrt{ {\bold C} } \vert \vec y_i \rangle \right]
\prod_{j=1}^l C( \vec x_{n-2l+2j-1} , \vec x_{n-2l+2j} )
  \label{BtraceofOrderk}
\end{eqnarray}

%%%%%%%%%%%%%%%%%%%%%%%%%%%%

\subsection{ Interpretation of the Ito integral $I_n(F^{(n)}) $ in terms of the initial field $\phi$ } 

The  functions $f^{(n-2l)} $ of Eq. \ref{BtraceofOrderk}
also appear in the inversion formula of Eq. \ref{StratofiniteInversion}
\begin{eqnarray} 
I_n(F^{(n)})  && 
= {\cal B}^{(n)} + \sum_{1 \leq l \leq \frac{n}{2} }  (-1)^l \frac{n!}{ l! (n-2l)! 2^l  }  
\langle f^{(n-2l)} \vert W^{\otimes (n-2l)} \rangle
  \label{BStratofiniteInversionf}
\end{eqnarray}
where the scalar products $\langle f^{(n-2l)} \vert W^{\otimes (n-2l)} \rangle $
 involving the white noise $W(.)$
\begin{eqnarray} 
&& \langle f^{(n-2l)} \vert W^{\otimes (n-2l)} \rangle
 = \int d^d \vec y_1 ...  \int d^d \vec y_{n-2l} 
 f^{(n-2l)}( \vec y_1, ... ,\vec y_{n-2l})
\prod_{i=1}^{n-2l} W( \vec y_i )
\nonumber \\
&& = \int d^d \vec x_1 ...  \int d^d \vec x_n B^{(n)}( \vec x_1,..,\vec x_n) 
\left[\prod_{j=1}^l C( \vec x_{n-2l+2j-1} , \vec x_{n-2l+2j} ) \right]
\left[ \prod_{i=1}^{n-2l} \int d^d \vec y_i \langle \vec x_i \vert \sqrt{ {\bold C} } \vert \vec y_i \rangle 
\langle \vec y_i \vert W \rangle \right]
\nonumber \\
&& = \int d^d \vec x_1 ...  \int d^d \vec x_n B^{(n)}( \vec x_1,..,\vec x_n) 
\left[\prod_{j=1}^l C( \vec x_{n-2l+2j-1} , \vec x_{n-2l+2j} ) \right]
\left[ \prod_{i=1}^{n-2l} \phi( \vec x_i ) \right]
\equiv \langle b^{(n-2l)} \vert \phi^{\otimes (n-2l)} \rangle
  \label{fWbphi}
\end{eqnarray}
can be rewritten as the scalar product $\langle b^{(n-2l)} \vert \phi^{\otimes (n-2l)} \rangle $ involving the field $\vert \phi \rangle =  \sqrt{ {\bold C} } \vert W \rangle $
and the functions
\begin{eqnarray} 
b^{(n-2l)}( \vec x_1, ... ,\vec x_{n-2l}) 
= \int d^d \vec x_{n-2l+1} ...  \int d^d \vec x_n B^{(n)}( \vec x_1,..,\vec x_n) 
\left[\prod_{j=1}^l C( \vec x_{n-2l+2j-1} , \vec x_{n-2l+2j} ) \right]
  \label{fWbphibb}
\end{eqnarray}
Plugging Eq. \ref{fWbphi}
into Eq. \ref{BStratofiniteInversionf}
yields the interpretation of the Ito integrals $I_n(F^{(n)}) $
in terms of observables concerning the scale invariant field $\phi$
\begin{small}
\begin{eqnarray} 
&& I_n(F^{(n)})   
 = \langle B^{(n)} \vert \phi^{\otimes n} \rangle 
+ \sum_{1 \leq l \leq \frac{n}{2} }  (-1)^l \frac{n!}{ l! (n-2l)! 2^l  }  \langle b^{(n-2l)} \vert \phi^{\otimes (n-2l)} \rangle
  \label{BStratofiniteInversion}
 \\
&& = \int d^d \vec x_1 ...  \int d^d \vec x_n B^{(n)}( \vec x_1,..,\vec x_n) 
\left( \phi(\vec x_1) ... \phi(\vec x_n)
+ \sum_{1 \leq l \leq \frac{n}{2} }  (-1)^l \frac{n!}{ l! (n-2l)! 2^l  } 
\phi(\vec x_1) ... \phi(\vec x_{n-2l}) \left[\prod_{j=1}^l C( \vec x_{n-2l+2j-1} , \vec x_{n-2l+2j} ) \right]
\right)
\nonumber \\
&& \equiv \langle B^{(n)} \vert \psi^{(n)} \rangle = \langle B^{(n)} \vert \psi^{(n)}_{sym} \rangle
\end{eqnarray}
\end{small}
with the tensor
\begin{eqnarray} 
\psi^{(n)}( \vec x_1,..,\vec x_n) 
\equiv \phi(\vec x_1) ... \phi(\vec x_n)
+ \sum_{1 \leq l \leq \frac{n}{2} }  (-1)^l \frac{n!}{ l! (n-2l)! 2^l  } 
\phi(\vec x_1) ... \phi(\vec x_{n-2l}) \left[\prod_{j=1}^l C( \vec x_{n-2l+2j-1} , \vec x_{n-2l+2j} ) \right]
\label{tensorpsin}
\end{eqnarray}
that can also be rewritten in a symmetric form $\vert \psi^{(n)}_{sym} \rangle $
as a consequence of the symmetry of the n-tensor $B^{(n)} $.

%%%%%%%%%%%%%%%%%%%%%%%%%%%%%%%%

\subsubsection{ Example of the cubic Ito integral $I_3(F^{(3)})  $ } 

For $n=3$ concerning cubic observables ${\cal B}^{(3)}  =\langle B^{(3)}  \vert \phi^{\otimes 3} \rangle$,
the Ito integral of Eq. \ref{BStratofiniteInversion} 
\begin{eqnarray} 
&& I_3(F^{(3)})   
 = \int d^d \vec x_1 \int d^d \vec x_2  \int d^d \vec x_3 B^{(3)}( \vec x_1,\vec x_2,\vec x_3) 
\left[ \phi(\vec x_1) \phi(\vec x_2) \phi(\vec x_3) - 3 \phi(\vec x_1)  C( \vec x_2, \vec x_3 )  \right]
\equiv \langle B^{(3)} \vert \psi^{(3)} \rangle
\nonumber \\
&& = \int d^d \vec x_1 \int d^d \vec x_2  \int d^d \vec x_3 B^{(3)}( \vec x_1,\vec x_2,\vec x_3) 
\left[ \phi(\vec x_1) \phi(\vec x_2) \phi(\vec x_3) -  \phi(\vec x_1)  C( \vec x_2, \vec x_3 ) 
-   \phi(\vec x_2)  C( \vec x_1, \vec x_3 )
-  \phi(\vec x_3)  C( \vec x_1, \vec x_2 ) \right]
\nonumber \\
&&\equiv \langle B^{(3)} \vert \psi_{sym}^{(3)} \rangle
  \label{BStratofiniteInversion33}
\end{eqnarray}
involves the tensors 
\begin{eqnarray} 
 \psi^{(3)}( \vec x_1,\vec x_2,\vec x_3) && \equiv
 \phi(\vec x_1) \phi(\vec x_2) \phi(\vec x_3) - 3 \phi(\vec x_1)  C( \vec x_2, \vec x_3 )
\nonumber \\
 \psi^{(3)}_{sym}( \vec x_1,\vec x_2,\vec x_3)
&& \equiv \phi(\vec x_1) \phi(\vec x_2) \phi(\vec x_3) -  \phi(\vec x_1)  C( \vec x_2, \vec x_3 ) 
-   \phi(\vec x_2)  C( \vec x_1, \vec x_3 )
-  \phi(\vec x_3)  C( \vec x_1, \vec x_2 ) 
  \label{psi3}
\end{eqnarray}

%%%%%%%%%%%%%%%%%%%%%%%%%%%%%

\subsubsection{ Example of the quartic Ito integral $I_4(F^{(4)})  $ }

For $n=4$ concerning quartic observables ${\cal B}^{(4)}  =\langle B^{(4)}  \vert \phi^{\otimes 4} \rangle
$, 
the Ito integral of Eq. \ref{BStratofiniteInversion}
\begin{small}
\begin{eqnarray} 
&& I_4(F^{(4)})   
 = \int d^d \vec x_1 \int d^d \vec x_2  \int d^d \vec x_3 \int d^d \vec x_4
 B^{(4)}( \vec x_1,\vec x_2,\vec x_3, \vec x_4) 
% \nonumber \\ && \times 
\left[ \phi(\vec x_1) \phi(\vec x_2) \phi(\vec x_3)\phi(\vec x_4)
 - 6 \phi(\vec x_1) \phi(\vec x_2) C( \vec x_3, \vec x_4 ) 
 +3 C( \vec x_1, \vec x_2 )C( \vec x_3, \vec x_4 ) \right]
\nonumber \\
&& \equiv \langle B^{(4)} \vert \psi^{(4)} \rangle = \langle B^{(4)} \vert \psi_{sym}^{(4)} \rangle
  \label{BStratofiniteInversion44}
\end{eqnarray}
\end{small}
involves the tensors
\begin{eqnarray} 
 \psi^{(4)}( \vec x_1,\vec x_2,\vec x_3,\vec x_4)
&& \equiv \phi(\vec x_1) \phi(\vec x_2) \phi(\vec x_3)\phi(\vec x_4)
 - 6 \phi(\vec x_1) \phi(\vec x_2) C( \vec x_3, \vec x_4 ) 
 +3 C( \vec x_1, \vec x_2 )C( \vec x_3, \vec x_4 )
\nonumber \\
 \psi^{(4)}_{sym}( \vec x_1,\vec x_2,\vec x_3,\vec x_4)
&& \equiv \phi(\vec x_1) \phi(\vec x_2) \phi(\vec x_3) \phi(\vec x_3)
 -  \phi(\vec x_1) \phi(\vec x_2) C( \vec x_3, \vec x_4 )  
-   \phi(\vec x_1) \phi(\vec x_3) C( \vec x_2, \vec x_4 ) 
-  \phi(\vec x_1) \phi(\vec x_4) C( \vec x_2, \vec x_3 ) 
\nonumber \\
&& -  \phi(\vec x_2) \phi(\vec x_3) C( \vec x_1, \vec x_4 )  
-   \phi(\vec x_2) \phi(\vec x_4) C( \vec x_1, \vec x_3 ) 
-  \phi(\vec x_3) \phi(\vec x_4) C( \vec x_1, \vec x_2 ) 
\nonumber \\
&&+ C( \vec x_1, \vec x_2 )C( \vec x_3, \vec x_4 )
+C( \vec x_1, \vec x_3 )C( \vec x_2, \vec x_4 )
+C( \vec x_1, \vec x_4 )C( \vec x_2, \vec x_3 )
  \label{psi4}
\end{eqnarray}

%%%%%%%%%%%%%%%%%%%%%%%%%%%%%%%%%%%%%%%%

\subsection{ Application to the spatial-average of the finite part $\phi_n(\vec x)$ of the composite operator $\phi^n(\vec x)$ } 

The naive spatial-average with the kernel $A_R(\vec x) $ of Eq. \ref{ARkernel}
of the ill-defined composite operator $\phi^n(\vec x)$
\begin{eqnarray} 
{\cal B}^{(n)}_R   = \int d^d \vec x A_R(\vec x) \phi^n(\vec x) 
  \label{Compositespatialav}
\end{eqnarray}
corresponds to the observable of Eq. \ref{Bnf}
for the choice of the n-tensor
\begin{eqnarray} 
 B_R^{(n)} (\vec x_1, \vec x_2,..,\vec x_n ) = A_R(\vec x_1) \prod_{j=2 }^n \delta^{(d)}( \vec x_j- \vec x_1) 
% =  \frac{1}{R^{nd} } A \left( \frac{ \vec x_1}{R} \right) \prod_{j=2 }^n \delta^{(d)}\left( \frac{ \vec x_j -\vec x_1}{R} \right)
 \label{BnSpace}
\end{eqnarray}
with the associated n-tensor of Eq. \ref{TensorFn}
\begin{eqnarray} 
F^{(n)}_R( \vec y_1,..,\vec y_n) 
&& =  \int d^d \vec x_1 ...  \int d^d \vec x_n B_R^{(n)}( \vec x_1,..,\vec x_n) 
\prod_{i=1}^n \langle \vec x_i \vert \sqrt{ {\bold C} } \vert \vec y_i \rangle
\nonumber \\
&& = \int d^d \vec x   A_R(\vec x) 
\prod_{i=1}^n \langle \vec x \vert \sqrt{ {\bold C} } \vert \vec y_i \rangle
  \label{TensorFnComposite}
\end{eqnarray}

Then the functions of Eq. \ref{fWbphibb}
\begin{eqnarray} 
b^{(n-2l)}_R( \vec x_1, ... ,\vec x_{n-2l}) 
&& = \int d^d \vec x_{n-2l+1} ...  \int d^d \vec x_n B_R^{(n)}( \vec x_1,..,\vec x_n) 
\left[\prod_{j=1}^l C( \vec x_{n-2l+2j-1} , \vec x_{n-2l+2j} ) \right]
\nonumber \\
&& =  
 A_R(\vec x_1) \left[ \prod_{j=2 }^{n-2l} \delta^{(d)}( \vec x_j- \vec x_1) \right]
\left[C( \vec x_1 , \vec x_1 ) \right]^l  = + \infty \ \ \text{ for any} \ \ l \geq 1
  \label{fWbphibbcomposite}
\end{eqnarray}
diverge for any $l \geq 1$ as a consequence of the divergence of the real-space correlation 
$C( \vec x , \vec y ) $
at coinciding points $\vec y \to \vec x$.

So the Ito integral $I_{n}(F_R^{(n)}) $ of Eq. \ref{BStratofiniteInversion}
corresponds to the finite linear combination of observables that would all diverge individually
\begin{eqnarray} 
I_n(F_R^{(n)})   
&& = \langle B_R^{n} \vert \phi^{\otimes n} \rangle + \sum_{1 \leq l \leq \frac{n}{2} }  (-1)^l \frac{n!}{ l! (n-2l)! 2^l  }  \langle b_R^{(n-2l)} \vert \phi^{\otimes (n-2l)} \rangle
\nonumber \\
&& =  \int d^d \vec x A_R(\vec x) \left( \phi^n(\vec x) 
 + \sum_{1 \leq l \leq \frac{n}{2} }  (-1)^l \frac{n!}{ l! (n-2l)! 2^l  } 
  \phi^{n-2l}(\vec x) \left[C( \vec x , \vec x ) \right]^l \right)
  \equiv \langle A_R \vert \phi_n \rangle
  \label{BStratofiniteInversionComposite}
\end{eqnarray}
and represents the spatial-average with the kernel $A_R$ of
the finite part $\phi_n (\vec x)$ of the composite operator $\phi^n(\vec x)$ given by
\begin{eqnarray} 
\phi_n (\vec x) \equiv \phi^n(\vec x) 
 + \sum_{1 \leq l \leq \frac{n}{2} }  (-1)^l \frac{n!}{ l! (n-2l)! 2^l  } 
  \phi^{n-2l}(\vec x) \left[  {\mathbb E} \bigg( \phi^2(\vec x) \bigg) \right]^l 
  \label{phinComposite}
\end{eqnarray}
For $n=2$, one recovers the fluctuating part $\phi_2(\vec x)$ of the composite operator $\phi^2(\vec x)$
of Eq. \ref{phi2defcomposite}
\begin{eqnarray} 
\phi_2 (\vec x) \equiv \phi^2(\vec x) 
 -   \left[  {\mathbb E} \bigg( \phi^2(\vec x) \bigg) \right] 
  \label{phinComposite2}
\end{eqnarray}
that was discussed in detail in the previous section \ref{sec_quadratic}.
For $n=3$, Eq. \ref{phinComposite} yields
\begin{eqnarray} 
\phi_3 (\vec x) \equiv \phi^3(\vec x)  -  3   \phi(\vec x)   {\mathbb E} \bigg( \phi^2(\vec x) \bigg)  
= \lim_{ \vec x_i \to \vec x} \bigg( \psi^{(3)}( \vec x_1,\vec x_2,\vec x_3) \bigg)
=  \lim_{ \vec x_i \to \vec x} \bigg( \psi^{(3)}_{sym}( \vec x_1,\vec x_2,\vec x_3) \bigg)
  \label{phinComposite3}
\end{eqnarray}
that corresponds to the limit of three coinciding points $\vec x_i \to \vec x $ for $i=1,2,3$
in the 3-tensors $ \psi^{(3)}( \vec x_1,\vec x_2,\vec x_3) $ 
or $\psi^{(3)}_{sym}( \vec x_1,\vec x_2,\vec x_3) $ of Eq. \ref{psi3}.

For $n=4$, Eq. \ref{phinComposite} yields
\begin{eqnarray} 
\phi_4 (\vec x)&&  \equiv \phi^4(\vec x) 
 -6  \phi^2(\vec x) \left[  {\mathbb E} \bigg( \phi^2(\vec x) \bigg) \right]^2 
   +   3
   \left[  {\mathbb E} \bigg( \phi^2(\vec x) \bigg) \right]^2 
  \nonumber \\
  &&  =   \lim_{ \vec x_i \to \vec x} \bigg( \psi^{(4)}( \vec x_1,\vec x_2,\vec x_3, \vec x_4) \bigg)
   =   \lim_{ \vec x_i \to \vec x} \bigg( \psi^{(4)}_{sym}( \vec x_1,\vec x_2,\vec x_3, \vec x_4) \bigg)
  \label{phinComposite4}
\end{eqnarray}
that corresponds to the limit of four coinciding points $\vec x_i \to \vec x $ for $i=1,2,3,4$
in the 4-tensors $ \psi^{(4)}( \vec x_1,\vec x_2,\vec x_3,\vec x_4) $ 
or $ \psi^{(4)}_{sym}( \vec x_1,\vec x_2,\vec x_3,\vec x_4) $ of Eq. \ref{psi4}.

The variance of the Ito integral $I_{n}(F_R^{(n)}) $ of Eq. \ref{BStratofiniteInversionComposite}
is given by Eq. \ref{Itonbraketvar}
in terms of $F_R^{(n)} $ of Eq. \ref{TensorFnComposite}
\begin{eqnarray} 
 {\mathbb E} \bigg( \langle A_R \vert \phi_n \rangle \langle \phi_n \vert A_R \rangle\bigg)  
 && = {\mathbb E} \bigg( I_n^2(F_R^{(n)}) \bigg)  
 =  n! \langle F_R^{(n)} \vert F_R^{(n)} \rangle
 =  n! \int d^d \vec y_1  \int d^d \vec y_2 ... \int d^d \vec y_n \left[ F_R^{(n)}(\vec y_1,\vec y_2,...,\vec y_n) 
 \right]^2
 \nonumber \\
 && = n! \int d^d \vec y_1  \int d^d \vec y_2 ... \int d^d \vec y_n 
 \int d^d \vec x_1   A_R(\vec x_1) \int d^d \vec x_2   A_R(\vec x_2)
 \prod_{i=1}^n \left[ \langle \vec x_1 \vert \sqrt{ {\bold C} } \vert \vec y_i \rangle
 \langle \vec x_2 \vert \sqrt{ {\bold C} } \vert \vec y_i \rangle
 \right]
  \nonumber \\
 && = n! \int d^d \vec x_1   A_R(\vec x_1) \int d^d \vec x_2   A_R(\vec x_2)
 \prod_{i=1}^n \left[ \int d^d \vec y_i \langle \vec x_1 \vert \sqrt{ {\bold C} } \vert \vec y_i \rangle
 \langle \vec y_i \vert \sqrt{ {\bold C} } \vert \vec x_2 \rangle
 \right]
   \nonumber \\
 && = n! \int d^d \vec x_1   A_R(\vec x_1) \int d^d \vec x_2   A_R(\vec x_2)
 \left[ \langle \vec x_1 \vert  {\bold C}  \vert \vec x_2 \rangle \right]^n
   \nonumber \\
 && = n! \int d^d \vec x_1   \langle A_R \vert \vec x_1 \rangle \int d^d \vec x_2    
 \left[ C( \vec x_1 ,\vec x_2 ) \right]^n
 \langle \vec x_2 \vert A_R \rangle
  \label{Itonbraketvarcomposite}
\end{eqnarray}
As a consequence,
 the correlation of the finite part $\phi_n (\vec x)$ of the composite operator $\phi^n(\vec x)$
involves the power $n$ of the real-space correlation $C(\vec x_1, \vec x_2)= \frac{\kappa }{\vert \vec x_1 -\vec y_2 \vert^{-2H} }$ of Eq. \ref{CorreFracRealSpace}
concerning the initial field $\phi(.)$
\begin{eqnarray} 
{\mathbb E} \bigg( \phi_n(\vec x_1)  \phi_n(\vec x_2)\bigg)  
=   n!      \left[ C( \vec x_1 ,\vec x_2 ) \right]^n
=  \frac { n! \kappa^n  }{ \vert \vec x_1-\vec x_2 \vert^{- n 2 H}}
  \label{correphincomposite}
\end{eqnarray}
that can be considered as the direct generalization of
 Eq. \ref{Correphi2} concerning the special case $n=2$.

In conclusion, the finite part $\phi_n (\vec x)$ of the composite operator $\phi^n(\vec x)$
is a non-Gaussian scale-invariant field 
\begin{eqnarray} 
 \phi_n ( \vec x ) \opsim_{law} b^{H_n} \phi_n\left( \frac{ \vec x}{ b}  \right)    
 \ \ \ \text{ with the Hurst exponent } \ \ \ H_n = n H
  \label{RescalFieldRealSpaceHn}
\end{eqnarray}
The  convergence region of Eq. \ref{CorreFracRealSpace}
\begin{eqnarray} 
-\frac{d}{2} <H_n=nH <0 \ \ \ \ \text{ corresponding to } \ \ -\frac{d}{2n} <H <0
  \label{regioncompositen}
\end{eqnarray}
is shrinking as $n$ grows.

%%%%%%%%%%%%%%%%%%%%%%%%%%%%%%%%%%%%

\subsection{  Generalization to arbitrary observables of the Fractional-Gaussian-Field field $\phi$ }

Via the change of variables $\vert \phi \rangle =  \sqrt{ {\bold C} } \vert W \rangle $ of Eq. \ref{LoeveW} towards the white-noise $W$, any observable of the Fractional-Gaussian-Field field $\phi$
can be translated into an observable of the white-noise $W$.
As recalled in detail in Appendix \ref{app_WienerIto},
any functional of the white noise $W(.)$ can be expanded 
into a series of multiple Ito integrals via the Wiener-Ito chaos-expansion of Eq. \ref{ChaosExpansion},
that can be then reinterpreted in terms of the field $\vert \phi \rangle $ as explained
in detail in the present section for the case of observables of arbitrary order $n$.

%%%%%%%%%%%%%%%%%%%%%%%%%%%%%%%%%%%%%

%%%%%%%%%%%%%%%%%%%%%%%%%%%%%

\section{ Conclusions }

\label{sec_conclusion}

In this paper, we have revisited the statistical properties of non-linear observables of the fractional Gaussian field $\phi(\vec x)$ of negative Hurst exponent $H<0$ in dimension $d$ via pedestrian calculations for statistical physicists familiar with stochastic processes. In particular, we have
focused on spatial-averaging observables and on the properties of the finite parts $\phi_n(\vec x)$ of the ill-defined composite operators $\phi^n(\vec x) $.
For the special case $n=2$ of quadratic observables, many explicit results have been written, in particular the cumulants of arbitrary order, the L\'evy-Khintchine formula for the characteristic function and the anomalous large deviations properties. The case of observables of arbitrary order $n>2$ has been analyzed via the Wiener-Ito chaos-expansion for functionals of the white noise (with the self-contained reminder in Appendix \ref{app_WienerIto}) :  we have explained how the multiple stochastic Ito integrals are useful to identify the finite parts $\phi_n(\vec x)$ of the ill-defined composite operators $\phi^n(\vec x) $ and to compute their correlations involving the Hurst exponents $H_n=nH$.

%%%%%%%%%%%%%%%%%%%%%%%%%%%%%

\appendix

\section{ fractional Gaussian fields with positive or negative Hurst exponents in dimension $d=1$  }

\label{app_positiveH}

In this Appendix, we recall some well-known properties 
of the one-dimensional Brownian motion $B(x)$ of Hurst exponent $H=\frac{1}{2}$,
of the fractional Brownian motion $B_H(x)$ of Hurst exponent $0<H<1$,
and of the corresponding fractional Gaussian noise $W_H(x)= \frac{dB_H(x)}{dx}$ 
of negative Hurst exponent $H'=(H-1) \in ]-1,0[$,
in order to stress the differences between positive and negative Hurst exponents
that are mentioned at the beginning of the Introduction,
and in order to make the link with
the case of arbitrary dimension $d$ discussed in the main text.

%%%%%%%%%%%%%%%%%%%%%%%%%%%%%%%%%%%%%%%%%%%

\subsection{The Brownian motion $B( x)$ of positive Hurst exponent $H=\frac{1}{2}$ in dimension $d=1$}

The one-dimensional Brownian motion $B(x)$ can be defined 
by the fact that its derivative $\frac{dB(x)}{dx}$
coincides with the one-dimensional white noise $W(x)$ of Hurst exponent $(-\frac{1}{2})$  
\begin{eqnarray}
\frac{dB(x)}{dx} =W(x)
  \label{DeriBrownd1}
\end{eqnarray}
So one needs to choose the value at the origin, usually $B(x=0)=0$, in order to define $B(x)$ via the stochastic integral
\begin{eqnarray}
B(x)=\int_0^x dy W(y) 
  \label{IntegBrownd1}
\end{eqnarray}
with the vanishing averaged value $  {\mathbb E} \left( B(x) \right) =0 $.
The Brownian motion is of course not statistically invariant by translation
but its increments are
\begin{eqnarray}
B(x_2+x)-B(x_1+x) = \int_{x_1+x}^{x_2+x} dy W(y) \opsim_{law} \int_{x_1}^{x_2} dz W(z) = B(x_2)-B(x_1)
  \label{IntegBrowndincrement}
\end{eqnarray}
as a consequence of the statistical invariance of the white noise $W(.)$ by translation.

The Gaussian probability distribution of Eq. \ref{FracGausszerowhitenoise} for the White noise
in dimension $d=1$
\begin{eqnarray} 
&& e^{\displaystyle  - \frac{1}{2}   \int dx  W^2(x) }
 =  e^{\displaystyle  - \frac{1}{2}   \int dx  \left( \frac{dB(x)}{dx}\right)^2 }
\nonumber \\
&& = e^{\displaystyle  - \frac{1}{2}   \int dx  B(x) \left( - \frac{d^2}{dx^2} \right)  B(x) } 
= e^{\displaystyle  - \frac{1}{2}   \int dx  B(x) \left( - {\bold \Delta} \right)  B(x) }
 \propto G^{[d=1]}_{H=\frac{1}{2}} \left( B(.) \right) = P^{[d=1]}_{free}(B(.))
  \label{PBrown}
\end{eqnarray}
corresponds for the Brownian motion $B(x)$
to the Gaussian measure $G^{[d=1]}_{H=\frac{1}{2} } \left( B(.) \right) = P^{[d=1]}_{free}(B(.))$
of Eq. \ref{FracGaussFrees1}
for the free-field in dimension $d=1$ that involves
the one-dimensional Laplacian $\left( - {\bold \Delta} \right) =
\left( - \frac{d^2}{dx^2} \right) $.
In Fourier space, the Gaussian probability distribution of Eq. \ref{PBrown} becomes
\begin{eqnarray} 
 e^{\displaystyle  - \frac{1}{2}   \int dq {\hat W}^* (q)   {\hat W} (q) }
 =  e^{\displaystyle  - \frac{1}{2}  \int d q (4 \pi^2 q^2 )  {\hat B}^*( q)  {\hat B}( q)   }
 \propto G^{[d=1]}_{H=\frac{1}{2}} \left( {\hat B}(.) \right)
  \label{PBrownFourier}
\end{eqnarray}
that involves the eigenvalues $(4 \pi^2 q^2)$ of the opposite Laplacian $\left( - {\bold \Delta} \right) =
\left( - \frac{d^2}{dx^2} \right) $.
Finally, Eq. \ref{IntegBrownd1} yields the spectral representation of $B(x)$ in terms of 
the Fourier-space white noise ${\hat W}(q) $
  \begin{eqnarray}
B(x)=\int_0^x dy  \int_{-\infty}^{+\infty} dq e^{i 2 \pi q y} {\hat W}(q) = \int_{-\infty}^{+\infty} dq \frac{ e^{i 2 \pi q x} -1}{ i 2 \pi q}  {\hat W}(q)
  \label{IntegBrownd1spectral}
\end{eqnarray}

%%%%%%%%%%%%%%%%%%%%%%%%%%%%%%%%%%%%%%%%%%%

\subsection{ The fractional Brownian motion $B_H( x)$ of positive Hurst exponent $0<H<1$ in dimension $d=1$}

The fractional Brownian motion $B_H( x)$
 of Hurst exponent $0<H<1$ is a Gaussian process that can be defined via its two-point correlation 
\begin{eqnarray}
 {\mathbb E} \left( B_H( x) B_H( y) \right)  = \frac{1}{2} \left( \vert  x \vert^{2H}
 +\vert  y \vert^{2H} - \vert  x - y\vert^{2H}
 \right)
  \label{FracBrowniand}
\end{eqnarray}

The special case $ x= y$ yields that the variance of $B_H( x)$ 
\begin{eqnarray}
 {\mathbb E} \left( B_H^2(  x)  \right)   =  \vert   x \vert^{2H} 
  \label{FracBrownianvar}
\end{eqnarray}
grows as the power-law $\vert   x \vert^{2H} $,
while its vanishing at $ x=0$ means that fractional Brownian motion $B_H(  x)$ itself has to vanish at $ x= 0$ 
\begin{eqnarray}
B_H( x= 0)=0
  \label{BFzeroAtx0}
\end{eqnarray}
The variance of the increment $[B_H(  x) - B_H(  y) ] $
\begin{eqnarray}
 {\mathbb E} \left( [B_H(  x) - B_H(  y) ]^2 \right)   && 
 =  {\mathbb E} \bigg( [B_H(   x)]^2+[B_H(   y)]^2  -2 B_H(   x) B_H(   y) \bigg)
 =  \vert  x - y\vert^{2H}
  \label{FracBrowniandincrements}
\end{eqnarray}
depends only on the distance $\vert   x - y \vert$, i.e. the increments of $B_H(.)$
are statistically invariant via translations.
It is thus useful to introduce the corresponding
fractional Gaussian noise $W_H(x) $ as recalled in the next subsection.

%%%%%%%%%%%%%%%%%%%%%%%%%%%%%%%%%%%%%%%%%%%

\subsection{ The fractional Gaussian noise $W_H(x)= \frac{dB_H(x)}{dx}$ of negative Hurst exponent $H'=(H-1) \in ]-1,0[$}

The fractional Gaussian noise $W_H(x)$
is defined via the derivative generalizing Eq. \ref{DeriBrownd1}
\begin{eqnarray}
W_H(x)\equiv \frac{dB_H(x)}{dx} \ \ \ \text {with the negative Hurst exponent $H'=(H-1) \in ]-1,0[ $ }
  \label{DeriBrownd1H}
\end{eqnarray}
Its correlation can be computed from the correlations ${\mathbb E} \left( B_H( x) B_H( y) \right) $ 
of Eq. \ref{FracBrowniand} via the double derivative
\begin{eqnarray}
 {\mathbb E} \left( W_H( x) W_H( y) \right)   && =  {\mathbb E} \left( \frac{dB_H(x)}{dx} \frac{dB_H(y)}{dy} \right) 
= \frac{\partial^2}{\partial x \partial y}  {\mathbb E} \left( B_H( x) B_H( y) \right)  
= - \frac{1}{2}  \frac{\partial^2}{\partial x \partial y} \vert  x - y\vert^{2H}
 \nonumber \\ &&
=  \frac{1}{2}  \frac{\partial^2}{\partial x^2 } \vert  x - y\vert^{2H} = H \frac{\partial}{\partial x } \left(  {\rm sgn}(x-y)  \vert  x - y\vert^{2H-1} \right) 
\nonumber \\
&&= 2H \delta(x-y) \vert  x - y\vert^{2H-1} + \frac{ H (2H-1) }{ \vert  x - y\vert^{2(1-H)} }
  \label{NoiseFracBrowniand1}
\end{eqnarray}
This correlation depends only on the distance $\vert x-y \vert$, 
in agreement with the statistical invariance by translation of the fractional Gaussian noise $W_H( .) $.

It is useful to distinguish the three following cases with very different properties.

\subsubsection{ Case $H=\frac{1}{2}$ }

For $H=\frac{1}{2}$, Eq. \ref{NoiseFracBrowniand1} reduces
 to the delta-function
\begin{eqnarray}
 {\mathbb E} \left( W_{H=\frac{1}{2}}( x) W_{H=\frac{1}{2}}( y) \right)    =  \delta(x-y) 
  \label{NoiseFracBrowniand1demi}
\end{eqnarray}
as it should to recover the one-dimensional white noise $W(x)$ of Hurst exponent $H'=H-1=-\frac{1}{2}$.
In the main text we have discussed the white noise $W(\vec x)$ in dimension $d$ 
with its correlations of Eqs \ref{CorreWhiteNoise} \ref{CorreWhiteNoiseOp}
and its probability distribution of Eq. \ref{FracGausszerowhitenoise}.

%%%%%%%%%%%%%%%%%%%%%%%%%%%%%%%%%%%

\subsubsection{ Region $\frac{1}{2}<H<1 $ }

In the region $\frac{1}{2}<H<1 $, the first contribution of Eq. \ref{NoiseFracBrowniand1}
involves the delta function $\delta(x-y)$ multiplied
by the vanishing factor $ \vert  x - y\vert^{2H-1} $ at coinciding points,
so that
Eq. \ref{NoiseFracBrowniand1} reduces to the second contribution corresponding to the power-law
\begin{eqnarray}
\text{Region }\frac{1}{2}<H<1 : \ \  {\mathbb E} \left( W_H( x) W_H( y) \right) 
 =  \frac{ H (2H-1) }{ \vert  x - y\vert^{2(1-H)} } =  \frac{ (1+H') (1+2H') }{ \vert  x - y\vert^{-2 H'} } 
  \label{NoiseFracBrowniand1power}
\end{eqnarray}
where the rewriting in terms of the Hurst exponent $H'=H-1 \in ]-\frac{1}{2},0  [$ 
is useful to make the link with the power-law decaying correlations of Eq. \ref{CorreFracRealSpace}
for the case of arbitrary dimension $d$ with negative Hurst exponent in the region $H \in ]-\frac{d}{2},0  [$
discussed in the main text.

\subsubsection{ Region $0<H<\frac{1}{2}$ }

In the region $0<H<\frac{1}{2}$, the last line of Eq. \ref{NoiseFracBrowniand1} 
is rather singular since the delta function $\delta(x-y)$ is multiplied
by the diverging factor $ \vert  x - y\vert^{2H-1} $ at coinciding points,
while the long-ranged power-law contribution $\frac{ H (2H-1) }{ \vert  x - y\vert^{2(1-H)} } $
has a negative amplitude : the fractional Gaussian noise $W_H( x) $ of Hurst exponent 
$H'=H-1 \in ]-1,-\frac{1}{2}  [$
is thus 'anti-correlated' on large distances in order to be able to produce 
the fractional Brownian motion $B_H(  x)$ of Hurst exponent $H<\frac{1}{2}$
smaller than in the standard Brownian motion associated to the white noise.

%%%%%%%%%%%%%%%%%%%%%%%%%%%%%%%

\subsection { Conclusion on $B_H(  x)=\int_0^x dy W_H(y)$ with
$0<H<1$ and on $W_H(x)= \frac{dB_H(x)}{dx}$ 
with $H'=(H-1) \in ]-1,0[$ }

In conclusion, the fractional Brownian motion $B_H(  x)=\int_0^x dy W_H(y)$ of positive Hurst exponent
$0<H<1$ is a continuous process with stationary increments,
while the corresponding fractional Gaussian noise $W_H(x)= \frac{dB_H(x)}{dx}$ 
of negative Hurst exponent $H'=(H-1) \in ]-1,0[$ is stationary  
but cannot be pointwise defined and should be considered as a Schwartz tempered distribution.
As described above, it is actually useful to consider both together to better understand their properties.
More generally via integration or derivation, one can construct other processes
of bigger Hurst exponents $H>1$ or smaller Hurst exponents $H'<-1$.

%%%%%%%%%%%%%%%%%%%%%%%%%%%%%%%

\subsection { Generalizations to higher dimension $d>1$ }

In dimension $d>1$, various types of generalizations are interesting to consider,
so let us mention three directions :

(a) The most well-known generalization
is the Brownian particle $\vec B(t) = \{ B^{(1)}(t),B^{(2)}(t) ,...,B^{(d)}(t) \}$ 
moving in a space of dimension $d$ as a function of the one-dimensional time $t$, 
where the $d$ components $B^{(\mu)}(t) $ for $\mu=1,2,..,d$ are independent 
 one-dimensional Brownian motions.

(b) The fractional Brownian field $B_H(\vec x) \in {\mathbb R}$
 of Hurst exponent $0<H<1$ in a space $\vec x \in {\mathbb R}^d$ 
 can be defined via 
its two-point correlation of Eq. \ref{FracBrowniandd}
that corresponds to the direct generalization of Eq. \ref{FracBrowniand}, 
and that produces the variance of Eq. \ref{FracBrowniandincrementsd} for the increments
that generalizes Eq. \ref{FracBrowniandincrements}.

(c) Another point of view is that the ratio
\begin{eqnarray}
 \frac{B_H(  x)}{x} = \frac{1}{x} \int_0^x dy W_H(y)
  \label{FracBrownianAsAverage}
\end{eqnarray}
can be considered as the spatial-average of the fractional noise $ W_H(y)$
over the spatial interval $[0,x]$. Its variance obtained from Eq. \ref{FracBrownianvar}
\begin{eqnarray}
 {\mathbb E} \left( \left(\frac{B_H(  x)}{x} \right)^2 \right)   =  \vert   x \vert^{2H-2} \equiv \vert   x \vert^{2H'}
  \label{FracBrownianvarRatioo}
\end{eqnarray}
involves the same negative Hurst exponent $H'=H-1\in ]-1,0[ $ as the fractional noise $W_H(.)$.
When the field $\phi$ is scale-invariant with a negative Hurst exponent $H<0$ in dimension $d$,
the generalization of the ratio of Eq. \ref{FracBrownianAsAverage}
 then corresponds to the empirical magnetization $m_e$ of Eq. \ref{empime} 
 and ${\cal M}_R$ of Eq. \ref{EmpiricalMagnetization}
 when one uses the more general spatial-averaging-kernel $A_R(\vec x)$ 
of Eq. \ref{ARkernel}.

%%%%%%%%%%%%%%%%%%%%%%%%%%%%%%%%%%%%%%%%%%%%

\section{ Reminder on the Wiener-Ito orthogonal basis for functionals of the White Noise}

\label{app_WienerIto}

Since the Wiener-Ito chaos expansion based on multiple stochastic integrals 
has a long history in the mathematical literature \cite{MeyerYan,HuMeyer1988,ItoStrato,ReviewWick1993,ReviewWick2009}
but is not well-known in the physics literature,
it seems useful in the present Appendix to give a self-contained pedestrian introduction
for statistical physicists.

\subsection{ Multiple stochastic integrals of order $n$ involving the white noise $W(.)$  } 

For a real function $f_n(\vec x_1,\vec x_2,...,\vec x_n)$ symmetric with respect to its $n$ variables,
one wishes to analyze the properties of the stochastic integral
\begin{eqnarray} 
S_n(f_n)  \equiv  \langle f_n \vert W^{\otimes n} \rangle
= \int d^d \vec x_1   ... \int d^d \vec x_n f_n(\vec x_1,...,\vec x_n) 
 W(\vec x_1)   ... W(\vec x_n) 
  \label{Snbraket}
\end{eqnarray}
and of the Ito integral
\begin{eqnarray} 
I_n(f_n)  \equiv   \langle f_n \vert W^{[Ito]}_n \rangle
= \int d^d \vec x_1   ... \int d^d \vec x_n f_n(\vec x_1,...,\vec x_n) 
\left( \prod_{1 \leq i<j \leq n}  \theta( \vec x_i \ne \vec x_j) \right)
 W(\vec x_1)   ... W(\vec x_n)
  \label{Itonbraket}
\end{eqnarray}
where the additional functions $\theta( \vec x_i \ne \vec x_j) $ 
remove the possibility of coinciding points
\begin{eqnarray} 
W^{[Ito]}_n( \vec x_1, \vec x_2,.. \vec x_n) 
\equiv \left( \prod_{1 \leq i<j \leq n}  \theta( \vec x_i \ne \vec x_j) \right)
 W( \vec x_1) W( \vec x_2) ... W( \vec x_n)   \ \ \ \text{ for } \ n \geq 1
  \label{tensorsWn}
\end{eqnarray}

The difference between the two types of integrals can already be seen as the level of averaged values :
 the averaged value of the Ito integral $I_n(f_n) $ vanishes by construction for any $n \geq 1$
\begin{eqnarray} 
 {\mathbb E} \bigg( I_n(f_n) \bigg)  =  0
  \label{Itonbraketav}
\end{eqnarray}
while for even order $n=2l \geq 2 $, the averaged value of $S_{2l}(f_{2l})  $ of Eq. \ref{Snbraket} 
does not vanish : in the Wick theorem, there are 
$(2l-1)! (2l-3)! ... 1= \frac{(2l)!}{ 2^l \times l!}$ different possible pairings between the $(2l)$ positions
$(\vec x_1,..,\vec x_{2l})$ that produce all the same result as a consequence of the symmetry of the function $f_{2l}(\vec x_1,..,\vec x_{2l})$ and one obtains
\begin{eqnarray} 
 {\mathbb E} \bigg(S_{2l}(f_{2l}) \bigg)  
 =  \frac{(2l)!}{ l! 2^l  }
 \int d^d \vec x_1  \int d^d \vec x_2 ... \int d^d \vec x_l f_{2l}(\vec x_1,\vec x_1,\vec x_2,\vec x_2,...,\vec x_l,\vec x_l) 
  \label{Snbraketav}
\end{eqnarray}
Let us recall the simple examples $n=1,2$ before returning to the case of arbitrary $n$.

%%%%%%%%%%%%%%%%%%%%%%%%%%%%%%%%%%%%%

\subsubsection{ Stochastic integrals of order $n=1$} 

For $n=1$, there is no difference between Eqs \ref{Snbraket} and \ref{Itonbraket}
\begin{eqnarray} 
S_1(f_1) = I_1(f_1) = \langle f_1 \vert W \rangle = \int d^d \vec x f_1(\vec x) W(\vec x) 
  \label{Ito1braket}
\end{eqnarray}
and the variance reduces to
\begin{eqnarray} 
{\mathbb E} \bigg( I_1^2(f_1) \bigg) ={\mathbb E} \bigg( \langle f_1 \vert W \rangle \langle W \vert f_1 \rangle \bigg)
= \langle f_1  \vert f_1 \rangle = \int d^d \vec x f^2_1(\vec x)  
  \label{Ito1braketvar}
\end{eqnarray}
More generally, the correlation between the two integrals $I_1(f_1) $ and $I_1(g_1) $
associated to the two functions $f_1$ and $g_1$ is given by the scalar product
\begin{eqnarray} 
{\mathbb E} \bigg( I_1(f_1)I_1(g_1) \bigg) ={\mathbb E} \bigg( \langle f_1 \vert W \rangle \langle W \vert g_1 \rangle \bigg)
= \langle f_1  \vert g_1 \rangle = \int d^d \vec x f_1(\vec x)  g_1(\vec x) 
  \label{Ito1braketvarcorr}
\end{eqnarray}

%%%%%%%%%%%%%%%%%%%%%%%%%%%%%%

\subsubsection{ Stochastic integrals of order $n=2$} 

For $n=2$, one sees the difference between $I_2(f_2)$ with its vanishing averaged value of Eq \ref{Itonbraketav} and $S_2(f_2)$ with its non-vanishing averaged value of Eq. \ref{Snbraketav} for $l=1$
\begin{eqnarray} 
 {\mathbb E} \bigg(S_{2}(f_{2}) \bigg)  
 =   \int d^d \vec x_1   f_2(\vec x_1,\vec x_1) = {\rm Tr}[f_2]
  \label{S2braketav}
\end{eqnarray}
that reduces to the trace of the function $f_2$.

The correlation between the Ito integrals $I_2(f_2) $ and $I_2(g_2) $
associated to the two symmetric functions $f_2(\vec x_1,\vec x_2)=f_2(\vec x_2,\vec x_1)$ 
and $g_2(\vec x_1,\vec x_2)=g_2(\vec x_2,\vec x_1)$ 
 reads using the Wick theorem 
\begin{small}
\begin{eqnarray} 
&& {\mathbb E} \bigg( I_2(f_2)I_2(g_2) \bigg)   =  \int d^d \vec x_1 \int d^d \vec x_2 f_2(\vec x_1,\vec x_2) 
\theta( \vec x_1 \ne \vec x_2)
\int d^d \vec y_1 \int d^d \vec y_2 g_2(\vec y_1,\vec y_2)\theta( \vec y_1 \ne \vec y_2)
{\mathbb E} \bigg( W(\vec x_1) W(\vec x_2) W(\vec y_1) W(\vec y_2) 
\bigg)
\nonumber \\
&&  =  \int d^d \vec x_1 \int d^d \vec x_2 f_2(\vec x_1,\vec x_2) \theta( \vec x_1 \ne \vec x_2)
\int d^d \vec y_1 \int d^d \vec y_2 g_2(\vec y_1,\vec y_2)\theta( \vec y_1 \ne \vec y_2)
\left[ \delta^{(d)}(\vec x_1-\vec y_1) \delta^{(d)}(\vec x_2-\vec y_2) 
+ \delta^{(d)}(\vec x_1-\vec y_2) \delta^{(d)}(\vec x_2-\vec y_1) 
\right]
\nonumber \\
&&  = 2 \int d^d \vec x_1 \int d^d \vec x_2 \theta( \vec x_1 \ne \vec x_2) f_2(\vec x_1,\vec x_2) 
g_2(\vec x_1,\vec x_2)
= 2 \int d^d \vec x_1 \int d^d \vec x_2  f_2(\vec x_1,\vec x_2)  g_2(\vec x_1,\vec x_2)
= 2 \langle f_2 \vert g_2 \rangle
  \label{Ito2braketcorr}
\end{eqnarray}
\end{small}
where on the last line, one can forget the constraint $\theta( \vec x_1 \ne \vec x_2) $ 
that has zero-measure
for usual integrals that do not contain the white noise anymore.

In particular, the 
 variance of the Ito integral $I_2(f_2) $ reduces to
\begin{eqnarray} 
&& {\mathbb E} \bigg( I_2^2(f_2) \bigg)   
= 2 \int d^d \vec x_1 \int d^d \vec x_2  f^2_2(\vec x_1,\vec x_2)  
= 2 \langle f_2 \vert f_2 \rangle
  \label{Ito2braketvar}
\end{eqnarray}

%%%%%%%%%%%%%%%%%%%%%%%%%%%%%%%%%%%%%%%%%%%%%%

\subsection{ Orthogonality properties of the multiple Ito integrals $I_n(.)$}

\subsubsection{ Correlation between two Ito integrals $I_n(f_n)$ and $I_n(g_n)$ of arbitrary order $n$ }

The calculation of Eq. \ref{Ito2braketcorr} can be directly generalized to evaluate
the correlation between two Ito integrals $I_n(f_n)$ and $I_n(g_n)$
associated to two symmetric functions $f_n$ and $g_n$
of arbitrary order $n$ as follows
\begin{eqnarray} 
&& {\mathbb E} \bigg( I_n(f_n) I_n(g_n)\bigg)  
 = \int d^d \vec x_1  \int d^d \vec x_2 ... \int d^d \vec x_n f_n(\vec x_1,\vec x_2,...,\vec x_n) 
 \int d^d \vec y_1  \int d^d \vec y_2 ... \int d^d \vec y_n g_n(\vec y_1,\vec y_2,...,\vec y_n) 
\nonumber \\
&& \times \left( \prod_{1 \leq i<j \leq n}  \theta( \vec x_i \ne \vec x_j) \right)
\left( \prod_{1 \leq i<j \leq n}  \theta( \vec y_i \ne \vec y_j) \right)
{\mathbb E} \bigg( W(\vec x_1) W(\vec x_2)  ... W(\vec x_n)
 W(\vec y_1) W(\vec y_2)  ... W(\vec y_n) \bigg) 
  \label{Itonbraketvarcalcul}
\end{eqnarray}
In the evaluation of the averaged value ${\mathbb E} \bigg( W(\vec x_1) W(\vec x_2)  ... W(\vec x_n)
 W(\vec y_1) W(\vec y_2)  ... W(\vec y_n) \bigg) $ via the Wick theorem,
the constraints $\vec x_i \ne \vec x_j $ and $\vec y_i \ne \vec y_j $
yield that the only possible pairings are of the type $(x_i,y_{\sigma(i)})$ 
where $\sigma \in {\cal P}_n$ is one of the $n!$ permutations of $\{1,2,..,n\}$
and since $ f_n$ is symmetric, Eq. \ref{Itonbraketvarcalcul}
reduces to
\begin{small}
\begin{eqnarray} 
 {\mathbb E} \bigg(I_n(f_n) I_n(g_n) \bigg)  
&& =\int d^d \vec x_1   ... \int d^d \vec x_n f_n(\vec x_1,...,\vec x_n) 
 \int d^d \vec y_1   ... \int d^d \vec y_n g_n(\vec y_1,...,\vec y_n) 
 \sum_{ \sigma \in  {\cal P}_n} \prod_{i=1}^n \delta( \vec x_i - \vec y_{\sigma(i) })
 \nonumber \\
 && = \sum_{ \sigma \in  {\cal P}_n} 
 \int d^d \vec y_1  \int d^d \vec y_2 ... \int d^d \vec y_n f_n(\vec y_1,\vec y_2,...,\vec y_n) 
 g_n(\vec  y_{\sigma(1) },\vec  y_{\sigma(2) },...,\vec  y_{\sigma(n) }) 
\nonumber \\
&& =  n! \int d^d \vec y_1  \int d^d \vec y_2 ... \int d^d \vec y_n f_n(\vec y_1,\vec y_2,...,\vec y_n) 
g_n(\vec y_1,\vec y_2,...,\vec y_n)
 =  n! \langle f_n \vert g_n \rangle
  \label{Itonbraketcorr}
\end{eqnarray}
\end{small}
that generalizes Eq. \ref{Ito2braketcorr}.
In particular, the variance of $I_n(f_n) $ reduces to
\begin{eqnarray} 
 {\mathbb E} \bigg( I_n^2(f_n) \bigg)  
&&  =  n! \int d^d \vec y_1  \int d^d \vec y_2 ... \int d^d \vec y_n f_n^2(\vec y_1,\vec y_2,...,\vec y_n) 
 =  n! \langle f_n \vert f_n \rangle
  \label{Itonbraketvar}
\end{eqnarray}

%%%%%%%%%%%%%%%%%%%%%%%%%%%%%%%%%%%%%%%%%%%%%%

\subsubsection{ Vanishing correlations between Ito integrals $I_n(.)$ and $I_m(.)$ of different orders $n \ne m$ } 

From the previous computation, it is clear that
two Ito integrals $I_n(f_n) $ and $I_m(g_m) $ of different orders $n \ne m$ have zero correlation 
\begin{eqnarray} 
 {\mathbb E} \bigg( I_n(f_n)  I_m(g_m) \bigg)  
&& = \int d^d \vec x_1  \int d^d \vec x_2 ... \int d^d \vec x_n f_n(\vec x_1,\vec x_2,...,\vec x_n) 
 \int d^d \vec y_1  \int d^d \vec y_2 ... \int d^d \vec y_m g_m(\vec y_1,\vec y_2,...,\vec y_m) 
\nonumber \\
&& \times \left( \prod_{1 \leq i<j \leq n}  \theta( \vec x_i \ne \vec x_j) \right)
\left( \prod_{1 \leq i<j \leq m}  \theta( \vec y_i \ne \vec y_j) \right)
{\mathbb E} \bigg( W(\vec x_1) W(\vec x_2)  ... W(\vec x_n)
 W(\vec y_1) W(\vec y_2)  ... W(\vec y_m) \bigg) 
 \nonumber \\
 && 
 = 0 \ \ \text{ for} \ n \ne m
   \label{Itonm}
\end{eqnarray}
since there is no possible pairing in the Wick theorem for $n \ne m$ 
as a consequence of the constraints $\vec x_i \ne \vec x_j $ and $\vec y_i \ne \vec y_j $.

%%%%%%%%%%%%%%%%%%%%%%%%%%%%%%%%%%%%%%%%%%%%%%%

\subsubsection{ Conclusion on the orthogonality properties of the multiple Ito integrals $I_n(.)$}

It is convenient to supplement the Ito integrals $I_n(.)$ associated to symmetric functions $f_n$
of order $n=1,..+\infty$
by the values $I_{n=0}(f_0) $ where the function $f_0$ of zero variables reduces to a constant $f_0$
\begin{eqnarray} 
 I_{n=0}(f_0)=f_0
  \label{Itozero}
\end{eqnarray}
Then one can summarize the properties of vanishing averages in Eq. \ref{Itonbraketav} of $I_n(.)$ for any $n \geq 1$, of vanishing correlations between two Ito integrals of different orders of Eq. \ref{Itonm}
and of the correlations between two Ito integrals of the same order ofEq. \ref{Itonbraketcorr} by
\begin{eqnarray} 
 {\mathbb E} \bigg( I_n(f_n)  I_m(g_m) \bigg)  
 = \delta_{n,m} \ n! \langle f_n \vert g_n \rangle  
   \label{ItoOrthog}
\end{eqnarray}

%%%%%%%%%%%%%%%%%%%%%%%%%%%%%%%%%%%%%

\subsection{ Rephrasing as the orthogonality of the family $\vert W^{[Ito]}_n \rangle  $
of functionals of the white noise $W$}

It is useful to supplement the family $W^{[Ito]}_n( \vec x_1, \vec x_2,.. \vec x_n) $
of Eq. \ref{tensorsWn}
 by the constant unity for $n=0$
\begin{eqnarray} 
W^{[Ito]}_{n=0} =1 
  \label{tensorsW0}
\end{eqnarray}
Then the property of Eq. \ref{ItoOrthog} for the Ito integrals $I_.(.)$ associated to arbitrary functions $f_n$
and $g_m$ 
\begin{eqnarray} 
n! \langle f_n \vert g_n \rangle \ \delta_{n,m} 
= {\mathbb E} \bigg( \langle f_n \vert W^{[Ito]}_n \rangle \langle W^{[Ito]}_m \vert g_m \rangle  \bigg)  
 =\langle f_n \vert  {\mathbb E} \bigg(  \vert W^{[Ito]}_n \rangle \langle W^{[Ito]}_m \vert   \bigg) \vert g_m \rangle 
   \label{ItoOrthogWf}
\end{eqnarray}
can be rephrased into
\begin{eqnarray} 
n!  \ \delta_{n,m} =  {\mathbb E} \bigg(  \vert W^{[Ito]}_n \rangle \langle W^{[Ito]}_m \vert   \bigg) 
   \label{ItoOrthogW}
\end{eqnarray}
for the family $\vert W^{[Ito]}_n \rangle $
of functionals of the white noise $W$.

%%%%%%%%%%%%%%%%%%%%%%%%%%%%%%%%%%%%%%%%%%

\subsection{ Expansion of an arbitrary functional $F[W(.)]  $ of the white noise $W(.)$
on the orthogonal family $\vert W^{[Ito]}_n \rangle $ }

A functional $F[W(.)]$ of the white noise $W(.)$
can be expanded on the orthogonal complete family of functionals $\vert W^{[Ito]}_n \rangle  $
via a series of Ito integrals $ I_n(F_n)\equiv \langle F_n \vert W^{[Ito]}_n \rangle$
\begin{eqnarray} 
F[W(.)] && =  \sum_{n=0}^{+\infty} I_n(F_n) =    \sum_{n=0}^{+\infty} \langle F_n \vert W^{[Ito]}_n \rangle 
\nonumber \\
&& =  \sum_{n=0}^{+\infty} 
\int d^d \vec x_1 ... \vec x_n F_n( \vec x_1, \vec x_2,.. \vec x_n) 
\left( \prod_{1 \leq i<j \leq n}  \theta( \vec x_i \ne \vec x_j) \right)
 W(\vec x_1) W(\vec x_2)  ... W(\vec x_n)
  \label{ChaosExpansion}
\end{eqnarray}
 where the coefficients $\langle F_n \vert $ of this expansion
 can be computed in terms of the functional $F[W(.)] $
 using the orthogonal property of 
 Eq. \ref{ItoOrthogW}
 \begin{eqnarray} 
  {\mathbb E} \bigg(  F[W(.)]  \langle W^{[Ito]}_m \vert   \bigg) = 
  \sum_{n=0}^{+\infty} \langle F_n \vert {\mathbb E} \bigg(  \vert W^{[Ito]}_n \rangle \langle W^{[Ito]}_m \vert   \bigg) 
  = m! \langle F_m \vert
   \label{FItoOrthogW}
\end{eqnarray}
i.e.  the real-space symmetric functions $F_m( \vec y_1, \vec y_2,.. \vec y_m)$
 can be computed from the correlations between the functional $F[W(.)] $ 
 and the functional $W^{[Ito]}_m( \vec y_1, \vec y_2,.. \vec y_m)$ of Eq. \ref{tensorsWn}
 via
 \begin{eqnarray} 
F_m( \vec y_1, \vec y_2,.. \vec y_m)
&& = \frac{1}{m!}  {\mathbb E} \bigg( F[W(.)] W^{[Ito]}_m(\vec y_1, \vec y_2,.. \vec y_m) \bigg) 
\nonumber \\
&& =\frac{1}{m!} \left( \prod_{1 \leq i<j \leq m}  
 \theta( \vec y_i \ne \vec y_j) 
\right)  {\mathbb E} \left( F[W(.)] 
  W( \vec y_1) W( \vec y_2) ... W( \vec y_m)  \right)
  \label{ChaosExpansionCoefs}
\end{eqnarray}
while $F_{n=0}$ reduces to the averaged value of the functional $F[W(.)] $
 \begin{eqnarray} 
F_0=  {\mathbb E} \bigg( F[W(.)] \bigg) 
  \label{ChaosExpansionCoefsn0}
\end{eqnarray}

In particular, the Wiener-Ito chaos-expansion of Eq. \ref{ChaosExpansion} can be used to 
compute the variance of the functional $F[W(.)] $
using Eq. \ref{ItoOrthog}
\begin{eqnarray} 
 {\mathbb E} \bigg( \left(F[W(.)] -{\mathbb E}(F[W(.)]) \right)^2 \bigg)
&& =   \sum_{n=1}^{+\infty}\sum_{m=1}^{+\infty}  {\mathbb E} \bigg(I_n(F_n) I_m(F_m)\bigg)
\nonumber \\
&&=  \sum_{n=1}^{+\infty} n! \langle F_n \vert F_n \rangle 
  \label{ChaosExpansionVariance}
\end{eqnarray}
in terms of the coefficients $F_n$ of Eq. \ref{ChaosExpansionCoefs}

%%%%%%%%%%%%%%%%%%%%%%%%

\subsection{ Expansion of the stochastic integral $S_n(f_n)$ of order $n$ in terms of Ito integrals 
$I_{n-2l}(. )$ with $0 \leq l \leq \frac{n}{2}$ }

Let us write the Wiener-Ito chaos-expansion of Eq. \ref{ChaosExpansion}
for the case where the functional  $F[W(.)]$ is the stochastic integral $S_n(f_n)$ of order $n$ 
of Eq. \ref{Snbraket}
\begin{eqnarray} 
S_n(f_n)  && \equiv \int d^d \vec x_1  \int d^d \vec x_2 ... \int d^d \vec x_n f_n(\vec x_1,\vec x_2,...,\vec x_n) 
 W(\vec x_1) W(\vec x_2)  ... W(\vec x_n) 
\nonumber \\
&& =  \sum_{m=0}^{+\infty} I_m( s^{[f_n]}_m)
\equiv \sum_{m=0}^{+\infty} \langle s^{[f_n]}_m \vert W^{[Ito]}_m \rangle
  \label{Strato}
\end{eqnarray}
where the coefficients
$s^{[f_n]}_m$ are given by Eq. \ref{ChaosExpansionCoefs}
 \begin{eqnarray} 
&& s^{[f_n]}_m( \vec y_1, \vec y_2,.. \vec y_m) 
   =\frac{1}{m!} \left( \prod_{1 \leq i<j \leq m}  
 \theta( \vec y_i \ne \vec y_j) 
\right)  {\mathbb E} \left( S_n(f_n)
  W( \vec y_1) W( \vec y_2) ... W( \vec y_m)  \right)
   \label{StratoCoefs} \\
&&  =\frac{1}{m!} \left( \prod_{1 \leq i<j \leq m}   \theta( \vec y_i \ne \vec y_j) \right) 
\int d^d \vec x_1  \int d^d \vec x_2 ... \int d^d \vec x_n f_n(\vec x_1,\vec x_2,...,\vec x_n)
 {\mathbb E} \left( W(\vec x_1) W(\vec x_2)  ... W(\vec x_n)
  W( \vec y_1) W( \vec y_2) ... W( \vec y_m)  \right) 
  \nonumber  
\end{eqnarray}

%%%%%%%%%%%%%%%%%%%%%%%%%%%%%%

\subsubsection{ Computation of the coefficients $s^{[f_n]}_m( \vec y_1, \vec y_2,.. \vec y_m) $ of Eq. \ref{StratoCoefs}} 

In Eq. \ref{StratoCoefs},
 the averaged value involves the white noise $W(.)$ at $(n+m)$ 
 positions $x_{1\leq i \leq n}$ and $y_{1 \leq j \leq m}$,
where the $n$ positions $y_{1 \leq j \leq m}$ are given and distincts $( \vec y_i \ne \vec y_j)  $, 
while the $n$ positions $x_{1\leq i \leq n}$ are integrated over without any constraints,
with the symmetric function $f_n( \vec x_1, \vec x_2,.. \vec x_n) $.
The Wick theorem thus gives a non-vanishing result only if the difference  $n-m=2l \geq 0$ is even and positive :

(i) The $m=n-2l$ distincts positions $(y_1,..,y_m)$ 
can be paired with $m$ values $(x_{i_1},..,x_{i_m})$ that have to be chosen among the $n$ values
$x_{1\leq i \leq n}$ with the binomial coefficient $\frac{n!}{m!(n-m)!} $, but
there are $m!$ equivalent permutations, and one can use the symmetry of $f_n$
to relabel them $(x_1,..,x_m)$ in order to summarize these contractions by $x_i=y_i$ pour $1 \leq i \leq m$.

(ii) The remaining even number $n-m=2l$ of positions $x_{m+1}...,x_{n=m+2l}$ have to be 
paired into $l$ pairs,
so there are $(2l-1)! (2l-3)! ... 1= \frac{(2l)!}{ 2^l \times l!}$ possibilities,
and one can use the symmetry of $f_n$
to relabel them in order to summarize these contractions by
$x_{m+2k}=x_{m+2k-1}$ pour $k=1,2,..,l$.

Putting everything together,
the coefficients $s^{[f_n]} _{m=n-2l}( \vec y_1, \vec y_2,.. \vec y_{n-2l})$ of Eq. \ref{StratoCoefs}
can be computed in terms of the function $f_n$ via
  \begin{eqnarray} 
 s^{[f_n]}_{n-2l}( \vec y_1, ... ,\vec y_{n-2l}) 
&& =    \frac{n!}{ l! (n-2l)! 2^l  }  \left( \prod_{1 \leq i<j \leq {n-2l}}   \theta( \vec y_i \ne \vec y_j) \right)
\int d^d \vec z_1 ... \int d^d\vec z_l f_n( \vec y_1, \vec y_2,.. \vec y_{n-2l},  
\vec z_1,\vec z_1, \vec z_2,\vec z_2,...,\vec z_l, \vec z_l) 
\nonumber \\
&& \equiv    \frac{n!}{ l! (n-2l)! 2^l  } \left( \prod_{1 \leq i<j \leq {n-2l}}   \theta( \vec y_i \ne \vec y_j) \right)
{\rm Tr}_{2l}[f_n]( \vec y_1, ... ,\vec y_{n-2l}) 
  \label{StratoCoefsCalcul}
\end{eqnarray}
where the notation $ {\rm Tr}_{2l}[f_n]$ represents the function of $(n-2l)$ variables
  \begin{eqnarray} 
&& {\rm Tr}_{2l}[f_n]( \vec x_1, ... ,\vec x_{n-2l}) 
 \equiv  \left( \prod_{1 \leq i<j \leq {n-2l}}   \theta( \vec x_i \ne \vec x_j) \right)
\int d^d \vec z_1 ... \int d^d\vec z_l f_n( \vec x_1, \vec x_2,.. \vec x_{n-2l},  
\vec z_1,\vec z_1, \vec z_2,\vec z_2,...,\vec z_l, \vec z_l) 
 \nonumber \\
 && = \left( \prod_{1 \leq i<j \leq {n-2l}}   \theta( \vec x_i \ne \vec x_j) \right)
\int d^d \vec x_{(n-2l)+1} ...  \int d^d\vec x_n 
f_n( \vec x_1, \vec x_2,.. \vec x_{n-2l},  
\vec x_{n-2l+1},... \vec x_n) \prod_{k=0}^{l-1} \delta^{(d)} (\vec x_{n-2k} - \vec x_{n-2k-1} ) 
\ \ \ \ \ \ 
  \label{traceofOrderk}
\end{eqnarray}
where the last $(2l)$ variables of the symmetric function $f_n$ of n variables are 
'traced over' into $l$ pairs.

Since the coefficients $s^{[f_n]}_{n-2l}( \vec y_1, ... ,\vec y_{n-2l}) $ will be used to compute Ito integrals $ I_m( s^{[f_n]}_m) $ that already contain the non-coinciding constraints $\left( \prod_{1 \leq i<j \leq {n-2l}}   \theta( \vec y_i \ne \vec y_j) \right) $, one can drop these constraints in Eq. \ref{StratoCoefsCalcul}
to obtain
 \begin{eqnarray} 
\langle s^{[f_n]}_m \vert =    \frac{n!}{ l! (n-2l)! 2^l  } \langle {\rm Tr}_{2l}[f_n] \vert
  \label{StratoCoefsbra}
\end{eqnarray}

%%%%%%%%%%%%%%%%%%%%%%%%%%%%%

\subsubsection{ Explicit Wiener-Ito chaos-expansion of the stochastic integral $S_n(f_n)$  in terms of Ito integrals 
$I_{n-2l}(. )$ with $0 \leq l \leq \frac{n}{2}$} 

Plugging the coefficients computed in Eq. \ref{StratoCoefsbra}
into the expansion of Eq. \ref{Strato}
yields the finite number of terms parametrized by $0 \leq l \leq \frac{n}{2} $
\begin{eqnarray} 
S_n(f_n)  && =  \sum_{0 \leq l \leq \frac{n}{2} }   I_{n-2l}( s^{[f_n]}_{n-2l})
=  \sum_{0 \leq l \leq \frac{n}{2} } \langle s^{[f_n]}_{n-2l} \vert W^{[Ito]}_{n-2l} \rangle
\nonumber \\
&& =  \sum_{0 \leq l \leq \frac{n}{2} }   \frac{n!}{ l! (n-2l)! 2^l  }  
\langle {\rm Tr}_{2l}[f_n] \vert W^{[Ito]}_{n-2l} \rangle  
=  \sum_{0 \leq l \leq \frac{n}{2} }   \frac{n!}{ l! (n-2l)! 2^l  }  
 I_{n-2l}(  {\rm Tr}_{2l}[f_n])
  \label{Stratofinite}
\end{eqnarray}
where the first term $l=0$ is the Ito integral $I_{n}( f_n) $ of order $n$ associated to $f_n$,
while the other terms $l=1,2,..$ involve the Ito integrals $I_{n-2l}(  {\rm Tr}_{2l}[f_n]) $ of lower orders associated to  
partial traces of $f_n$ introduced in Eq. \ref{traceofOrderk}
\begin{eqnarray} 
S_n(f_n) &&  = \langle f_n \vert W^{[Ito]}_{n} \rangle
+ \frac{n (n-1)}{  2  } \langle {\rm Tr}_{2}[f_n] \vert W^{[Ito]}_{n-2} \rangle
+  \frac{n (n-1) (n-2) (n-3)}{ 8  } \langle {\rm Tr}_{4}[f_n] \vert W^{[Ito]}_{n-4} \rangle+...
\nonumber \\
&& = I_n( f_n)+ \frac{n (n-1)}{  2  } I_{n-2}( {\rm Tr}_2[f_n])
+  \frac{n (n-1) (n-2) (n-3)}{ 8  }I_{n-4}(  {\rm Tr}_{4}[f_n])+...
  \label{Stratofinitefirst}
\end{eqnarray}

When $n$ is even, the last contribution associated to $l=\frac{n}{2}$ 
corresponds to the constant $I_0( s^{[f_n]}_{0}) $ that coincides with the averaged value ${\mathbb E} \bigg(S_{n}(f_{n}) \bigg)  $
already mentioned in Eq. \ref{Snbraketav}
\begin{eqnarray} 
l=\frac{n}{2} : \ \  I_0( s^{[f_n]}_{0})
&& =   \frac{n!}{ \left(\frac{n}{2} \right)!  2^{\frac{n}{2} } }  I_{0}(  {\rm Tr}_{n}[f_n])
=  \frac{n!}{ \left(\frac{n}{2} \right)!  2^{\frac{n}{2} } } 
 \int d^d \vec x_1  \int d^d \vec x_2 ... \int d^d \vec x_{\frac{n}{2}} f_{n}(\vec x_1,\vec x_1,\vec x_2,\vec x_2,...,\vec x_{\frac{n}{2}},\vec x_{\frac{n}{2}}) 
\nonumber \\
&& =  {\mathbb E} \bigg(S_{n}(f_{n}) \bigg) 
  \label{Stratofinitezero}
\end{eqnarray}
as it should for consistency since the averaged value of all the other Ito integrals that appear in the expansion
of Eq. \ref{Stratofinite} vanish.
So one can also rewrite Eq. \ref{Stratofinite} for the difference between $S_n(f_n) $ and its averaged value
${\mathbb E} \bigg(S_{n}(f_{n}) \bigg) $
\begin{eqnarray} 
S_n(f_n) -  {\mathbb E} \bigg(S_{n}(f_{n}) \bigg)   && 
% =  \sum_{0 \leq l < \frac{n}{2} }   I_{n-2l}( s^{[f_n]}_{n-2l})
=  \sum_{0 \leq l < \frac{n}{2} }   \frac{n!}{ l! (n-2l)! 2^l  }  I_{n-2l}(  {\rm Tr}_{2l}[f_n])
\nonumber \\
&& = I_n( f_n)+ \frac{n (n-1)}{  2  } I_{n-2}( {\rm Tr}_1[f_n])
+  \frac{n (n-1) (n-2) (n-3)}{ 8  }I_{n-4}(  {\rm Tr}_{4}[f_n])+...
  \label{Stratofinitediff}
\end{eqnarray}
The variance formula of Eq. \ref{ChaosExpansionVariance} 
then reads for the present case $F[W(.)]=S_n(f_n) $
\begin{eqnarray} 
 {\mathbb E} \bigg( \left(S_n(f_n) -{\mathbb E}(S_n(f_n)) \right)^2 \bigg)
&& =   \sum_{m=1}^{+\infty} m! \langle s^{[f_n]}_m \vert s^{[f_n]}_m \rangle 
=    \sum_{0 \leq l < \frac{n}{2} } (n-2l)! \left( \frac{n!}{ l! (n-2l)! 2^l  } \right)^2 
\langle {\rm Tr}_{2l}[f_n] \vert {\rm Tr}_{2l}[f_n] \rangle 
\nonumber \\
&&=  \sum_{0 \leq l < \frac{n}{2} } \frac{1}{(n-2l)! } \left( \frac{n!}{ l!  2^l  } \right)^2 
\langle {\rm Tr}_{2l}[f_n] \vert {\rm Tr}_{2l}[f_n] \rangle 
  \label{ChaosExpansionVarianceSn}
\end{eqnarray}

%%%%%%%%%%%%%%%%%%%%%%%%%%%%%%%%%%%

\subsubsection{ Inversion formula : Ito integral $I_n(f_n)$ in terms of
stochastic integrals $S_{n-2l}(. )$ with $0 \leq l \leq \frac{n}{2}$ } 

The inclusion–exclusion principle yields that the Wiener-Ito chaos expansion of Eq. \ref{Stratofinite}
can be inverted  to obtain the Ito integral $I_n(f_n) $ of order $n$
\begin{eqnarray} 
I_n(f_n)  && 
=  \sum_{0 \leq l \leq \frac{n}{2} }  (-1)^l \frac{n!}{ l! (n-2l)! 2^l  }  S_{n-2l}(  {\rm Tr}_{2l}[f_n])
=  \sum_{0 \leq l \leq \frac{n}{2} }  (-1)^l \frac{n!}{ l! (n-2l)! 2^l  } 
\langle {\rm Tr}_{2l}[f_n] \vert W^{\otimes(n-2l) } \rangle
\nonumber \\
&& = S_n( f_n)- \frac{n (n-1)}{  2  } S_{n-2}( {\rm Tr}_1[f_n])
+  \frac{n (n-1) (n-2) (n-3)}{ 8  }S_{n-4}(  {\rm Tr}_{4}[f_n])+...
  \label{StratofiniteInversion}
\end{eqnarray}
in terms of the
stochastic integrals $S_{n-2l}({\rm Tr}_{2l}[f_n])$ with $0 \leq l \leq \frac{n}{2}$,
where the coefficients are the same as in Eq. \ref{Stratofinite}
except for the additional factors $(-1)^l$.

Note that the coefficients that appear in Eq \ref{Stratofinite}
and \ref{StratofiniteInversion} are exactly the same as 
those that appear between the 'probabilist' Hermite polynomials $H_n(u) $
that are orthogonal with respect to the Gaussian probability $\frac{e^{- \frac{u^2}{2} }}{\sqrt{2 \pi}} $
\begin{eqnarray} 
\int_{-\infty}^{+\infty} du \frac{e^{- \frac{u^2}{2} }}{\sqrt{2 \pi}} H_n(u) H_m(u)  = \delta_{n,m} n!
  \label{HermiteOrtho}
\end{eqnarray}
and the powers $u^n$ 
\begin{eqnarray} 
u^n  &&   =  \sum_{0 \leq l \leq \frac{n}{2} }   \frac{n!}{ l! (n-2l)! 2^l  }   H_{n-2l}(  u)
\nonumber \\
H_n(u) && = \sum_{0 \leq l \leq \frac{n}{2} }  (-1)^l \frac{n!}{ l! (n-2l)! 2^l  }  u^{n-2l}
  \label{PowerHermite}
\end{eqnarray}
with the first polynomials
\begin{eqnarray} 
H_0(u)  &&   = 1
\nonumber \\
H_1(u) && = u
\nonumber \\
H_2(u)  &&   = u^2-1
\nonumber \\
H_3(u) && = u^3-3 u
\nonumber \\
H_4(u) && = u^4-6 u^2+3
  \label{Hermite}
\end{eqnarray}

%%%%%%%%%%%%%%%%%%%%%%%%%%%%%%%%%%%

\subsubsection{ Examples for small values of $n=2,3,4$ }

For $n=1$, Eq. \ref{Stratofinite} yields that the two integrals coincide $S_1(f_1) =I_1( f_1) $
as already mentioned in Eq. \ref{Ito1braket}.

For $n=2$, Eq. \ref{Stratofinitediff} and Eq. \ref{StratofiniteInversion}
reduce to
\begin{eqnarray} 
 S_2(f_2)  -  {\mathbb E} \bigg(S_{2}(f_{2}) \bigg)  = I_2( f_2) 
  \label{Stratofinitediff2}
\end{eqnarray}
so that the Ito integral $ I_2( f_2)$ directly represents the difference
between $S_2(f_2) $ and its averaged value $  {\mathbb E} \bigg(S_{2}(f_{2}) \bigg)$
 given in Eq. \ref{S2braketav}.
 As a consequence, their variances coincide and are given by Eq. \ref{Ito2braketvar}.

For $n=3$, Eqs \ref{Stratofinite} and \ref{StratofiniteInversion}
involve only two terms $l=0,1$ and agree since $S_1=I_1$
\begin{eqnarray} 
S_3(f_3)  && = I_3( f_3)+   3  I_{1}(  {\rm Tr}_2[f_3]) \equiv  I_3( f_3)+   3  I_{1}(  f_1)
\nonumber \\
I_3(f_3)  &&  = S_3( f_3)- 3 S_1( {\rm Tr}_2[f_3]) \equiv  S_3( f_3)-3   S_{1}(  f_1)
  \label{Stratofinite3}
\end{eqnarray}
where  
the function $f_1\equiv  {\rm Tr}_2[f_3]$ of the single variable $\vec y$
can be computed from $f_3$ via Eq. \ref{traceofOrderk}
  \begin{eqnarray} 
f_1(\vec y) \equiv  {\rm Tr}_2[f_3]( \vec y) 
=\int d^d \vec z  f_3( \vec y, \vec z, \vec z) 
  \label{traceofOrderk31}
\end{eqnarray}
The variance formula of Eq. \ref{ChaosExpansionVariance} or \ref{ChaosExpansionVarianceSn} yields
\begin{eqnarray} 
 {\mathbb E} \bigg( \left(S_3(f_3) -{\mathbb E}(S_3(f_3)) \right)^2 \bigg)
&& =   6  \langle f_3 \vert f_3 \rangle + 9 \langle f_1 \vert f_1 \rangle
\nonumber \\
&& = 6 \int d^d \vec x_1   \int d^d \vec x_2  \int d^d \vec x_3 f_3^2(\vec x_1 , \vec x_2  , \vec x_3 )
+ 9 \int d^d \vec x \left[ \int d^d \vec z  f_3( \vec x, \vec z, \vec z)\right]^2
  \label{ChaosExpansionVariance3}
\end{eqnarray}
in agreement with a direct calculation based on the possible pairings in the Wick theorem between 6 positions.

For $n=4$, Eq. \ref{Stratofinitediff} and Eq. \ref{StratofiniteInversion}
involves only three terms $l=0,1,2$
\begin{eqnarray} 
S_4(f_4) && = I_4( f_4)+ 6 I_{2}( {\rm Tr}_2[f_4]) +  3I_0(  {\rm Tr}_{4}[f_4])
\equiv    I_4( f_4)+ 6 I_{2}( f_2) +3 f_0
\nonumber \\
I_4(f_4)  &&  = S_4( f_4)- 6 S_2( {\rm Tr}_2[f_4])+  3S_0(  {\rm Tr}_{4}[f_4])
\equiv S_4( f_4)- 6 S_2( f_2)+  3 f_0
  \label{Stratofinite4}
\end{eqnarray}
where  
the function $f_2 \equiv  {\rm Tr}_2[f_4]$ of two variables
can be computed from $f_4$ via Eq. \ref{traceofOrderk}
  \begin{eqnarray} 
f_2(\vec y_1,\vec y_2) \equiv  {\rm Tr}_2[f_4](\vec y_1,\vec y_2) 
=\int d^d \vec z  f_4( \vec y_1,\vec y_2, \vec z, \vec z) 
  \label{traceofOrderk42}
\end{eqnarray}
while the constant $f_0\equiv  {\rm Tr}_1[f_4] $ reads
  \begin{eqnarray} 
f_0 \equiv  {\rm Tr}_4[f_4]
=\int d^d \vec y \int d^d \vec z  f_4( (\vec y,\vec y), \vec z, \vec z) 
  \label{traceofOrderk44}
\end{eqnarray}

The variance formula of Eq. \ref{ChaosExpansionVariance} or \ref{ChaosExpansionVarianceSn} yields
\begin{eqnarray} 
&& {\mathbb E} \bigg( \left(S_4(f_4) -{\mathbb E}(S_4(f_4)) \right)^2 \bigg)
= 
 24 \langle f_4 \vert f_4 \rangle  
 +  72 \langle f_2 \vert f_2 \rangle
 \nonumber \\
&& = 24 \int d^d \vec x_1   \int d^d \vec x_2  \int d^d \vec x_3 \int d^d \vec x_3
f_4^2(\vec x_1 , \vec x_2  , \vec x_3, \vec x_4 )
+ 72 \int d^d \vec x_1  \int d^d \vec x_2 \left[ \int d^d \vec z  f_4( \vec x_1,\vec x_2, \vec z, \vec z)\right]^2
  \label{ChaosExpansionVarianceS4}
\end{eqnarray}
in agreement with a direct calculation based on the possible pairings in the Wick theorem between 8 positions.

%%%%%%%%%%%%%%%%%%%%%%%%

\subsection{ Wiener-Ito chaos-expansion of the exponential functional 
 $e^{\lambda I_1(h)}=e^{\lambda \langle h \vert W \rangle} = e^{ \lambda\int d^d \vec x h(\vec x) W(\vec x) }  $ 
} 

When the functional $F [W(.)] $ of the white noise $W(.)$
is the exponential functional
\begin{eqnarray} 
e^{\lambda I_1(h)} = e^{\lambda \langle h \vert W \rangle} = e^{\lambda \int d^d \vec x h(\vec x) W(\vec x) } 
  \label{ExpoExample}
\end{eqnarray}
the averaged value corresponds to the generating function of the random variable $ I_1(h)$
with its simple explicit expression
\begin{eqnarray} 
   {\mathbb E} \bigg( e^{\lambda I_1(h)} \bigg)= {\mathbb E} \bigg( e^{\lambda \langle h \vert W \rangle} \bigg)
= e^{ \frac{\lambda^2}{2} \langle h \vert h \rangle  } =   e^{ \frac{\lambda^2}{2} \int d^d x h^2(\vec x)  }  
     \label{ChaosExpansionExpmomentk}
\end{eqnarray}
It is thus interesting to compare the usual expansion in terms of stochastic integrals $S_n(.)$
and the Wiener-Ito chaos-expansion in terms of Ito integrals $I_n(.)$ in the two following subsections.

%%%%%%%%%%%%%%%%%%%%%%%%%%%%%%%%%%%%%%

\subsubsection{ Usual expansion of the exponential functional $e^{\lambda I_1(h)} $ in terms of stochastic integrals $S_n(.)$}

The usual expansion of the exponential functional  
\begin{eqnarray} 
e^{\lambda I_1(h)} = e^{\lambda \int d^d \vec x h(\vec x) W(\vec x) } 
&& =  1+ \sum_{n=1}^{+\infty} \frac{\lambda^n}{n!} \int d^d \vec x_1 ... \int d^d \vec x_n
h(\vec x_1)  ...  h(\vec x_n) W(\vec x_1) ... W(\vec x_n)
\nonumber \\
&& = 1+ \sum_{n=1}^{+\infty} \frac{\lambda^n}{n!} S_n( h^{\otimes n} )
  \label{UsualExpansionExp}
\end{eqnarray}
involves the stochastic integrals $S_n(.)$ of Eq. \ref{Snbraket}.
So here the generating function ${\mathbb E} \bigg( e^{\lambda I_1(h)} \bigg) $
involves the resummation 
of the averaged values ${\mathbb E} \bigg(S_{2l}(f_{2l}) \bigg)  $
of Eq. \ref{Snbraketav} for the even stochastic integrals
\begin{eqnarray} 
{\mathbb E} \bigg( e^{\lambda I_1(h)} \bigg)
 && =
 1+ \sum_{l=1}^{+\infty} \frac{\lambda^{2l}}{(2l)!}  {\mathbb E} \bigg(S_{2l}(h^{\otimes (2l)} )\bigg)
\nonumber \\
&& = 1+ \sum_{l=1}^{+\infty} 
 \frac{\lambda^{2l}}{ l! 2^l  }
 \int d^d \vec x_1  \int d^d \vec x_2 ... \int d^d \vec x_l h^2(\vec x_1)  h^2(\vec x_2)...h^2(\vec x_l)
 \nonumber \\
&& =\sum_{l=0}^{+\infty} 
 \frac{1}{ l!  } \left[ \frac{\lambda^2}{2} \int d^d \vec x   h^2(\vec x)  \right]^l 
   =   e^{ \frac{\lambda^2}{2} \int d^d x h^2(\vec x)  }  =  e^{ \frac{\lambda^2}{2} \langle h \vert h \rangle  } 
  \label{UsualExpansionExpAv}
\end{eqnarray}
in agreement with the direct evaluation of Eq. \ref{ChaosExpansionExpmomentk}.

%%%%%%%%%%%%%%%%%%%%%%%%%%%%%%%%%%%%%

\subsubsection{ Wiener-Ito Chaos-expansion of the exponential functional $e^{\lambda I_1(h)} $
in terms of Ito integrals $I_n(.)$ }

The stochastic integrals $S_n(h^{\otimes n} ) $ appearing in the expansion of Eq. \ref{UsualExpansionExp} can be rewritten in terms of Ito integrals via 
the Wiener-Ito Chaos-expansion of Eq. \ref{Stratofinite}
\begin{eqnarray} 
S_n(h^{\otimes n})  
=  \sum_{0 \leq l \leq \frac{n}{2} }   \frac{n!}{ l! (n-2l)! 2^l  }  I_{n-2l}( g_{n-2l} = {\rm Tr}_{2l}[h^{\otimes n}])
  \label{StratofiniteCasExp}
\end{eqnarray}
where the functions $g_{n-2l} ={\rm Tr}_{2l}[h^{\otimes n}] $ 
are given by Eq. \ref{traceofOrderk} 
  \begin{eqnarray} 
 g_{n-2l}( \vec y_1, ... ,\vec y_{n-2l}) 
&& = h( \vec y_1) h(\vec y_2)... h( \vec y_{n-2l})
\int d^d \vec z_1 ... \int d^d\vec z_l 
h^2(\vec z_1) h^2( \vec z_2)...h^2(\vec z_l) 
\nonumber \\
&& = h( \vec y_1) h(\vec y_2)... h( \vec y_{n-2l})
 \left[  \int d^d \vec x   h^2(\vec x)  \right]^l
\nonumber \\
&& = \left[  \langle h \vert h \rangle \right]^l h( \vec y_1) h(\vec y_2)... h( \vec y_{n-2l})
  \label{traceofOrderkExp}
\end{eqnarray}
i.e. one recognizes the function $ h^{\otimes (n-2l)}$ 
up to the multiplicative factor $\left[  \langle h \vert h \rangle \right]^l $
  \begin{eqnarray} 
 g_{n-2l}  = \left[  \langle h \vert h \rangle \right]^l  h^{\otimes (n-2l)}
  \label{traceofOrderkExpSimply}
\end{eqnarray}

The expansion of Eq. \ref{StratofiniteCasExp} becomes
\begin{eqnarray} 
S_n(h^{\otimes n})  
&& =  \sum_{0 \leq l \leq \frac{n}{2} }   \frac{n!}{ l! (n-2l)! 2^l  } 
 I_{n-2l}( g_{n-2l}  =  \left[  \langle h \vert h \rangle \right]^l  h^{\otimes (n-2l)} )
 \nonumber \\
 && =  \sum_{0 \leq l \leq \frac{n}{2} }   \frac{n!}{ l! (n-2l)! 2^l  }  \left[  \langle h \vert h \rangle \right]^l
 I_{n-2l}(  h^{\otimes (n-2l)} )
  \label{StratofiniteCasExpSimpli}
\end{eqnarray}
and can be plugged into Eq. \ref{UsualExpansionExp}
to obtain 
\begin{eqnarray} 
e^{\lambda I_1(h)}
&& =   \sum_{n=0}^{+\infty} \frac{\lambda^n}{n!} S_n( h^{\otimes n} )
=  \sum_{n=0}^{+\infty}   \sum_{0 \leq l \leq \frac{n}{2} }   \frac{\lambda^n \left[  \langle h \vert h \rangle \right]^l}{ l! (n-2l)! 2^l  }  
 I_{n-2l}(  h^{\otimes (n-2l)} )
 \nonumber \\
&& =  \sum_{l=0}^{+\infty}  \frac{\lambda^{2l} \left[  \langle h \vert h \rangle \right]^l}{ l!  2^l  }
\sum_{n=2l}^{+\infty}    \frac{\lambda^{n-2l} }{  (n-2l)!   }   I_{n-2l}(  h^{\otimes (n-2l)} )
 \nonumber \\
&& =  \sum_{l=0}^{+\infty}  \frac{ \left[ \frac{\lambda^2}{2} \langle h \vert h \rangle \right]^l}{ l!    }
\sum_{m=0}^{+\infty}    \frac{\lambda^{m} }{  m!   }   I_{m}(  h^{\otimes m} )
=  e^{ \frac{\lambda^2}{2} \langle h \vert h \rangle  } \sum_{m=0}^{+\infty}    \frac{\lambda^{m} }{  m!   }   I_{m}(  h^{\otimes m} )
  \label{UsualExpansionExpversIto}
\end{eqnarray}
Here the generating function ${\mathbb E} \bigg( e^{\lambda I_1(h)} \bigg) $
 corresponds to the contribution $m=0$ 
\begin{eqnarray} 
  {\mathbb E} \bigg( e^{\lambda I_1(h)} \bigg) =e^{ \frac{\lambda^2}{2} \langle h \vert h \rangle  } 
    I_{m=0}(  f_0=1 )
 =   e^{ \frac{\lambda^2}{2} \langle h \vert h \rangle  }  
     \label{ChaosExpansionExpmomentkm0}
\end{eqnarray}
that appears as a global prefactor in Eq. \ref{UsualExpansionExpversIto}.
So it is the ratio between the exponential functional $e^{\lambda I_1(h)}$ and its averaged value
${\mathbb E} \bigg( e^{\lambda I_1(h)} \bigg)$ that has a very simple 
Wiener-Ito chaos-expansion
\begin{eqnarray} 
\frac{ e^{\lambda I_1(h)} }{ {\mathbb E} \bigg( e^{\lambda I_1(h)} \bigg)}
=  e^{\lambda \langle h \vert W \rangle - \frac{\lambda^2}{2} \langle h \vert h \rangle } 
&& =   \sum_{m=0}^{+\infty}    \frac{\lambda^{m} }{  m!   }   I_{m}(  h^{\otimes m} )
  \label{UsualExpansionExpversItoRatio}
\end{eqnarray}

%%%%%%%%%%%%%%%%%%%%%%%%%%%%%%%%%%%%%%%%%%%

\end{document}